\documentclass[12pt,preprint]{aastex}

\shorttitle{Massive Binaries}

\shortauthors{Massey et al.}

\begin{document}

\title{Photometric and Spectroscopic Studies of Massive Binaries in the Large Magellanic Cloud. I. Introduction and Orbits for Two Detached Systems: Evidence for a Mass Discrepancy?\altaffilmark{1}}

\slugcomment{Accepted by the ApJ Jan 16, 2012}

\author{Philip Massey\altaffilmark{2}, Nidia I. Morrell\altaffilmark{3}, Kathryn F. Neugent\altaffilmark{2},
Laura R. Penny\altaffilmark{4}, Kathleen DeGioia-Eastwood\altaffilmark{5}, and  Douglas R. Gies\altaffilmark{6}}

\altaffiltext{1}{This paper includes data gathered with the 6.5 meter Magellan and 1.0 meter Swope Telescopes located at Las Campanas Observatory, Chile, as well as data obtained with the SMARTS Consortium 1.3 and 1.0 meter telescopes located at Cerro Tololo Inter-American Observatory, Chile.}
\altaffiltext{2}{Lowell Observatory, 1400 W. Mars Hill Road, Flagstaff, AZ 86001, USA; phil.massey@lowell.edu; kneugent@lowell.edu}
\altaffiltext{3}{Las Campanas Observatory, Carnegie Observatories, Casilla 601, La Serena, Chile; nmorrell@lco.cl}
\altaffiltext{4}{Department of Physics and Astronomy, The College of Charleston, Charleston, SC 29424, USA; pennyl@cofc.edu}
\altaffiltext{5}{Department of Physics and Astronomy, Northern Arizona University, P.O. Box 6010; Flagstaff, AZ 86011-6010; kathy.eastwood@nau.edu}
\altaffiltext{6}{Center for High Angular Resolution Astronomy and Department of Physics and Astronomy, Georgia State University, P.O. Box 4106, Atlanta, GA 30302, USA; gies@chara.gsu.edu}

\begin{abstract}
 
 The stellar mass-luminosity relation is poorly constrained by observations for high mass stars.  We describe our program to find eclipsing massive binaries in the Magellanic Clouds using
 photometry of regions rich in massive stars, and our spectroscopic follow-up to obtain radial velocities and orbits.   Our photometric campaign identified 48 early-type
 periodic variables, of which only 15 (31\%) were found as part of the microlensing surveys. Spectroscopy is now complete for 17 of these systems, and
 in this paper we present analysis of the first two, LMC 172231 and ST2-28, simple detached systems
 of late-type O dwarfs of relatively modest masses.  Our orbit analysis yields very precise masses ($\sim 2$\%),
 and we use tomography to separate the components and determine effective temperatures
 by model fitting, necessary for determining accurate (0.05-0.07 dex) bolometric luminosities 
 in combination with the
 light-curve analysis.  Our approach allows 
 more precise comparisons with evolutionary theory than previously possible.  
 To our 
 considerable surprise, we find a small, but significant, systematic discrepancy: all of the stars
 are slightly under-massive, by typically 11\% (or over-luminous by 0.2 dex) compared to that predicted by the evolutionary models.
 We examine our approach for systematic problems, but  find no satisfactory explanation.  The discrepancy is in the same
 sense as the long-discussed and elusive discrepancy between the masses measured from stellar atmosphere 
 analysis with the stellar evolutionary models, and might suggest that either increased rotation or
 convective overshooting is needed in the models.
 Additional  systems will be discussed in future papers of this series, and will hopefully confirm or refute this trend.
 
 \end{abstract}


\section{Introduction}
\label{Sec-intro}

\subsection{Motivation}

The mass of a star is arguably its most fundamental quantity; according to the Russell-Vogt theorem, it
is the mass of a star (along with the chemical composition) that uniquely determines a star's evolution.
We now know that the initial angular momentum also plays
an important role in determining the evolution of a star (Maeder \& Meynet 2000; Meynet \& Maeder 2000).    

For most stars, the simple way to estimate the mass ($m$)
of a star is by measuring the star's luminosity ($L$), 
as $L \sim m^x$ where the exponent $x$ is approximately 4 for solar-type stars.
For both lower mass
($<0.5M_\odot$) and higher mass ($>10M_\odot$) stars, the exponent becomes smaller, due to
the importance of convection in lower mass stars, and radiation pressure in higher mass stars.
For most stars,
the exponent in this mass-luminosity relationship (MLR) is well established both by evolutionary
theory and empirical measurements\footnote{Indeed, one
of the great vindications of stellar evolutionary theory was the fact that Eddington (1924) was able to derive
the exponent for solar-type stars purely from the physics of radiative diffusion.}.  

However, for high mass stars there are two additional complications that make applying the MLR difficult.
First, main-sequence massive stars (O- and early B-type) stars
are quite hot ($T_{\rm eff}$=25,000-50,000~K) and because of this
only a tiny fraction of their light leaks out in the visible.   In order to apply the MLR, one 
needs to know not
only the  distance and reddening of a star (in order to get the absolute visual magnitude, $M_V$), but also
an accurate value for the effective temperature, as the correction to $M_V$ needed to obtain the bolometric luminosity
is a steep function of effective temperature.   The second complication is that the MLR is really a function
of age.  This is true for all stars, as their luminosities increase slightly as they evolve, but massive stars
also {\it lose mass} as they evolve, due to radiatively-driven stellar winds. 
For massive stars these issues can be solved by first modeling the optical spectra using 
 non-LTE stellar atmosphere codes, such as CMFGEN
(Hillier \& Miller 1998) or FASTWIND (Puls et al.\ 2005).  This then provides both the effective
temperature and the bolometric luminosity, allowing placement of the stars on the H-R diagram (HRD). 
Reference to stellar evolutionary models then allow  a determination of age and current mass (referred to often as the ``evolutionary mass")
without reference to a MLR {\it per se}.

However, this method is no better than the stellar atmosphere and evolutionary models on which they are based.
A worrisome issue that remains is that of the so-called ``mass discrepancy", a systematic difference
between the masses one obtains from a stellar atmosphere analysis, and that inferred from evolutionary
theory.
 Modeling the star's spectra with a stellar atmosphere code produces  a measurement of the mass via the surface gravity $g$, since
 $g\sim m/R^2$, and the radius $R$ is known once the effective temperature is known {\it if} the distance and hence the luminosity $L$
are known, since $R^2\sim L/T^4_{\rm eff}$ by the Stefan-Boltzmann equation.  Herrero
et al.\ (1992) called attention to the fact that the masses derived from spectroscopic analysis are systematically
lower than those found from evolutionary models.   This mass discrepancy has never been fully resolved,
despite significant improvements in both the evolution and stellar atmosphere models.  (See, for example,
Massey et al.\ 2005.)  But it would be of great interest to measure masses in some more direct way to
test the validity of the models.   Such an opportunity is granted to us by binaries, where Newtonian physics
and Kepler's 3rd law provides us (in principle) with direct mass measurements.

This is the first of a series of papers presenting dynamical masses and bolometric luminosities for massive stars in the Large Magellanic Cloud (LMC).  The title of our series is intended to pay homage to the ``Spectroscopic Studies of O-type Binaries"  series by Peter S. Conti and collaborators, 
which appeared 30-35 years ago (e.g., Bohannan \& Conti 1976; 
Conti \& Walborn 1976; Massey \& Conti 1977; Morrison \& Conti 1978, 1980; Conti et al.\ 1980).  Although
telescopes have gotten larger since those days, and instruments and data analysis methods have improved, 
the basic need to test theory with fundamental mass determinations remains.
Work over the past several decades has helped improve the situation\footnote{We particularly want to acknowledge the many contributions in the field by our late friend and colleague Virpi Niemela.} but the
paucity of systems with well-determined parameters is underscored by our continuing poor 
knowledge of the MLR and
the persistence of the mass discrepancy. 

\subsection{Our Approach}

Historically, studies of massive binaries have usually begun by the accidental 
discovery of double lines
in the spectrum as a result of an observation taken when the system just happened to be near 
orbital quadrature\footnote{Throughout this paper we refer refer to the phases
where eclipses occur (conjunction) as phases 0 and 0.5, and to the phases that are most double-lined (quadrature) as phases 0.25 and 0.75.}.
 Several dozen subsequent spectra are then usually needed (obtained over a period
of months or years) to determine the orbital period and the orbital parameters.  Usually it is only then
that the system is monitored to detect the light variations that would indicate eclipses 
that allow an orbital inclination ($i$) to be determined.  
Without knowledge of the inclination, a double-lined spectroscopic orbit solution tells us only
the {\it minimum} masses for each component, $m_1 \sin^3 i$ and $m_2 \sin^3 i$, since
$\sin^3 i \le 1$.

If eclipses are present, then analysis of the light curve will also yield the stellar radii and the flux
ratios of the two components, if the effective temperatures can be accurately determined.  For most
stars the effective temperatures can be determined by obtaining light curves in different bandpasses
(i.e., from the colors), but for main-sequence massive stars,  the colors are largely degenerate with effective temperatures
because of the high temperatures, and
it requires additional spectral analysis if good values for the effective temperatures are to be found.
With these, and with the knowledge of stellar radii, one can then compute
the bolometric luminosities of the components,
which can be used to refine our knowledge of the mass-luminosity relationship.  It is additionally useful to
know the distances to the system, as the comparison between the modeled absolute visual magnitude
and the observed absolute visual magnitudes provides assurance that the physical parameters are 
correct\footnote{We note that the courageous efforts of several groups 
who invert
this problem in order to determine distances to nearby galaxies by measuring the stellar radii from the eclipses and orbital parameters, and then using the effective temperatures to infer the absolute magnitude of the
system, which can then be compared to observed magnitude to derive a distance.  See, for example, the
review by Paczynski (1997), and recent work by Bonanos et al.\ (2006, 2011), Guinan et al.\ (1998); 
Harries et al.\ (2003),  Hilditch et al.\ (2005), Fitzpatrick et al.\ (2003), North et al.\ (2010), and Vilardell et
al.\ (2010), to name but a few such attempts.}.

Since eclipsing systems are necessary for the full analysis, we decided to instead start 
by searching for stars whose light curves
suggested they might be eclipsing massive stars.   We chose to concentrate on massive stars in the
Magellanic Clouds since their distances are well known, and they are bright enough for spectroscopic
followup, albeit with larger apertures.  In addition, since their metallicities are relatively low, mass loss on the main-sequence should be relatively modest, and thus the connection
to the initial masses should be less dependent upon the assumed mass loss rates of the evolutionary models. 
By refining the periods and times of primary eclipses 
precisely using frequent photometric observations
prior to spectroscopic observations, we should know exactly when to observe these stars at
maximum velocity separation (orbital quadrature).   

This allowed us to measure the orbital semi-amplitudes
very efficiently for two reasons.  First, O-type stars have very broad spectral lines due to rotational broadening
($v\sin i$ typically greater than 100 km s$^{-1}$), and {\it most} spectral observations more than 0.1 phase 
away from quadrature are unlikely to show resolvable double lines and hence do not provide useful measurements. Secondly,
it is the velocities around quadrature that best define the orbital semi-amplitudes.  
And, since the phases would be known exactly, determining the
amplitude of the orbital motion could be done with very few observations, as we could fix
the phases in the orbital solutions to those determined from the light curve.  Thus we
 minimized the amount of large telescope time needed for the spectroscopy by utilizing small aperture
 telescopes to determine the light curves\footnote{Some studies, such as Gonzalez et al.\ (2005),
 Morrell et al. (2007), and Bonanos (2009)  have used massive eclipsing binaries found
 by way of the MACHO  (Alcock et al.\ 1997) and OGLE (Udalski et al.\ 1998) microlensing searches.  Here
 we decided to obtain our own photometry as the data needed to be nearly contemporaneous with the
 spectroscopy for precise phases, and also because we suspected such surveys might have missed
 many interesting systems. Based upon the scant number of our systems that were detected by
 MACHO and OGLE, as discussed below, our approach appears to have been justified.}.  
 
 One of the downsides to this approach is that it is easier to find
 short period systems as their light variabilities are more pronounced: for a given orbital inclination 
 eclipses will be deeper for shorter period systems. However, these shorter period systems are more likely to be in
 contact, and therefore their masses will be less pertinent to understanding single stars.  But for many
 such systems our program detected easily interpreted, detached systems.  Nevertheless, as we will
 see, even these ``simple"
 systems offer some unexpected surprises!
  
 We begin this series of papers with a study of two  such simple, detached system consisting of late O-type
 dwarfs of modest mass:
 [M2002] LMC 172231 (O9 V + O9.5 V) 
 and ST2-28 (O7 V + O8 V).   Subsequent papers will
 describe the other eclipsing systems we've identified, including
 early-type O binaries, Wolf-Rayet binaries, and contact systems.  We describe in some detail here the
 observational and reduction techniques, since these will be used in this future work, and illustrate
 the results we've been able to achieve.  It is our intent that these systems will allow us to make more 
 critical tests of stellar evolution models than usually possible, and to serve as 
 {\it linchpins} for the empirical mass-luminosity relationship. 

\section{Observations and Reductions}

\subsection{Photometry and the Identification of Eclipsing Systems}
\label{Sec-photometry}

Our photometric monitoring of stars in selected OB associations in the Magellanic Clouds began in 2003 July,
and continued through 2011 February.  The observing seasons generally extended from August
through January or February of the following year.  
The only observing season missed entirely in our photometric monitoring program
was 2008/2009, as we had begun
our spectroscopic observations, but as the analysis proceeded we realized we needed more 
supporting photometry.  The only observations during the 2010/2011 season were made in January-February 2011 of selected stars at targeted times in order to improve missing phase coverage in the light curves.

A variety of telescopes were used; their properties are listed in Table~\ref{tab:telescopes}.   The Swope
observations were made in ``classical" mode by N.I.M.\footnote{During the 2005/2006 observing season she was
aided by her co-workers in the Carnegie Supernova
Project (Hamuy et al.\ 2006).} and were often nightly during dark time.  
The SMARTS 1.0-m and 1.3-m observations
were obtained in queue mode 
usually with a cadence of once every other night throughout both bright
and dark time, with time primarily allocated through Georgia State University, although the first
season was obtained through NOAO (Proposal 2005B-0108).   In the case of the 1.0-m, the queue
observations were carried out for a few minutes each night by whatever astronomer happened to be
scheduled that night in classical mode; in the case of the 1.3-m, the data were obtained by dedicated
queue observers.
We list in Table~\ref{tab:clusters} the six Magellanic Cloud regions we observed, and
which telescopes were used in each observing seasons.  
This table is intended primarily to give
an overall view of  the scope of the project; the specific telescope and instrument will be given for each
photometric measurement in the individual photometry tables.  All observations were made through a
V filter, and exposure times were typically 10-30 sec.  Data were obtained under both photometric
and non-photometric conditions (in some cases, through several magnitudes of extinction!), but since
each field contained hundreds of reference stars, we were able to obtain good photometry nonetheless.

The reductions of the Swope data were straightforward.  Our tests on a series of exposures did not justify the use of the nonlinearity correction typically applied to these data (Hamuy et al.\ 2006), but inclusion or
not would have had negligible effect on our photometry.   Overscan columns were used to remove the bias, and
nightly bias frames were used to remove the (negligible) bias structure.   Flat fielding was achieved by
exposures of dome flats.  

The SMARTS 1.0-m Yale Y4KCam data proved a challenge to reduce.  
Our data were among the first to be taken with the instrument, and our reductions quickly identified
a number of issues.  The 4K$\times$4K chip
was read out using four amplifiers, and we found we had to  
mask out the central
rows and columns as good photometry could not be performed for stars that straddled
quadrants as the bias structure was highly unstable. 
Worse, tests showed that dome flats and twilight sky flats differed by nearly 10\% 
from center to edge.   We prevailed upon one of the observers,  David James, to obtain images with
the star moved from center to near the edge, and our photometry of these data showed that the twilight
flats gave good ($<$1\%) results while the use of dome flats would have resulted in a 10\% error.  
The lack of header information describing the data and bias sections of the four quadrants resulted in
our writing our own IRAF\footnote{IRAF is distributed by the National Optical Astronomy Observatory, which is operated by the Association of Universities for Research in Astronomy (AURA) under cooperative agreement with the National Science Foundation.} reduction scripts, which have been made available on the
SMARTS web site\footnote{http://www.astro.yale.edu/smarts/smarts1.0m.html} for others to use, along with the characterization of the instrument.
We hope this provided some partial compensation for the scheduled
observers who were asked to take 15 minutes of data for us every few nights.  

At the start of the 2007/2008 observing season,
one quadrant of the Y4KCam 
chip  died, and our program was switched to the SMARTS
1.3-m ANDICAM instrument.
Reduced data are kindly provided for all  ANDICAM observations by Suzanne Tourtellotte, and the reductions
are essentially identical to what we used for the Swope data.  

Photometry was carried out by a series of automatic IRAF scripts that characterized the data, identified
stars, and obtained aperture photometry.   These routines were loosely based upon the scripts written
for producing photometry for the Local Group Galaxies Survey (Massey et al.\ 2006, 2007).
Experiments to perform photometry via point spread function (PSF)
fitting did not yield improved results, and often were inferior, particularly on the SMARTS 1.0-m Yale data
where significant PSF variations were present across the field.  For the Swope and SMARTS 1.3-m data, an aperture with a radius of 3 pixels was used (1\farcs1-1\farcs3). 
For the SMARTS 1.0-m data, a similar value in arcseconds was used (5 pixels, or 1\farcs4).
The sky values were taken from the modal value in
an annulus located between radii of 10 and 18 pixels from each star.

For each telescope and cluster combination a ``master list" was created, 
consisting of pixel coordinates and an instrumental magnitude for each star from some frame 
obtained in
good seeing.  In addition, each master list included the celestial coordinates for each star, obtained
by using the {\it HST} Guide Star Catalog 1.1 catalog to calibrate the frame with a world coordinate system.
The masterlist also included our best estimate of the standard magnitude of each star, obtained
using the UBVRI photometry of Massey (2002) and a modest color correction determined for each
telescope from a $B$ image obtained for these purposes.  The photometry from each frame was cross-correlated against this master list in order to determine the offset, rotation, and a magnitude difference
for the ensemble.  The magnitude for each star was then determined by correcting for the magnitude
difference of the frame to the masterlist, and the correction between the masterlist's instrumental
magnitude and our estimate of the standard magnitude.  A single file of the photometry of each star
was then produced including the heliocentric Julian day (HJD) and the magnitude and instrumental
magnitude error.   The photometry from the various years and telescopes were then combined, applying a small
zero point shift (typically on the order of several thousandths of a magnitude) if needed to bring the
median out-of-eclipse data into accord.

We analyzed the data as we went along, and re-evaluated our target list
every year, with the goal of having a preliminary set
of targets for spectroscopy in late 2008.  It is for that reason that we dropped NGC 602c from our program
(see Table~\ref{tab:clusters}),
as none of the stars of interest showed significant variability.  

Various methods have been employed over the years to detect variability.  In our case, we were 
interested only in {\it periodic} variability.  We therefore adopted
the one-way analysis of variance (AOV) method of Schwarzenberg-Czerny (1989) which excels at
detecting periodic narrow events, inspired by the success
demonstrated by L\'{o}pez-Morales \& Clemens (2004).  (We are grateful to Mercedes L\'{o}pez-Morales
for helpful correspondence and for passing on a FORTRAN77 version of the Schwarzenberg-Czerny 1989 code.  Updated versions of the code can be found at Dr.\ Schwarzenberg-Czerny's web site\footnote{http://users.camk.edu.pl/alex/}.)  We added software that allowed the
automatic detection of the spikes in the 
resulting periodograms, and produced phased light curves that could be examined by eye for significance
and to distinguish likely periods from their aliases.  In all, several {\it thousand} such light curves were
examined multiple times as the data collection grew; this provided useful training material for a number
of undergraduates.

For the final periods, we checked our values using the Lafler \& Kinman (1965) technique, which involves
phasing the data by successive trial periods and determining the point-to-point scatter in the phased data.

We list in Table~\ref{tab:winners} {\it all} 48 of the stars in the SMC and LMC that we thought were interesting from
the light curves; follow up spectroscopy (described in the next section) caused us to drop many of these
from our program for one reason or another.  For some,
 the stars were not double-lined at the anticipated phase.  This is unlikely due to our having the wrong period,
 as in some cases the same period had been found from analysis of MACHO data (Faccioli et al.\ 2007,
 Derekas et al.\ 2007).  In other cases, double lines were seen but were too blended for reliable velocity
 information to be extracted.  In a few cases it was just not possible to obtain a sufficient number of observations at the needed orbital phases due to time spent on other systems.  In one very disappointing
 case, that of the early-type binary NGC~346 MPG 342 (Massey et al.\ 1989),  spectroscopy revealed that
 the system was triple, with the third component also shifting.  In the end we were left with no suitable
 stars in the SMC, and 17 in the LMC.
 
Wyrzykowski et al.\ (2004), Derekas et al.\ (2007), and Faccioli et al.\ (2007) have published lists of eclipsing binaries in the SMC and LMC from the OGLE (Udalski et al.\ 1998) and MACHO (Alcock et al.\ 1997) microlensing projects, and it is interesting to note that of the 48 periodic light variables we found, only 15  (31\%)
of them are also in these lists. The reasons for this are not obvious.  Saturation is likely not the sole
explanation: the brightest stars in the Wyrzykowski et al.\ (2004) and Derekas et al.\ (2007) catalogs have
 $V\sim 13$, while the average $V$ magnitude of stars in our sample have $V\sim14$.  Of the two stars
 discussed in the present paper, the $V$ magnitudes are nearly the same ($V$=14.04 and 14.15), and
 yet one was previously detected as a variable while the other one was not.
 None of the
stars in NGC 1910 or the R136 region are identified as eclipsing systems by these other studies. R136
is embedded in strong nebulosity, but the nebulosity around NGC 2074 (where numerous MACHO objects
are found)  is considerably stronger than that of NGC 1910.
Crowding may also be an issue, although one of the stars discussed in the current paper
(ST2-28) is relatively crowded and was correctly identified.
  It is also true that in some cases the periodic variability 
was of very low amplitude. For instance, LMC 171520, which will be discussed in Paper II,
has peak-to-trough variations of $<$0.1 mag. Still, other stars, such as one of the
stars discussed in this paper, LMC 172231, have very deep (0.6 mag) 
eclipses
but were not detected in these microlensing surveys. (It is also not located in
strong nebulosity.)  We believe this emphasizes
the need for such targeted studies such as ours, particularly for finding massive eclipsing systems where
nebulosity is a characteristic of the sample.

We do note that when MACHO or OGLE did detect our objects, the periods are in excellent agreement.
The one exception is LMC 164717, where the MACHO period is half of ours.

\subsection{Spectroscopy and Radial Velocities}

All spectroscopy intended for radial velocities was carried out on the Magellan 
Clay and Baade 6.5-m telescopes,
although a few classification spectra were also obtained on the DuPont 2.5-m telescope.
The large aperture of the Magellan telescopes allowed us to reach the very high signal-to-noise (S/N)
we desired (200 or more per 1~\AA\ spectral resolution element) in exposures of about an hour for our
faintest targets, and a few minutes for our brightest.  Such high S/N is needed given the general weakness
of the spectral features in early-type stars.

All in all, we collected data at the Magellan telescopes on 21 nights between 2008 December 8 and 2010 November 27.  The individual runs varied in length from 2 to 6 consecutive nights.  Roughly half of the time was allocated through Carnegie, and the other half through the University of Arizona. Nine of the
nights were on the Baade with the Inamori-Magellan Areal Camera and Spectrograph (IMACS), while
the other 12 nights were on the Clay telescope with the Magellan Echellette (MagE) spectrograph.  
Generally, the first three authors (P. M., N. I. M., and K. F. N.) 
were present for all of the observing, although for a few runs either P. M. or N. I. M.
were the only ones observing.

For all of the observing, orbital ephemerides were computed based upon the accumulated photometry prior to the observing run, and observations were planned to take place near orbital phases of 0.25 and 
0.75, the times of maximum velocity separations.  Quick-look was performed in real time  to
assess the overall quality of the data, and (in the first few spectroscopic runs) to determine spectral types
and confirm that double lines were present.

For IMACS, observations were obtained with the 1200 line mm$^{-1}$ grating with a 0\farcs9 slit.  The
resulting spectral resolution was about 1.0~\AA\ (5.0 pixels on the detector).  Spectral coverage was nominally
3650-5250~\AA\ over the four CCDs, but in practice only the central two chips contained useful spectral
features and were used, covering  4020-4410~\AA\ and 4430-4820~\AA.   Because IMACS uses a very long
slit, it was often possible to rotate the instrument to a position angle that allowed two objects of interest to be placed on the slit and observed at the same time.  A series of three integrations (ranging from 3x300 s to 3x750 s depending upon
the brightness of the object and the sky conditions) was  obtained of
the program object, followed by a 150 s exposure for the HeNeAr.  
The observing procedure for such spectroscopy is relatively straight forward but time
consuming for a program such as ours, as there is no provision for viewing the slit directly.
Despite our experience, it was never
possible to lower the overhead (setup plus comparison) to less than 10 minutes for each new object.
Flat field data were obtained by exposure
of a calibration screen during the afternoon.  Typical signal-to-noise was 150-200 per 5 pixel resolution element.

Data reduction for IMACS used the standard IRAF tasks.
The overscan at the top of each chip was used, as this removed most of the bias structure.  The
remainder was removed using a series of bias exposures obtained each night.  Flat-fielding was
accomplished by the normalized  calibration screen exposures.  Spectra of the sources were extracted
using an optimal extraction ``clean" algorithm, with sky chosen adjacent to the object,
and combined at the end.  Each of the two chips were
treated separately.  Typical wavelength errors (root mean square) from fitting the comparison spectra
were 0.03-0.05~\AA\ (2-3 km s$^{-1}$).

For MagE, spectral coverage is nominally from 3100~\AA\ to 1$\mu$m covering 15 orders,
with a spectral resolution of 4100, almost
identical to the resolution we achieved with IMACS.  We did not reduce either the bluest or reddest parts of the
spectra, but kept only the central 13 orders (orders 7-19), covering 3150-9400~\AA.
Acquisition is by means of a slit-viewing TV,  so
overhead was minimal.  Although the short slit (10\arcsec) precluded multiplexing by obtaining
two objects, the high throughput and lack of overhead made this a more efficient instrument for our
program.  Exposure times were shorter than with IMACS by about 30\%, and resulted in higher signal-to-noise than with IMACS.
Wavelength calibration was by means of a 3 s ThAr exposure.  

Flat-fielding of MagE is complicated by the very large wavelength coverage.  Traditionally users have
employed a combination of flat-field quartz
lamp exposures in the red and in- and out-of-focus exposures of a Xe lamp
in the violet.   Following a suggestion made one night at dinner by Ian Thompson, we experimented
and demonstrated to our satisfaction that we could meet our very high signal-to-noise requirement
{\it without} flat fielding: that the chip really is quite uniform.  Instead, we dithered the star to three
positions along the slit.  In the end we typically achieved a S/N of 400 per 4-pixel spectral resolution element
at 5000~\AA\footnote{We note that many users require significantly lower S/N than ours, sometimes in the
single digits.  Flat-fielding is unlikely to help such data.  See discussion in Massey \& Hanson (2012).}.

The reductions of the MagE data were complicated by the curvatures of each order (due to the anamorphic
distortions of the cross-dispersing prisms), and spectral features are also tilted, with a tilt that
varies with wavelength along each order.   Although various MagE pipelines have been developed
by several groups, we found we did very well with the  
``mtools" IRAF package written by Jack Baldwin for dealing with data from
another instrument, the
Magellan Inamori Kyocera Echelle (MIKE)\footnote{http://www.lco.cl/telescopes-information/magellan/instruments/mike/iraf-tools}, in combination with the standard IRAF Echelle reduction
tools.  Wavelength calibration consisted of  a 2-dimensional fit to all of the orders,
with RMS residuals of 0.05-0.06~\AA\ (3-4 km s$^{-1}$).

Radial velocities were measured from the normalized spectra using interactive fitting of two Gaussians
to resolve double lines, a standard technique with double-lined O-type 
binaries (see, for example,
Burkholder et al.\ 1997, Rauw et al.\ 2001, Massey et al.\ 2002, Morrell et al.\ 2003, Niemela et al.\ 2006, and Mayer et al.\ 2008)\footnote{The scant number of spectral lines in the 
spectrum of O stars means that standard cross-correlation techniques are less useful
than for stars of later spectral types.}. 
If a spectrum did not reveal double lines, it was not used; an inspection of the
data revealed that there were no instances for our final sample 
where a star should have shown double lines at an
appropriate phase but did not, or vice versa, giving us additional confidence in our period determinations.
Anywhere from 1 to 17  lines were measured in a given spectrum (mainly the He I and He II although occasionally the Balmer lines were used if well separated) depending upon the phase and the
quality of the data, but the typical (median) number was 5.  The effective rest wavelengths were taken
primarily from Conti et al.\ (1977), supplemented by the updated version of Moore (1972) by 
Coluzzi (1993). 

\section{Analysis: Derivation of Orbital Parameters and Physical Parameters}

\subsection{Determining the orbital semi-amplitudes and the minimum masses}
\label{Sec:Orbits}

Typically one uses the radial velocities of each component of a spectroscopic binary to solve
for the orbital elements $P$ (the period), $K_1$ and $K_2$ (the orbital semi-amplitudes), $\gamma$ (the center-of-mass motion of the system),  $T$ (the time of
periastron passage of the primary), $e$ (the orbital eccentricity), and 
$\omega$ (the angular distance of periastron passage relative to the ascending node).  In general the radial
velocity $v_1$ of star 1 is related to the orbital parameters as
$$v_1=\gamma + K_1 e\cos{\omega} + K_1 \cos({\nu+\omega}),$$
where $\nu$ is the true anomaly, a function only of the orbital phase and $e$ (see, e.g., equations 64 and 65
in Binnendijk 1960).  The minimum masses
 are given by 
 \begin{equation}
 m_1 \sin^3 i=1.036\times10^{-7} (K_1+K_2)^2 K_2 P (1-e^2)^{3/2}
 \end{equation} 
 and
 \begin{equation}
 m_2 \sin^3 i=1.036\times10^{-7} (K_1+K_2)^2 K_1 P (1-e^2)^{3/2}
 \end{equation}
 when
 $P$ is expressed in days, $K$ in km s$^{-1}$, and $m$ is in solar masses.  The orbital inclination relative
 to the line of sight is $i$, and hence the expression $minimum$ mass, as $\sin^3 i \le 1$.   Note from
 these equations that
 mass ratio is inversely proportional to the ratio of the semi-amplitudes; i.e., $m_2/m_1 = K_1/K_2$.

In designing our spectroscopic observations, we decided to concentrate on the systems
with circular orbits, as one can then eliminate the need to determine $e$ and $\omega$,
and fewer observations will yield high accuracy for the orbital semi-amplitudes and
hence the masses.  Fortunately, we expect that to be the case for most massive binaries
with short periods, as  tidal forces circularize orbits very quickly.  (For instance,
of the 15 young detached massive binary systems listed by Gies 2003 with periods less than 5 days,
only 2, or possibly 3,  have been shown to have non circular orbits.)  
For eclipsing systems, most systems with non-circular orbits can be quickly spotted as
the primary and secondary eclipses will likely be separated by other than 0.5 phase. 
However, an eccentric system which just
happens to have $\omega$ near 90$^\circ$ or 270$^\circ$ will also have eclipses 0.5 phase apart, and so the definitive test will come from modeling the light curve.  An light curve with
eclipses 0.5 phase apart {\it and} the same eclipse widths means that the orbit is circular.

For massive binaries there can be a further complication.
Hot luminous stars have radiatively-driven stellar winds that, if strong enough, may result in
appreciable (tens of km s$^{-1}$) outward motion even down 
in the photosphere where the spectral lines are formed.  This was first
demonstrated in a series of papers by Hutchings (1968a, 1968b, 1969, 1970a, 1970b; see also 1979).
This results in the possibility that the two components in an early-type binary may have different
average velocities, equivalent to saying that separate ``center of mass" velocities may be needed
for each component, i.e., that actually $\gamma_1$ and $\gamma_2$ are needed.  Examples
include the Of-type binaries HDE 228766 (Massey \& Conti 1977, Rauw et al.\ 2002)
and Sk-67$^\circ$ 105 (Niemela \& Morrell 1986) where the ``systemic"
velocities of the two components differ by 30-40 km s$^{-1}$ or more.  In many cases,
however, the $\gamma$ velocities agree within the errors, particularly if the components are late O-type
dwarfs  (see, for example, Stickland et al.\ 1997, Penny et al.\ 2002) as these stars
will have weaker stellar winds than early O stars or Of-type supergiants.  This issue sometimes comes as
a surprise even to binary experts used to working on less massive stars, or is overlooked by 
newcomers to the field\footnote{Some hesitancy to accept the occasional need for two different values
of $\gamma$ may trace back to the realization by Petrie (1962) that poorly determined values for $e$ and $\omega$ will result primarily in erroneous values for $\gamma$; see discussion in Batten (1973).}.

In order to minimize the amount of spectroscopy time needed, and to put all of the radial
velocity information towards the most accurate determinations of the orbital semi-amplitudes, we adopt the values of $T$ and
$P$ from the light curve in order to compute the phase $\theta$.   Then for circular orbits
($e=0.00$) the relationship between 
the observed radial velocities $v$ and the orbital parameters $K$ and $\gamma$ become simple
linear equations, 
\begin{equation}
\label{Equ-v1}
v_1 = -K_1 \sin{2\pi\theta} + \gamma_1
\end{equation}
\begin{equation}
\label{Equ-v2}
v_2 =  K_2 \sin{2\pi\theta} + \gamma_2,
\end{equation}
where $\theta$ is the fractional part of $(t-T)/P$, where $t$ is the heliocentric Julian day (HJD) 
of the observation, and $T$ is
the HJD corresponding to primary conjunction, when the secondary is in front of the primary.
Thus one can compute the values for the $K's$ (and hence the masses)
 much more precisely than
in cases where the radial velocities are also used to determine $P$ and $T$ and hence $\theta$.

In practice, we solved Equations~\ref{Equ-v1} and \ref{Equ-v2} both with the individual $\gamma$'s and
with a single $\gamma$ (i.e., $\gamma_1=\gamma_2=\gamma$), and compared the residuals.  In only
a few cases did we find that using individual values for $\gamma$ were warranted.  (Both of the 
systems discussed here were well modeled with a single $\gamma$, consistent with their being late O-type
dwarfs, presumably with small mass loss-rates.)    We determined the best
values for the $K's$ and $\gamma$ by least-square fitting, we assigned $1/\sigma^2_\mu$ weight to each velocity,
where $\sigma_\mu$ is the standard deviation of the mean for each velocity.  If only one
spectral line was measured,  
we assumed a large velocity error (30 km~s$^{-1}$) so as to include the point but with
low weight.  If $\sigma_\mu$ was 
less than
5 km~s$^{-1}$ we assumed the actual error was 5 km~s$^{-1}$, so as not to overweight points where
the few individual velocities forming the average were fortuitously similar.  We also ran a
differential orbit fitting program (written by our colleague L. H. Wasserman and
described in more detail by Rosero et al.\ 2011) on the velocity data to confirm that the data were consistent with our assumption of a circular orbit.

A nearly independent check on our orbital semi-amplitudes is provided by the method of
Wilson (1941), who demonstrated that one could determine the ratios of the orbital semi-amplitudes
directly
from the radial velocities themselves without knowledge of the period or indeed any of the orbital parameters.  It is applicable for non-circular (as well as circular) orbits.
(In practice, this works reliably 
only if there are velocity
measurements from both sides of the orbit.)  
If one plots the radial velocity $v_2$ against $v_1$, the slope of the best fit line 
will be proportional to the ratio of the orbital semi-amplitudes,   $\Delta v_2/\Delta v_1$ = $-K_2/K_1$,
which is just the negative of the inverse mass ratio $-m_1/m_2$.

The intercept
 is $\gamma(1+K_2/K_1)$.  Massey \& Conti (1977) showed that,  in the case that 
two different $\gamma$'s were needed, the slope remains the same, but  the meaning of the
intercept changes to be $\gamma_2 + \gamma_1 (K_2/K_1)$.   ``Wilson's method" was  used
by Bohannan \& Conti (1976) and Massey \& Conti (1977) in their studies of high mass systems, and has been employed to good effect
recently by Lisa Prato and collaborators (e.g., Prato 2007, Schaefer et al.\ 2008,
Mace et al.\ 2009, Rosero et al.\ 2011) in their studies of low-mass binaries.  The resulting mass ratios
and the linear correlation coefficients $r$ from the Wilson diagrams are included in describing our
orbit solutions. 

The improvement in computer processing power and analytical techniques have made it possible to solve for both
light- and radial velocity curves simultaneously; these sophisticated programs model the ``center of light"
velocities, which are no longer the simple sine curves given in Equations \ref{Equ-v1} and \ref{Equ-v2} near
conjunction if the systems eclipse, or if tidal distortions are significant.  In the cases we consider here, the
stars are nearly perfectly round (as discussed in Section~\ref{Sec-results}), and none of our velocity measurements
are taken near conjunction.  Indeed, for massive O-type stars, with their broad spectral lines, double lines can
only be resolved near quadrature and so the use of these advanced tools of are limited benefit.

\subsection{Determining the orbital inclinations and other physical components}
\label{LightCurves}

Our light curves combined with the orbital parameters
allow us to determine the orbital inclination, and they also allow us to determine
the individual flux ratios between the two stars by adopting effective temperatures. 
 From first principles we expect that (in the absence of tidal distortions and the like) the same areas will be eclipsed at both 
primary and secondary eclipse.  Thus, the deeper eclipse 
will correspond to the hotter star being eclipsed as
the surface brightness of the hotter star will be greater (at
every wavelength) according to the Stefan-Boltzmann law.  However, this is not necessarily 
the {\it brighter} star, nor is it necessarily the star with the stronger spectral features.

Consider this from the standpoint of stellar evolution.  A binary might form consisting
of two stars with nearly equal masses.  The slightly more massive one will be the hotter
one, initially.  It will also have a slightly larger radius and bolometric luminosity.  It will
also be the brighter one visually.  It will also be losing mass at a slightly higher rate
(through radiatively driven stellar winds) than its companion.
 However, as stellar evolution proceeds its temperature will decrease.  Most stellar
 evolutionary models would have its bolometric luminosity increasing as well.  So,
 in the absence of complications it should remain the visually brighter star.  However,
 because of mass-loss, it may not remain the more massive of the two components.
 Thus a system that begins as an O5 V + O6 V  pair (with the O5 V star
the more massive component)  
 could evolve to an O7 III + O6.5 V pair, but either component could be the more massive.
 In such a system the light curve would show the deepest eclipse when the O6.5~V component
 was eclipsed, since it would be the hotter star.  We'll note that because the luminosity class
 ``V" is very broad in terms of its spectral features, the O5~V + O6~V pair could even be
 classified as O7~V+O6~V at a later time.  It is not clear which component
 would be the more massive given the fact that both stars will be losing mass, with the initially
 more luminous (massive) star having the larger mass-loss rate.
 Thus we should not expect that the hotter component (which we will call the primary) is
 necessarily the more massive component.

For modeling the eclipses, we used the light curve synthesis code 
GENSYN (Mochnacki \& Doughty 1972)
to produce model $V$-band differential light curves.  Our approach was to make a 
constrained fit using as much data as possible from the spectroscopic results. 
The orbital parameters were taken from the spectroscopic solution, and initial
effective temperatures were estimated from the spectral types of the stars, using
the calibrations of Massey et al. (2005) and Trundle et al. (2007) for O-type and B-type stars, respectively. We then estimated the physical fluxes and limb darkening coefficients from tables in the OSTAR2002 and BSTAR2006 models\footnote{From http://nova.astro.umd.edu/Tlusty2002/tlusty-frames-guides.html} based upon TLUSTY (Hubeny \& Lanz 1995),
and Claret (2000), respectively. Each trial run of GENSYN was set by three independent parameters, the system inclination $i$ and primary and secondary polar radii. For the initial run, we attempt to match three observables: the absolute visual magnitude of the system,  the eclipse depths, and the eclipse durations (widths)\footnote{Note that although the polar radii were entered, the program also reports the ``volume radius", i.e., the radius for a
sphere of the same volume as the tidally distorted star.  It is this volume radius that we will report, and use with the
effective temperatures to determine a bolometric luminosity.}.

We used the velocity curves 
and the flux ratios to separate the spectra
of the two stars using tomographic analysis (Bagnuolo \& Gies 1991, 1992; Gies 2004).
With these cleanly separated spectra, we next used the stellar atmosphere code FASTWIND
(Puls et al.\ 2005) to model the components, producing more accurate effective temperatures. 
Examination of the line depths, particularly of the Balmer lines, also allowed us to reevaluate
the flux ratios.  The new values of the effective temperatures were then used to remodel
the light curve.  As described below, the star ST2-28 turned out to be triple, with the third component stationary.
It was easy to extend our analysis to include this situation.

To briefly summarize,  we used the spectroscopic orbit to fix as many of the physical
properties of the system as possible, such as the orbital separation.  Our preliminary
light curve analysis then used a reasonable approximation for the effective temperatures 
based on the spectral types, and produced a flux ratio.  We used this flux ratio to 
separate the spectra using tomographic analysis, which we then modeled with 
FASTWIND, yielding
improved effective temperatures, and revealed any problem in the flux ratios, which we then
used to remodel the light curve.  If needed, a new tomographic extraction and remodeling
was then performed.   This complete an approach has not generally been used on
massive binaries, although Fitzpatrick et al.\ (2003), Bonanos (2009) and Bonanos et al.\ (2011)
did similar fitting using model atmospheres on tomographically separated spectra
to determine the effective temperatures.  
The values of the orbital inclinations were not much affected by this approach, as the value we derived
from the final light curve analysis was always within one or two sigma of the original analysis.
Rather, the strength of this
approach was that it resulted in far more accurate luminosity determinations than
 could have been achieved by simply trying to match the absolute
magnitude of the total system, as the latter is always uncertain by 0.3~mag or so due to
uncertainties in reddening, etc.  

The bolometric luminosities were computed using the effective temperatures and effective stellar radii
in the Stefan-Boltzmann equation.  We then found the absolute visual magnitude by subtracting the
bolometric correction:   $${\rm BC}=-6.90\log{T_{\rm eff}}+27.99,$$ from Massey et al. (2005).
All this assumes spherical geometry.  The light-curve analysis program
also produces an estimate of the absolute visual luminosity; this integrates over the (non-spherical)
surface, but replies upon inputting the monochromatic specific intensity from the stellar atmospheric models.  We feel
this method is less certain, but will include it as the ``tidal model", although as we will see, these stars
are not much distorted from spheres.  The tidal model bolometric luminosity is then found by adding
the bolometric correction.

\section{Results for Two Detached Systems}
\label{Sec-results}

The two systems whose orbit solutions and masses we present here were chosen in part because their
parameters were very well determined.  Higher mass systems, and more physically complex interpretations,
will be presented in future papers. 
Throughout we assume a distance modulus to the LMC of 18.50 (50 kpc), following
van den Bergh (2000).

\subsection{[M2002]LMC 172231, O9 V + O9.5 V}

LMC 172231 is located in a relatively uncrowded region of NGC 2074, also known as the Lucke-Hodge 101
OB association (Lucke \& Hodge 1970).  The star was first cataloged by Westerlund (1961) in
his photographic study of NGC 2074 ([W61] 3-9), and was included in 
Testor \& Niemela (1998)'s photometric and spectroscopic
study of the region ([ST92] 5-67)\footnote{The designation [ST92]5-67 is 
apparently an extension from Zones 1-4 photometered by Schild \& Testor (1992).}.  Additional photometry
was obtained by Massey (2002) in his survey of the SMC and LMC ([M2002] LMC 172231).

\begin{figure}
\epsscale{0.9}
\plotone{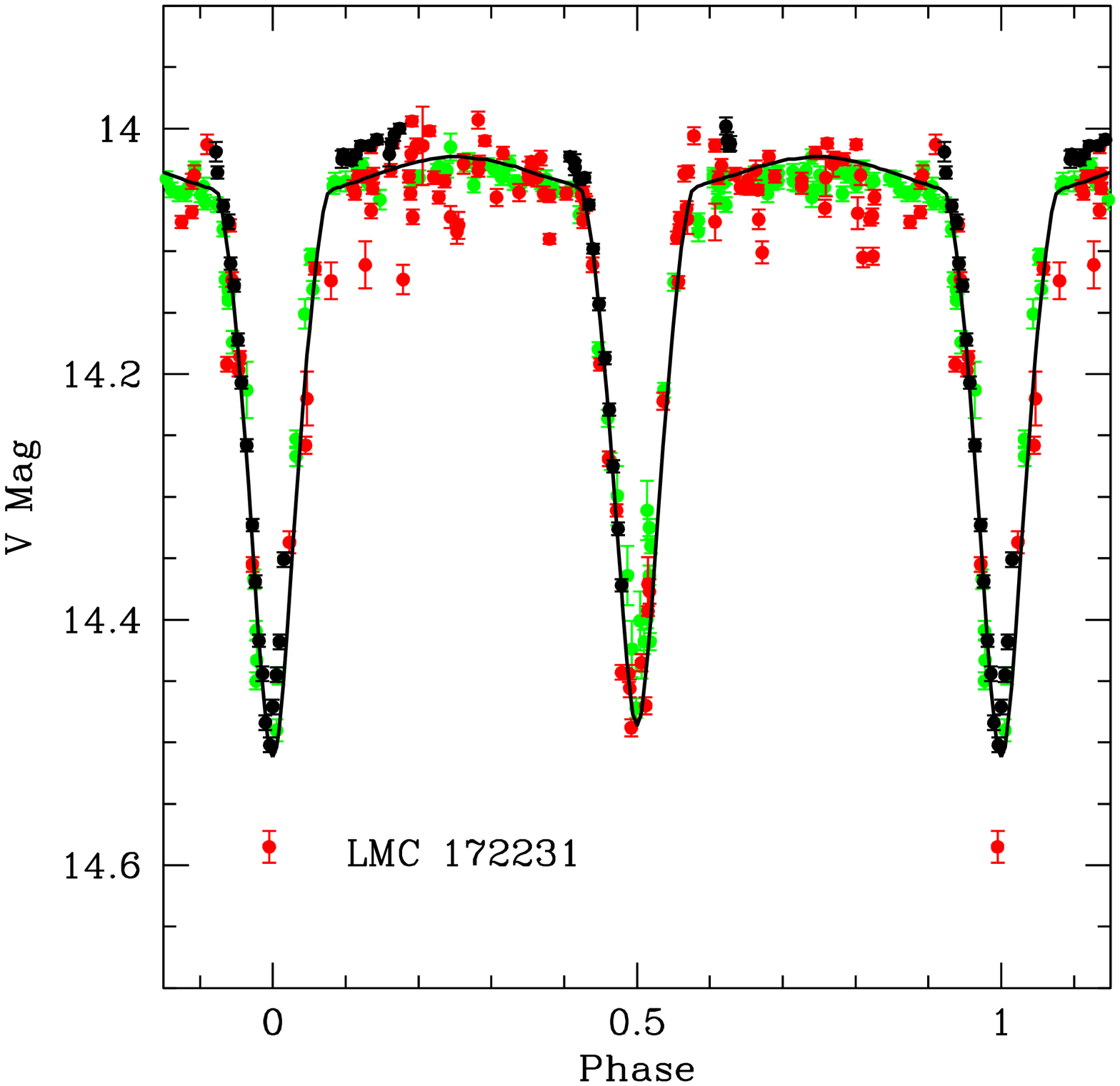}
\caption{\label{fig:lmc172231lc} Light curve of LMC 172231.  The data have been phased using a period of 
3.225414 days and a time of primary eclipse of HJD 2453591.469.  Black points denote data taken with
the Swope 1.0-m telescope, red points denote data taken with the SMARTS Yale 1.0-m telescope, and
green points denote data taken with the SMARTS 1.3-m telescope.  The light curve model is
shown in black. 
}
\end{figure}

LMC 172231 attracted our attention for its relatively long period (3.225414 days) and a light
curve that was characteristic of a detached system (Figure~\ref{fig:lmc172231lc}).  
A summary of the photometric data can be found in Table~\ref{tab:lcsummary}, and the 
photometry itself can 
be found in Table~\ref{tab:LMC172231phot}.  
Note from Figure~\ref{fig:lmc172231lc} the very
good agreement between the photometry from various telescopes and instruments.

\begin{figure}
\epsscale{0.48}
\plotone{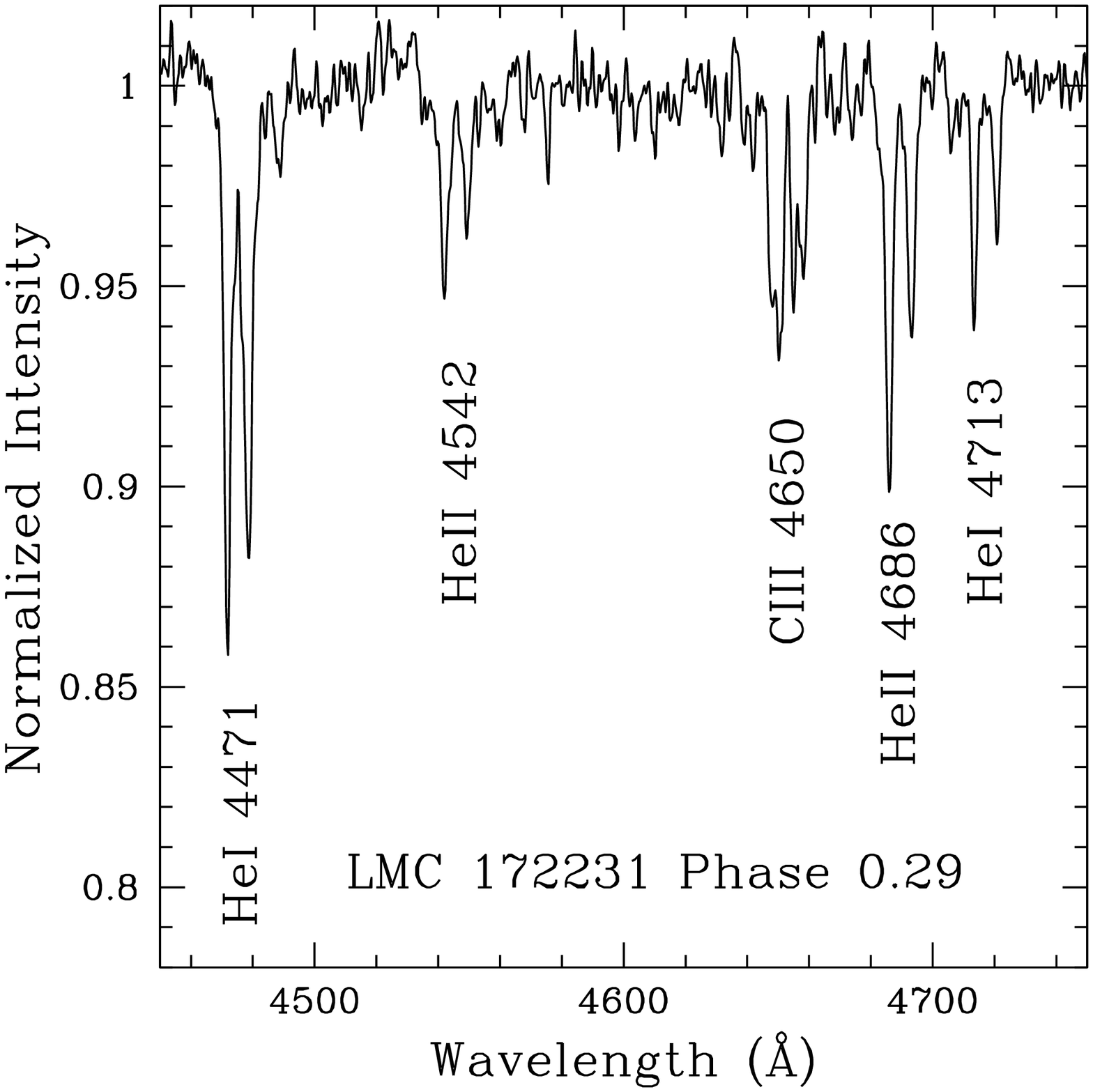}
\plotone{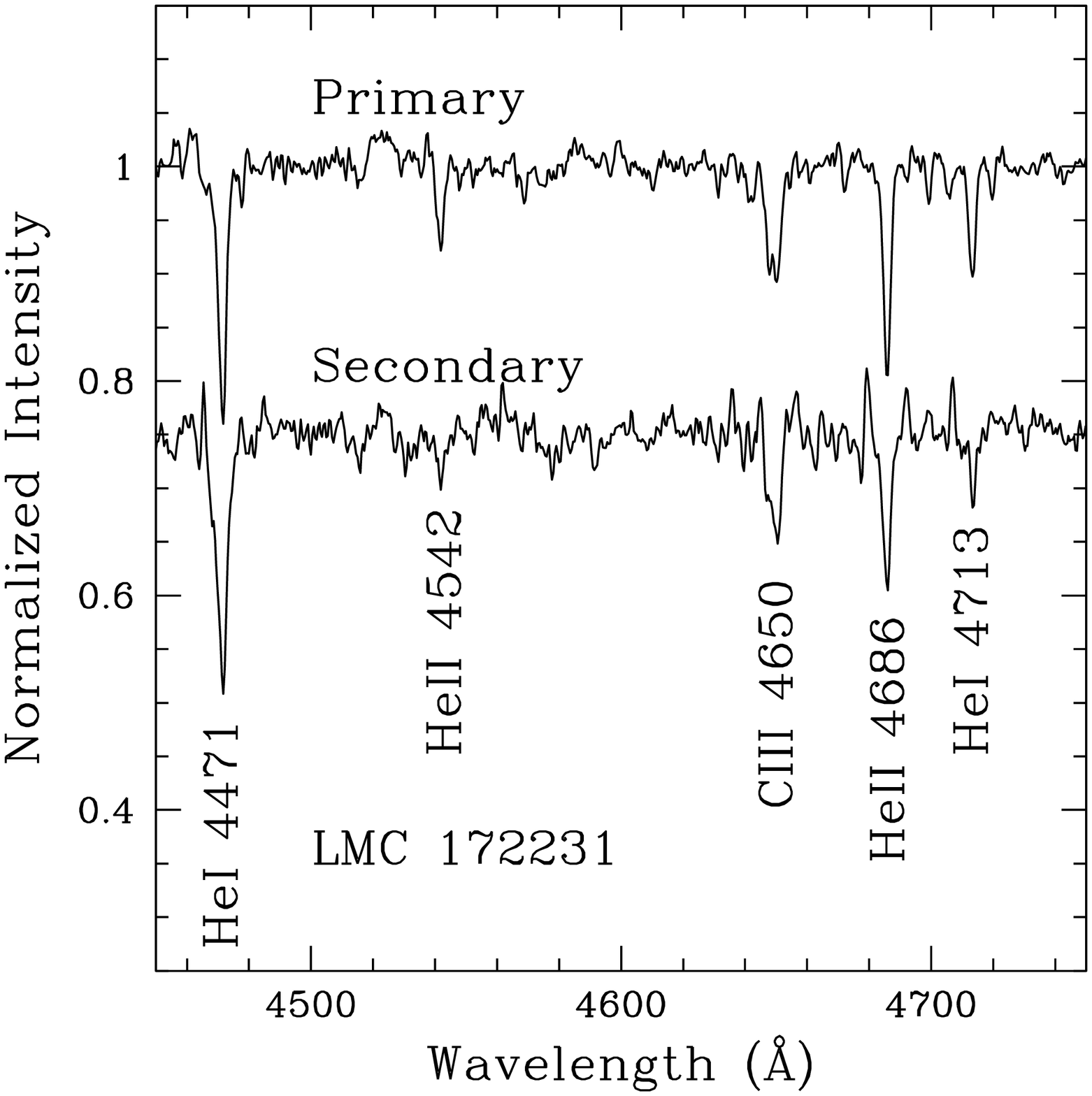}
\caption{\label{fig:lmc172231spec} Spectrum of the LMC 172231 system. A section of the spectral
region of primary importance for classification is shown. {\it Left:} The data were taken 
with MagE on HJD 2455143.825, at a phase of 0.29, i.e., just a little over 
a quarter of a cycle after primary eclipse.  The brighter and slightly earlier-type star is blue shifted, while
the fainter and slightly later type star is red shifted. The double lines are well separated, except for the CIII blend at 
$\lambda 4647-50-51$ which has a complex structure and was not used for radial velocities.
The original spectrum has a S/N of 400 per 4-pixel spectral resolution element, and has been
smoothed here by a 3-point boxcar average for display purposes.  {\it Right:} The spectra of the individual components separated by tomography.  The normalized spectrum of the secondary has been
shifted downwards by 0.25 normalized units.}
\end{figure}

Spectroscopy
revealed nicely separated double lines, characteristic of two late O-type dwarfs.  We show an
example of the star's spectrum at a double-lined phase in Figure~\ref{fig:lmc172231spec} {\it left}.
We initially classified the components as O8.5 V and O9~V, although the
spectra separated by tomography yield slightly later spectral types than the blended spectra 
indicated, as described below.
Radial velocities were measured primarily from He I and He II, although in a few cases some of the Balmer lines were also used.  The radial velocities are given in Table~\ref{tab:lmc172231vel}, along with the standard deviations of the means, and number of 
lines measured. 

The stars are roughly equally bright; the differences apparent 
in the spectra (Figure~\ref{fig:lmc172231spec}) are primarily due to slight differences
in spectral type (effective temperature) rather than luminosity.  
We will refer to the hotter star as the ``primary", keeping with the standard convention.

\subsubsection{Physical Parameters}

\begin{figure}
\epsscale{0.9}
\plotone{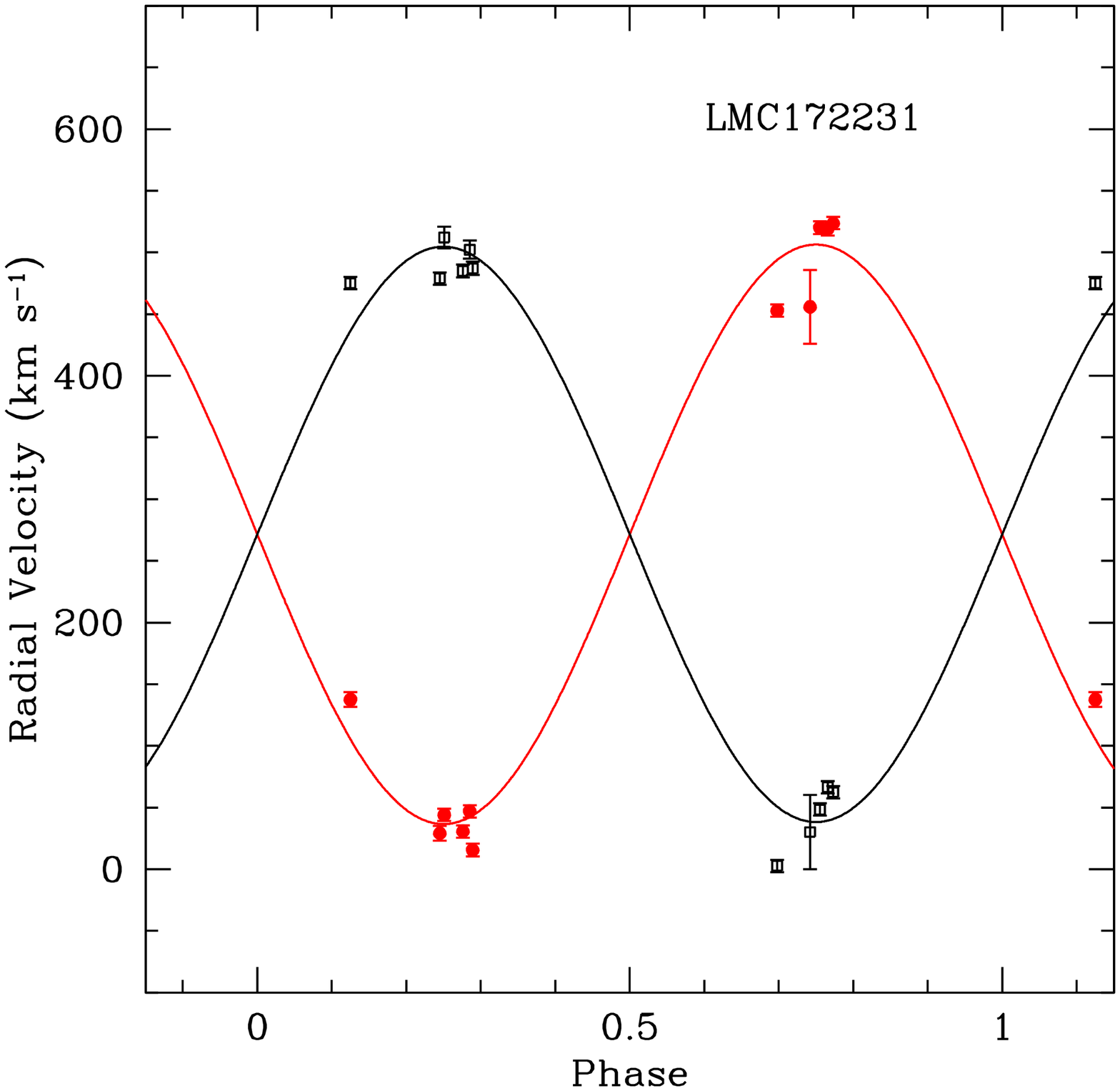}
\caption{\label{fig:lmc172231orbit} Velocity curve for LMC 172231.  The radial velocities of the primary (O9 V component) are
shown by filled red circles, and the radial velocities of the secondary (O9.5 V component) are shown by
black open squares.  The red and black  curves come for the best fit orbit solutions for the primary and
secondary, respectively.}

\end{figure}

If we allow the orbital eccentricity to be a free parameter, holding
only the period fixed, we find $e=0.000\pm0.032$, consistent with our expectation from the light
curve that the orbit is circular.  By adopting $e=0.00$ we can therefore solve for the optimal values of the orbital
semi-amplitudes $K$'s and center-of-mass motion $\gamma$, as described in Section~\ref{Sec:Orbits}.  The orbit velocity curves are shown in Figure~\ref{fig:lmc172231orbit}, and the physical
parameters are given in Table~\ref{tab:lmc172231physical}.    
The minimum masses are essentially
identical for the two components, $17.1\pm0.3 M_\odot$  and $17.2\pm0.3 M_\odot$.
The radial
velocity curves show good agreement with the data.

\begin{figure}
\epsscale{0.9}
\plotone{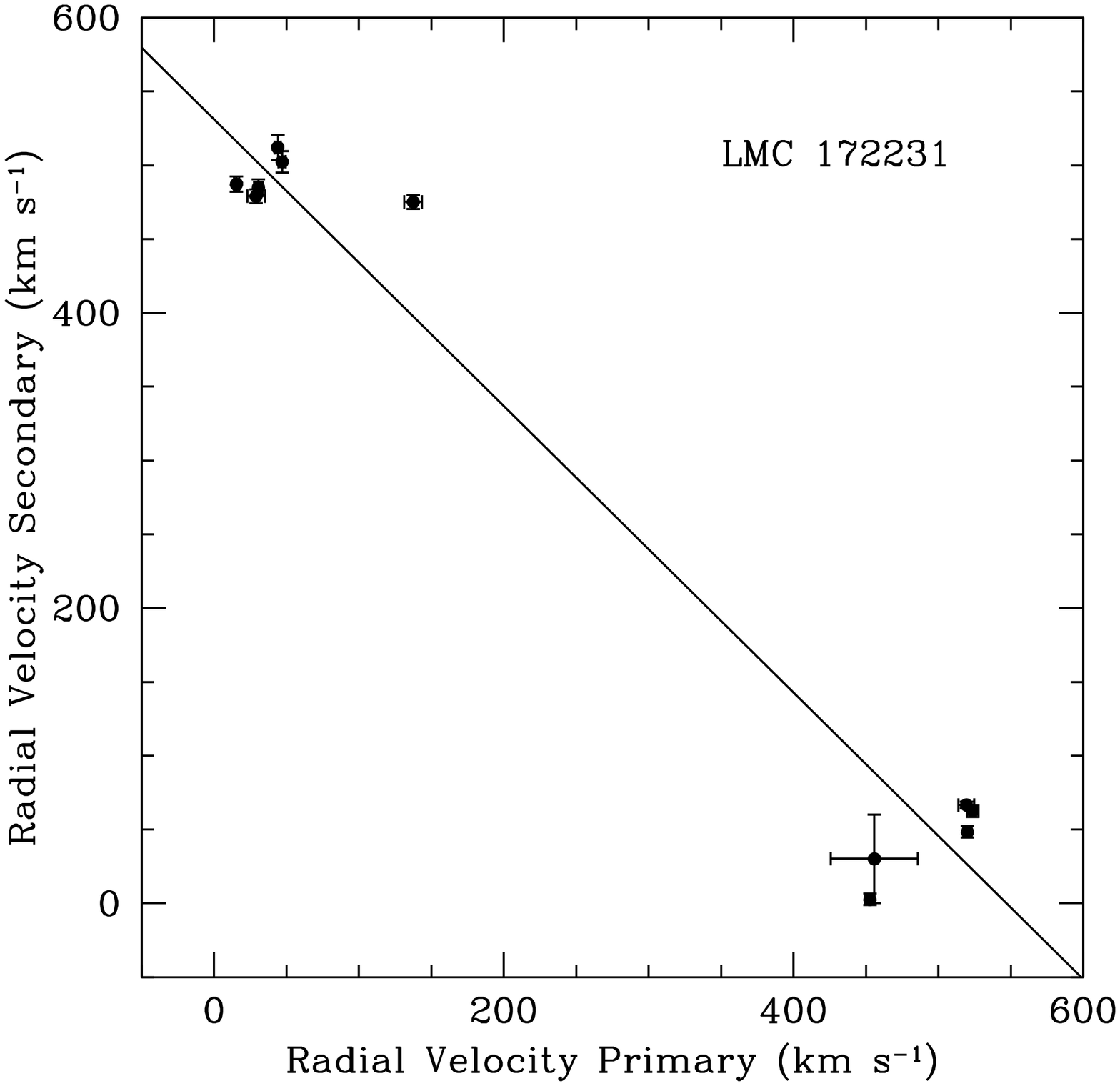}
\caption{\label{fig:lmc172231wil} Wilson diagram for LMC 172231.  The velocities of the two components are shown plotted against each other, along with the best line fit. The slope and intercept are consistent with those derived from the orbit solution shown in Figure~\ref{fig:lmc172231orbit}.}
\end{figure}

As a further check we have plotted the data in a Wilson diagram (Wilson 1941; see Section~\ref{Sec:Orbits}) 
shown in Figure~\ref{fig:lmc172231wil}.
In obtaining the ``best fit" it would be inappropriate to use standard linear least-squares fitting (e.g., Bevington 1969) as
there are errors associated with both the ordinate and abscissa.  Following Section 15.3 of Press et al.\ (1997)\footnote{See also
the excellent description at http://mathworld.wolfram.com/LeastSquaresFitting.html.}, we
determine the best ``straight line fits" and their associated errors explicitly including our error estimates for both $v_1$ and $v_2$.  We compare the orbital results with the Wilson's results
in Table~\ref{tab:wilson}.  Note the good agreement both in $\gamma$ and in $K_1/K_2$,
and recall that for Wilson's method to work we make no use of the phase information or
even the periods.

For the light curve analysis we started with
effective temperatures of 36,500 K and 35,000,
somewhat hotter than our spectral types would indicate (Massey et al.\ 2006)
 in order to better match the observed absolute magnitude for
the system.  However,   the  existent photometry for LMC 172231 is not
in good agreement.  The
out-of-eclipse light-curve magnitude ($V\sim 14.04$)
is consistent with Massey (2002)'s $V=13.96$ and Schild \& Testor (1992)'s  $V=14.02$.
The {\it average} reddening of LMC OB stars is $E(B-V)=0.13$ and so we might expect
colors for late O-type dwarfs of $B-V\sim-0.1$ and $U-B\sim-1.0$.
Massey (2001) obtains $B-V=+0.10$ and $U-B=-1.14$.  These values are inconsistent
with each other (the first is too red, and the second too blue),
 and suggest that the star may have been entering or exiting eclipse
when the data were taken.  Testor \& Niemela (1998) obtained  $B-V=-0.08$, close to our
expectations, and so for the purposes of computing $M_V$ for the system, we adopt
the Testor \& Niemela (1998) values\footnote{Testor \& Niemela (1998)'s $U-B=-0.88$ is
redder than expected for its colors, suggesting that the {\it U} might be in error.}.  Adopting the intrinsic color $(B-V)_0=-0.26$  of Martins \& Plez (2006),
we derive $M_V=-5.0\pm0.3$ for the system, where the error reflects a
modest uncertainty of 0.07 mag in $E(B-V)$, and hence $0.2$~mag in $A_V$, plus another 0.1~mag
uncertainty in the distance modulus of the LMC (see van den Bergh 2000).  

The preliminary light curve solution yielded an estimate of the flux ratio $F_{V_2}/F_{V_1}\sim 0.9$.  We used this with the radial velocity solution to perform a tomographic analysis using
seven MagE spectra covering both double-lined quadratures.   We fit these with FASTWIND,
finding significantly lower effective temperatures than had been first assumed.  It was also
clear from the model fitting that a flux ratio of 0.8 was more appropriate
than 0.9, constrained mostly by the Balmer line depths. 
 A single additional iteration was made fixing the effective temperatures to
the values derived from FASTWIND.  The flux ratio of 0.8 then yielded the excellent fit
to the light curve shown in Figure~\ref{fig:lmc172231lc}.   Note that the light curve model
is computed with $e=0.00$ and shows excellent agreement both the separation of the eclipses
{\it and} the durations of the eclipses, substantiating that the orbit is circular.

The tomographically separated spectra are shown in 
Figure~\ref{fig:lmc172231spec} {\it (right)}.
A casual inspection shows that one component is indeed slightly hotter than the other,
based upon the relative amounts of He I $\lambda 4471$ and He II $\lambda 4542$.   With the
tomographically separated spectra, we classify the components as a bit
later than we had originally with the blended spectra, and adopt here O9~V and O9.5~V.

\begin{figure}
\epsscale{0.32}

\plotone{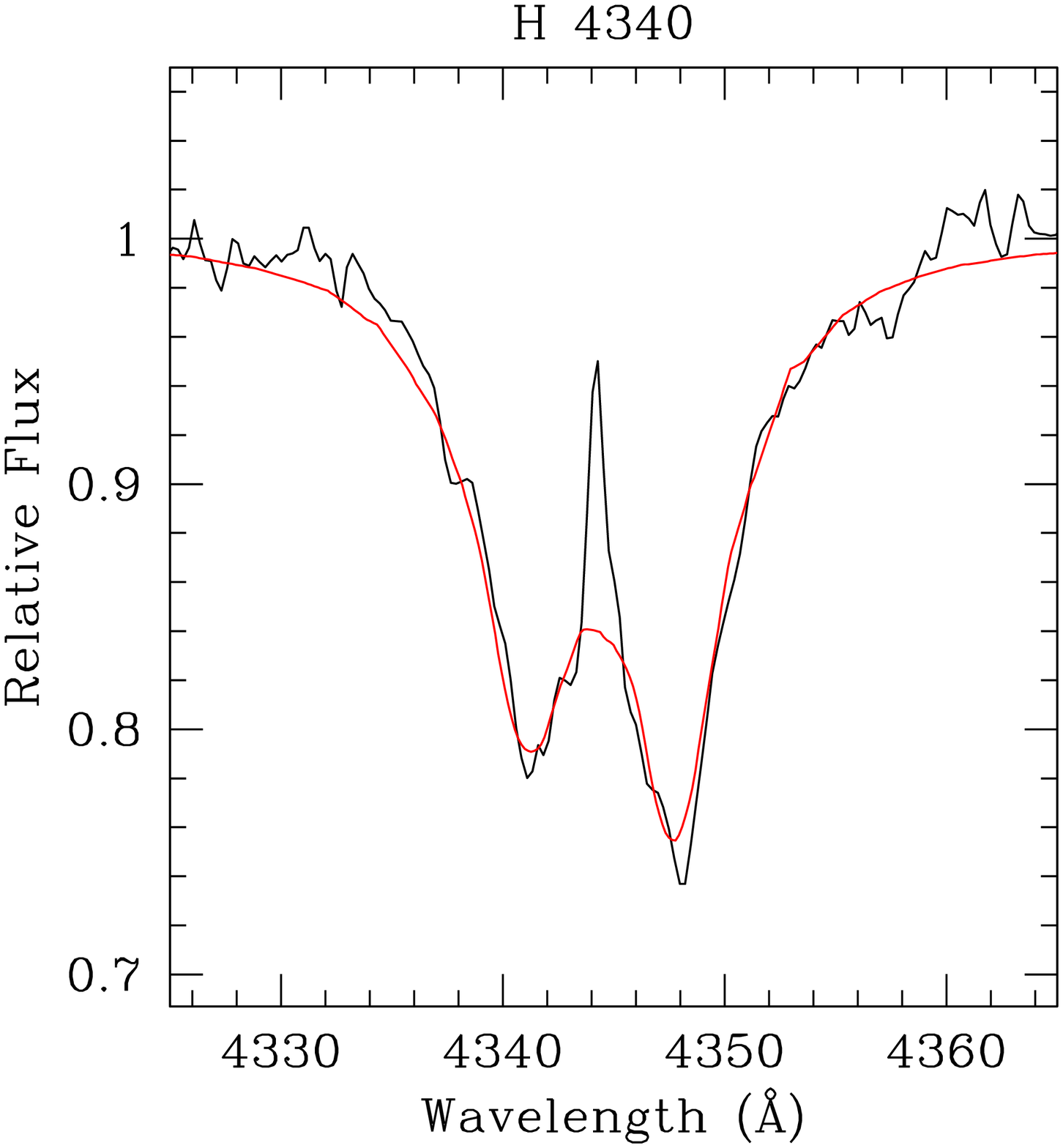}
\plotone{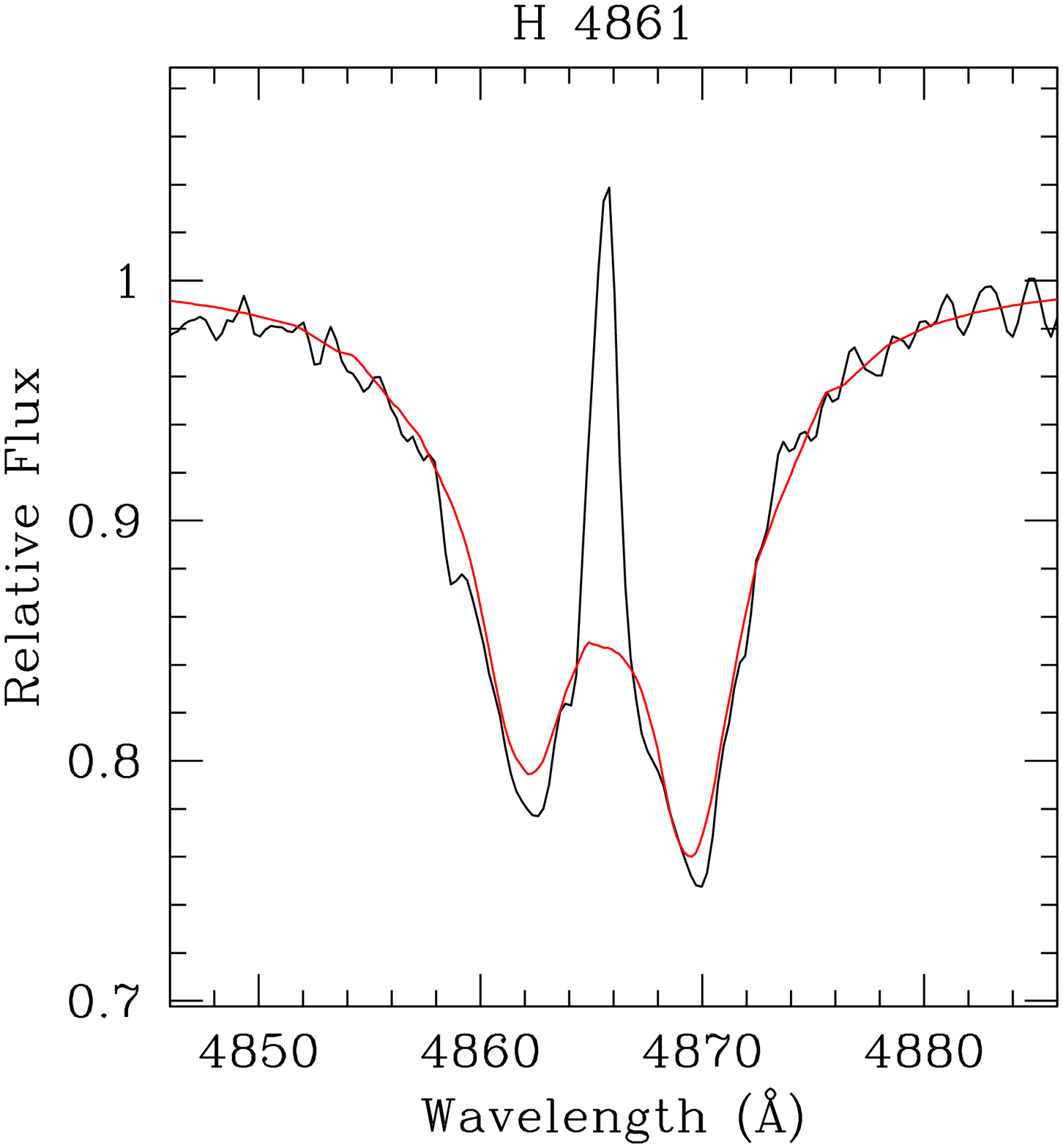}
\plotone{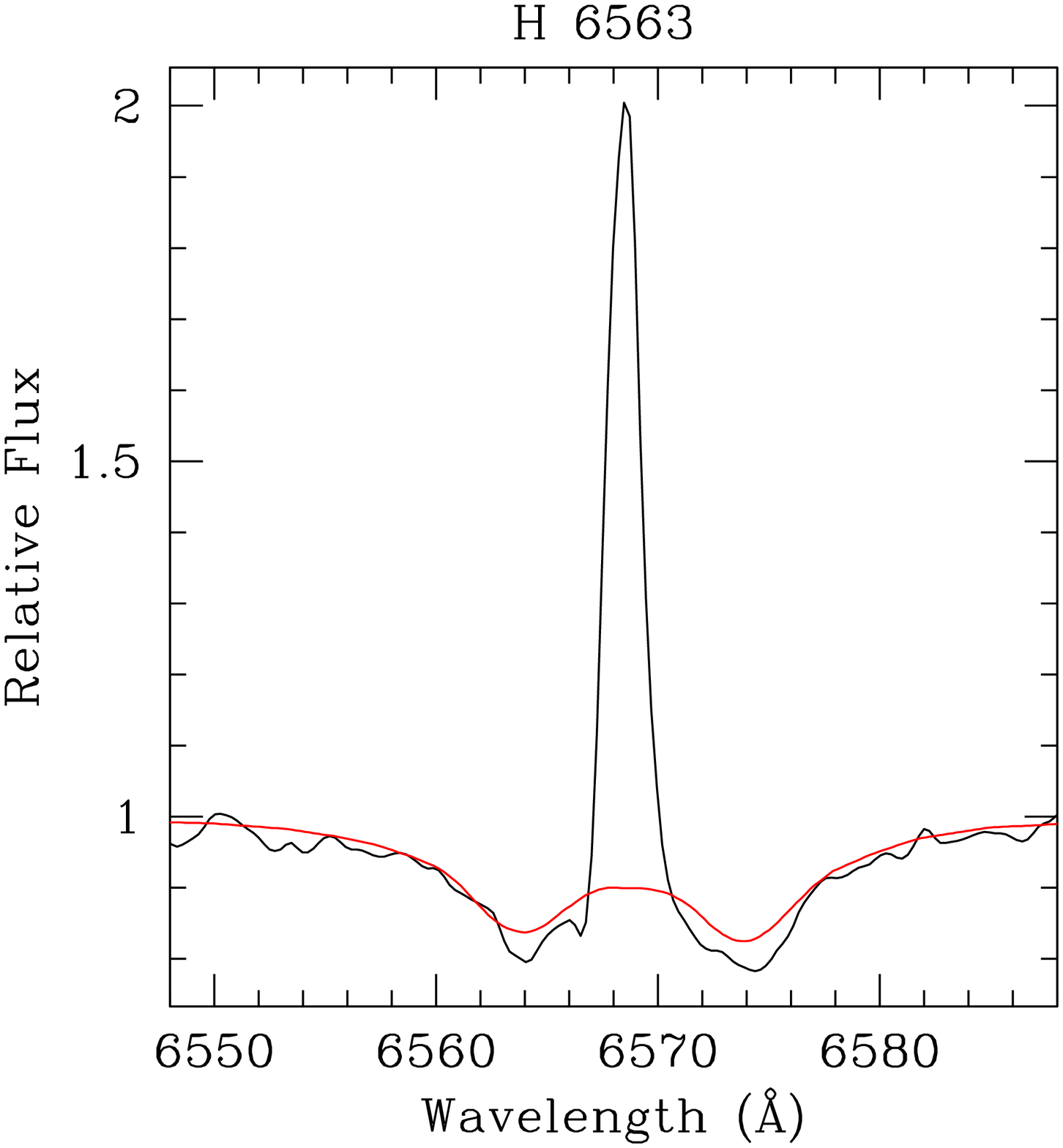}
\plotone{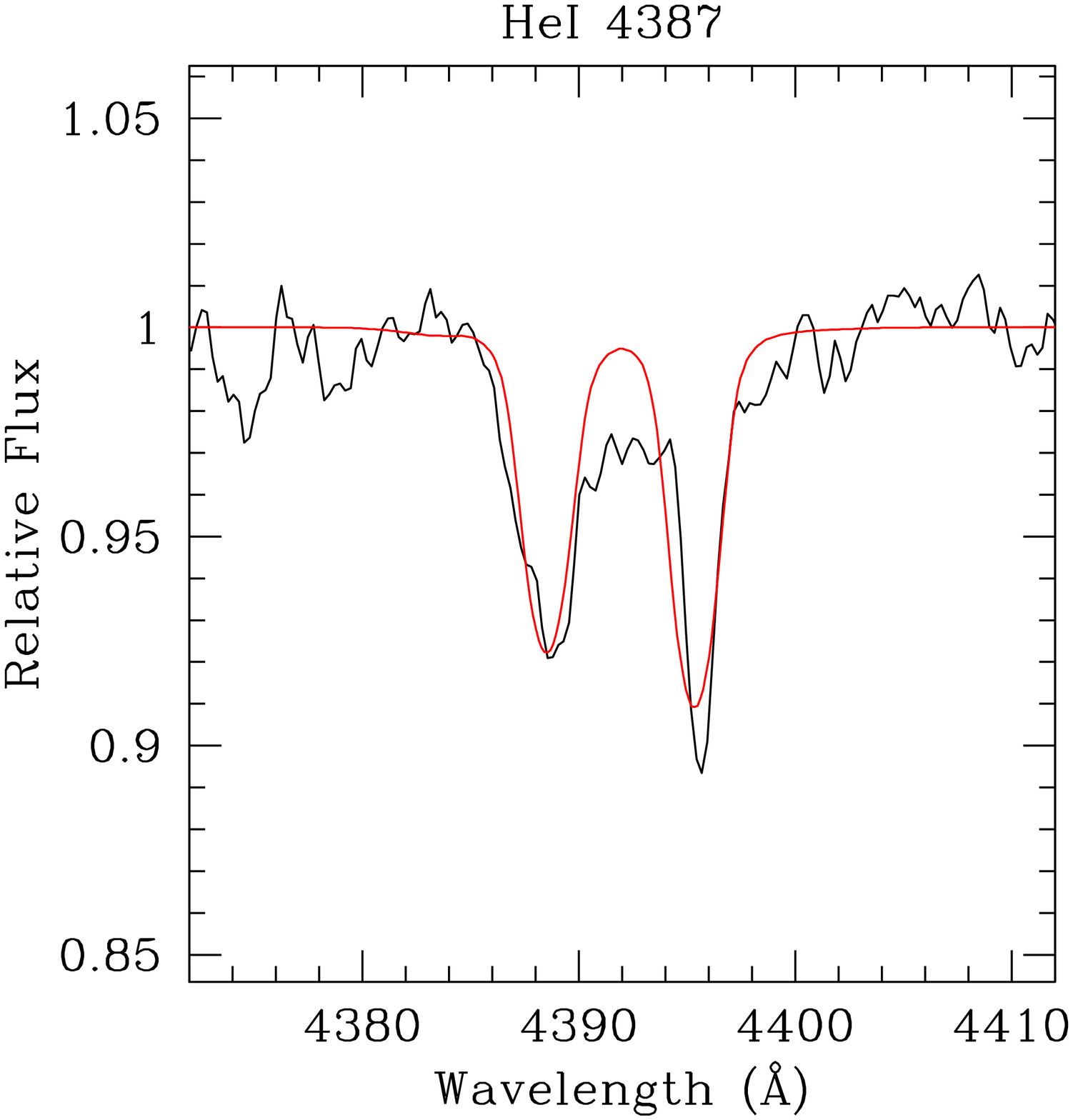}
\plotone{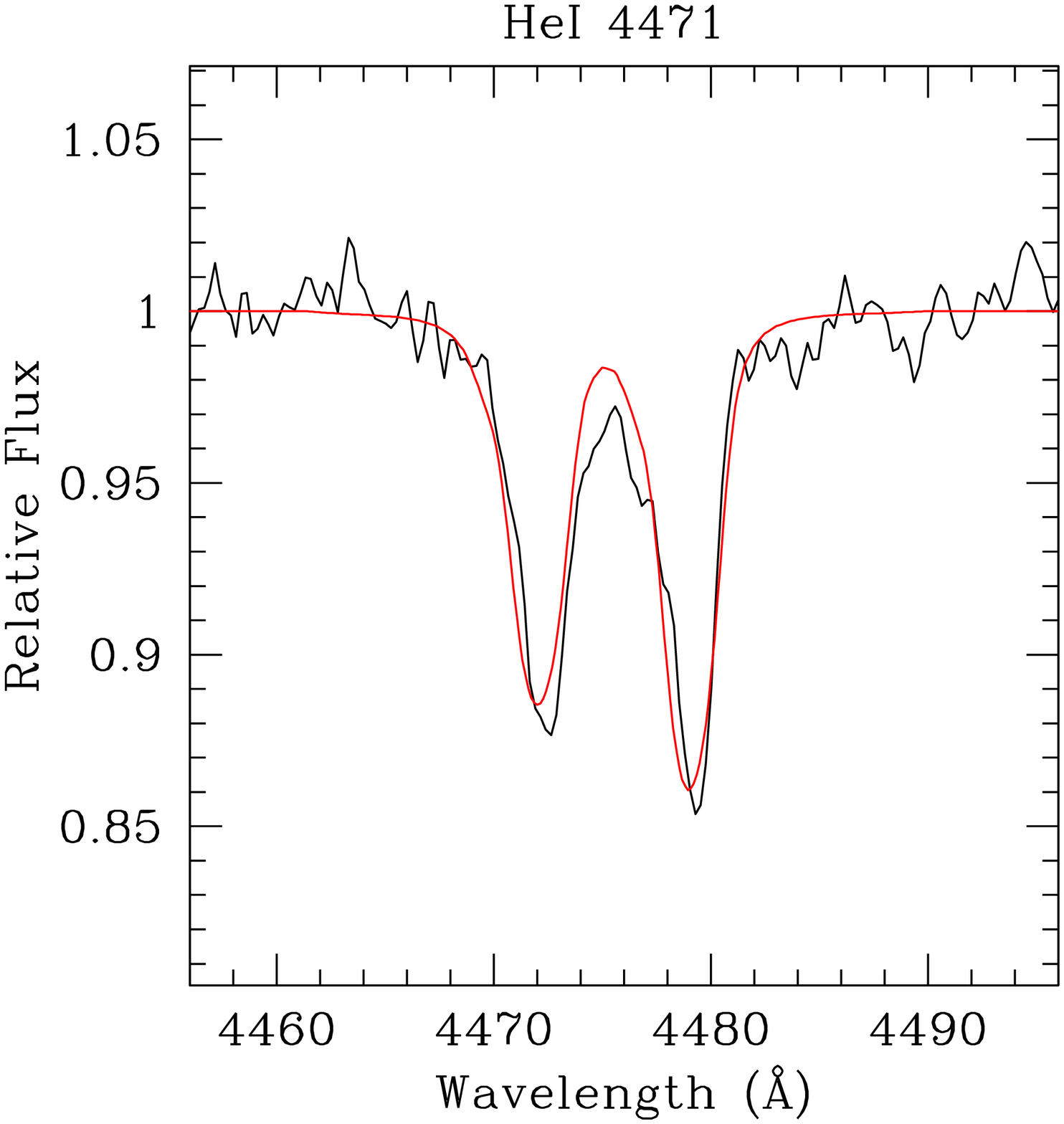}
\plotone{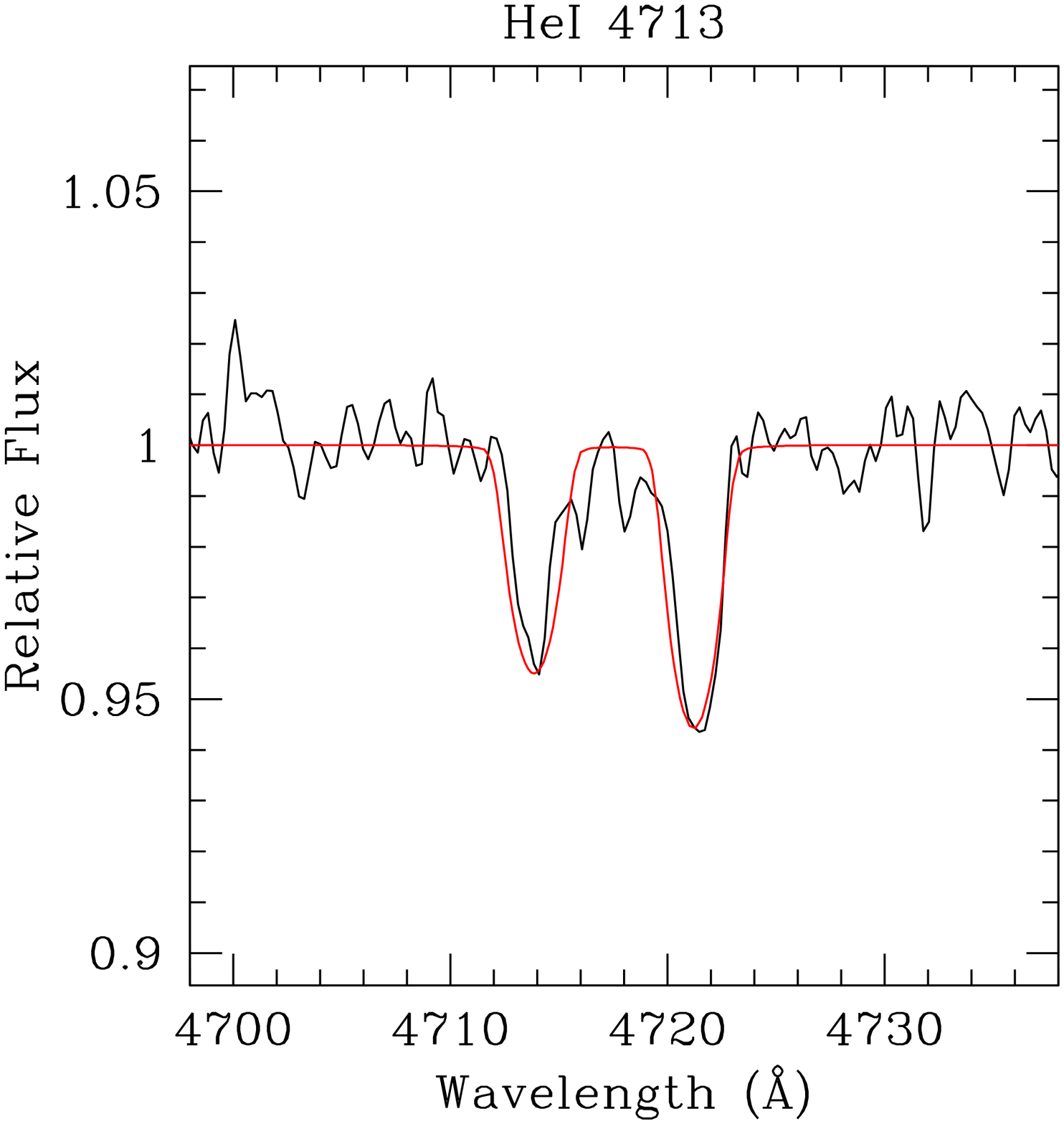}
\plotone{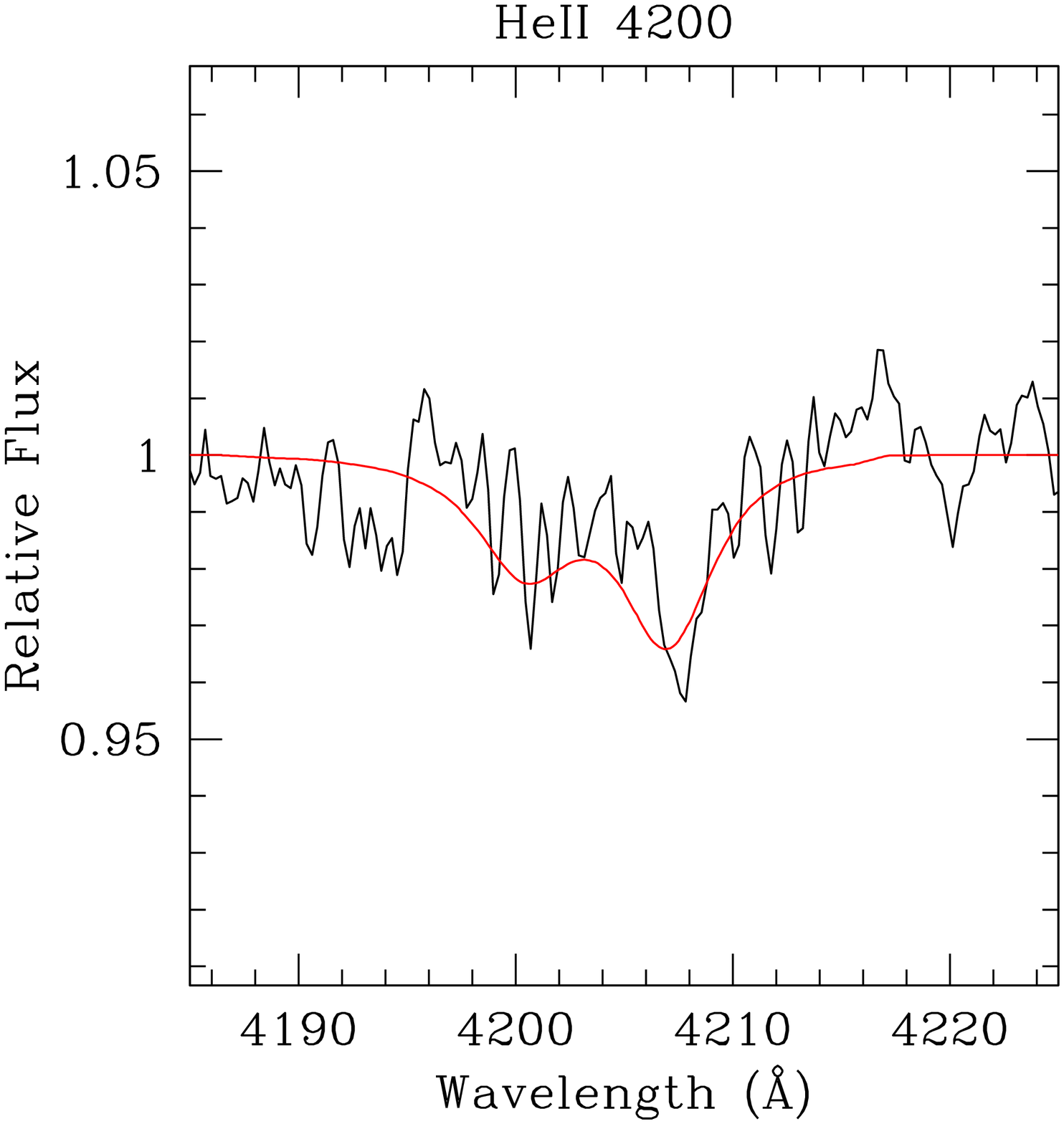}
\plotone{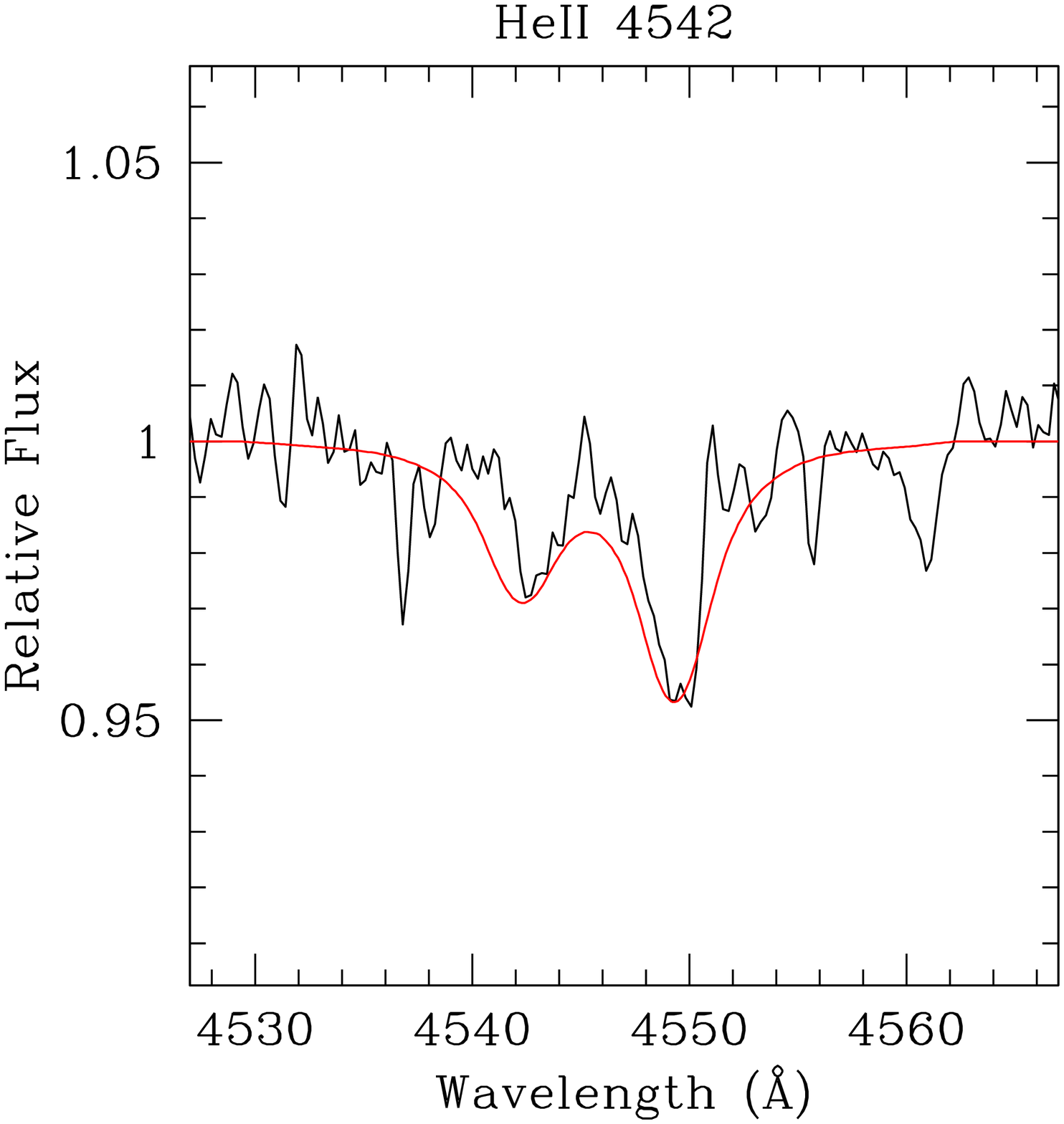}
\plotone{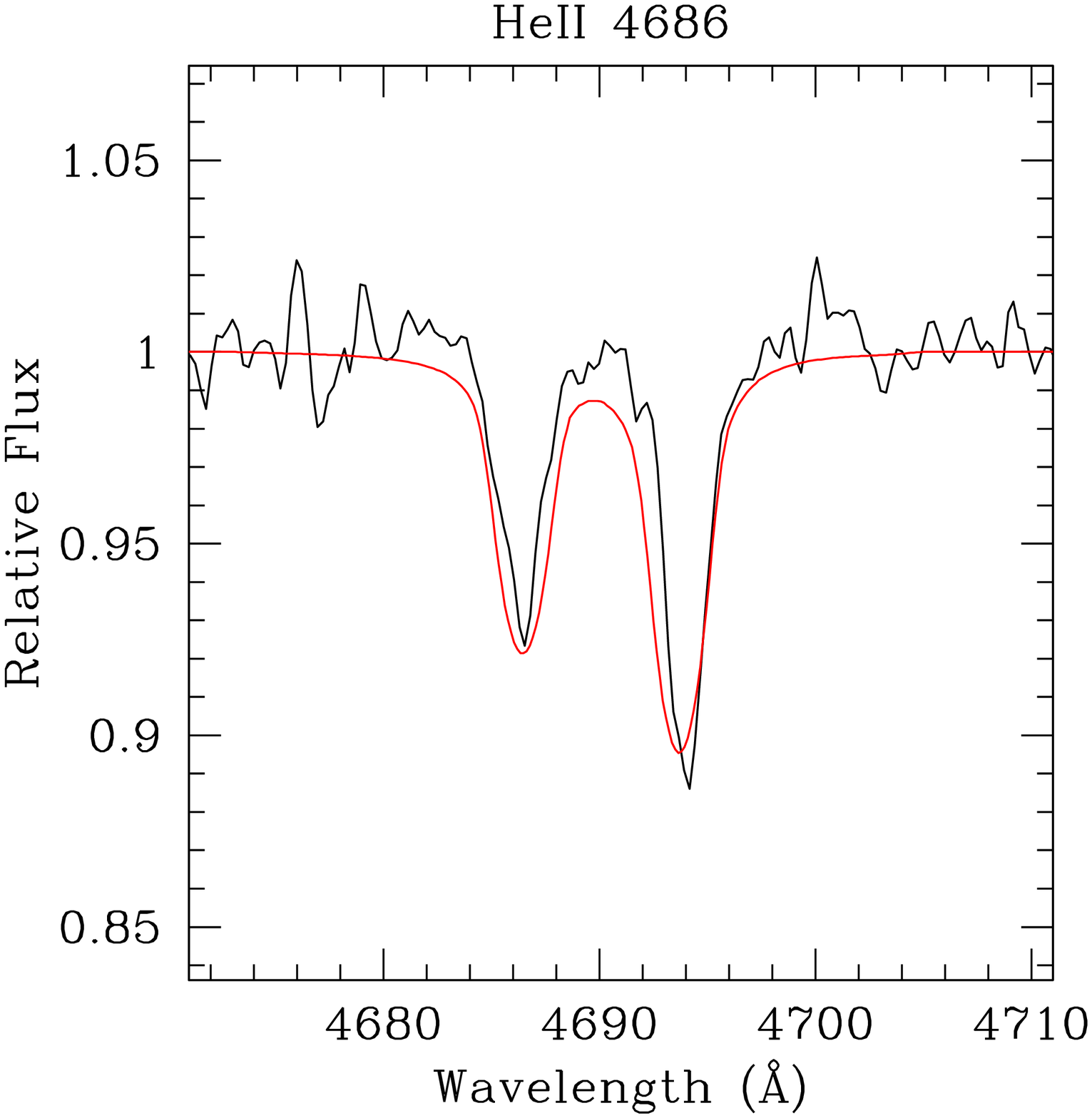}
\caption{\label{fig:LMC172231mods} LMC 172231 Comparison with Stellar Atmosphere Model.
We compare the FASTWIND models (smooth, red curves) with the observed spectrum.
The spectrum used was taken with MagE on HJD 2454877.619 and corresponds to a phase of 0.755.
The top row shows the H$\gamma$, H$\beta$, and H$\alpha$ lines; note the presence of some remaining nebular emission
at line center.  The second row shows the  He I $\lambda4387$, He I $\lambda$4471,
and He I $\lambda4713$ lines.  The bottom row shows the He II $\lambda$ 4200, He II $\lambda 4542$, and He II $\lambda 4686$ lines. The latter is primarily sensitive to the 
mass-loss rate and wind law.
}
\end{figure}

The adopted FASTWIND model fits are shown in 
 Figure~\ref{fig:LMC172231mods}.  We have opted to show the fits with the original
 (non-separated) data as this shows the good agreement not only with the adopted
 effective temperatures, but also the final flux ratio, and the radial
 velocity of the orbit solution.  We wound up adopting rotational velocities of 110 km s$^{-1}$ as these
 gave good fits to the lines, and we note that the synchronous rotation velocities of the two components
 are $109\pm5$ km$^{-1}$ and $101\pm5$ km$^{-1}$, in good agreement with these values.

The final masses we derive are $17.5\pm0.3M_\odot$ for the O9~V component and $17.6\pm0.3$ for the O9.5~V component.  The bolometric luminosities were computed as described above, and are $\log L/L_\odot \sim 4.8$ and 4.7 for the O9~V and O9.5~V components,
respectively.

It is interesting to note that the orbital {\it inclination} changed by only $1\sigma$ (from 
$82\fdg5\pm0.5$ to $83\fdg0\pm0.5$) by our iterative application of tomography and
light-curve analysis.  What {\it did} change significantly were the stellar radii and bolometric
luminosities.  Since masses are impossible to determine by ``inspection" one would expect
that the greatest uncertainties in the MLR are the masses.  But, this exercise emphasizes
that the luminosities of the individual components also require careful analysis.
In the end, the total visual luminosity we derive for the system {\it does} agree with the
``observed" values, within the errors.  But, the errors are large.  Had we instead eschewed
the tomographic analysis, and insisted on matching the absolute visual magnitude, we
would have adopted effective temperatures and radii that gave just as good a fit to the
visual light curve, but would have overestimated the luminosities (we believe) by  0.3~mag.
In practice our initial solution had a difference in effective temperatures larger than we 
adopted, and this led to a different flux ratio as well.   The uncertainties in ``L" in the MLR
may not always have been fully appreciated.  We note below (Section~\ref{Sec-MLR}) that
this has implications for using massive eclipsing binaries as distance indicators.

\subsection{ST2-28, O7~V + O8~V}

\begin{figure}
\epsscale{0.9}
\plotone{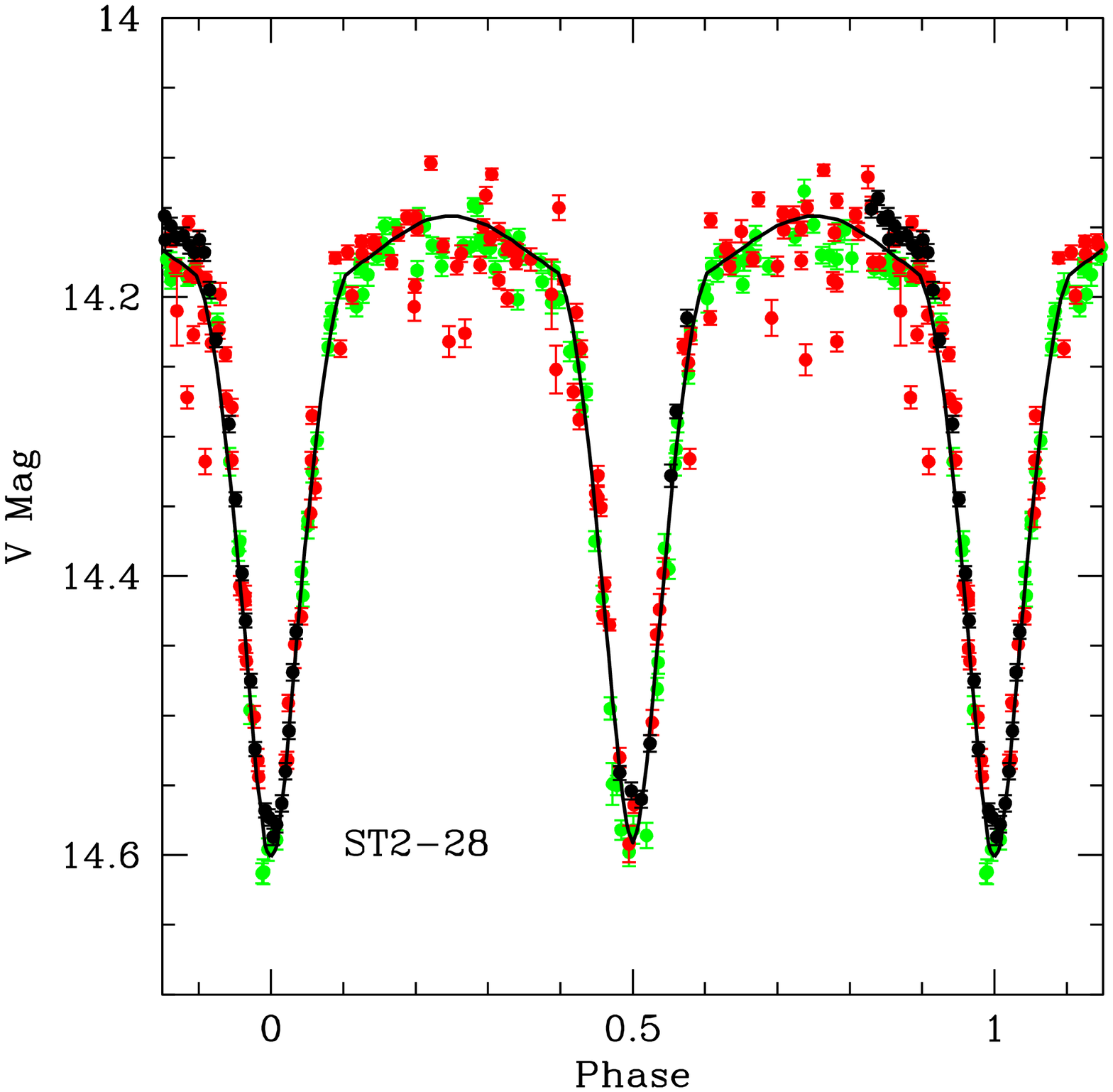}
\caption{\label{fig:ST2-28lc} Light curve of ST2-28.  The data have been phased using a period of 2.762456 days and a time of primary eclipse of HJD 2453590.217.
Black points denote data taken with
the Swope 1.0-m telescope, red points denote data taken with the SMARTS Yale 1.0-m telescope, and
green points denote data taken with the SMARTS 1.3-m telescope. The light curve model is
shown in black. }
\end{figure}

The designation ST2-28\footnote{Known to SIMBAD as [ST92]2-28.} comes from Schild \& Testor (1992), where the star is listed as number 28 in Zone 2, a region that covers the Lucke-Hodge 90 OB association (Lucke \& Hodge 1970). The star was also cataloged as [M2002] LMC 163763 by Massey (2002).  It was identified automatically
as a probable eclipsing binary (MACHO 82.9010.36) in the MACHO 
database by Alcock et al.\ (1997)\footnote{See VizieR II/247.} with a
period of 1.38122 days, half of the 2.762446 day period determined by 
Derekas et al.\ (2007) in their reanalysis of the LMC MACHO eclipsing binaries.  This
period is in good accord with the  2.762456 day period we find from our own data. Our own
photometry is summarized in Table~\ref{tab:lcsummary} and the data given in
Table~\ref{tab:ST2-28phot}. Our
light curve is shown in Figure~\ref{fig:ST2-28lc}.  The eclipse depths are 
nearly identical, as expected from the similarity in 
spectral types.     The phase difference between primary and secondary
eclipses appears to be indistinguishable from 0.5 with both eclipses showing the same duration, 
further validating our having adopted a circular orbit.
 It is clear from the spectra that the earlier component
is significantly brighter visually.

\begin{figure}
\epsscale{0.48}
\plotone{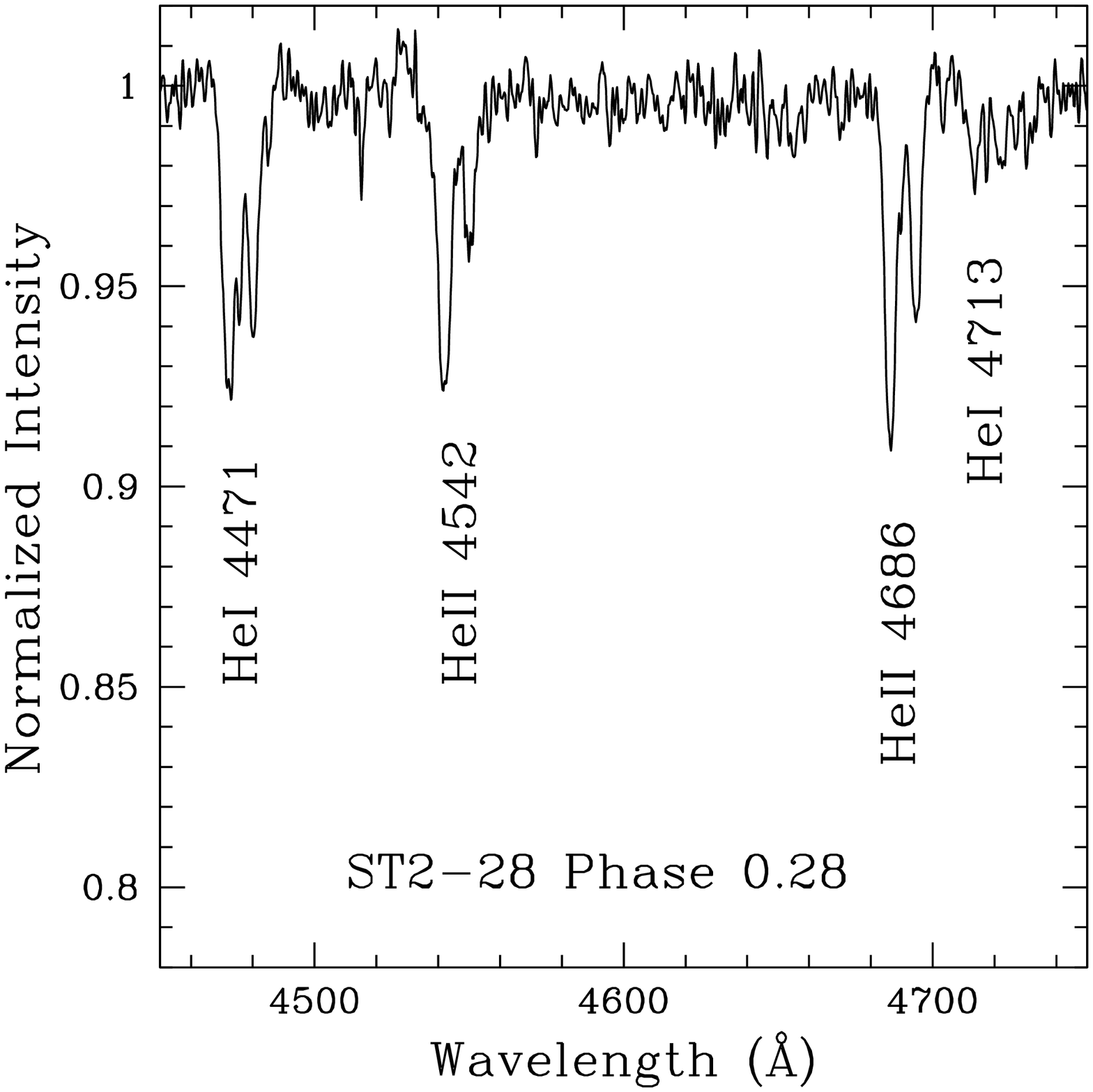}
\plotone{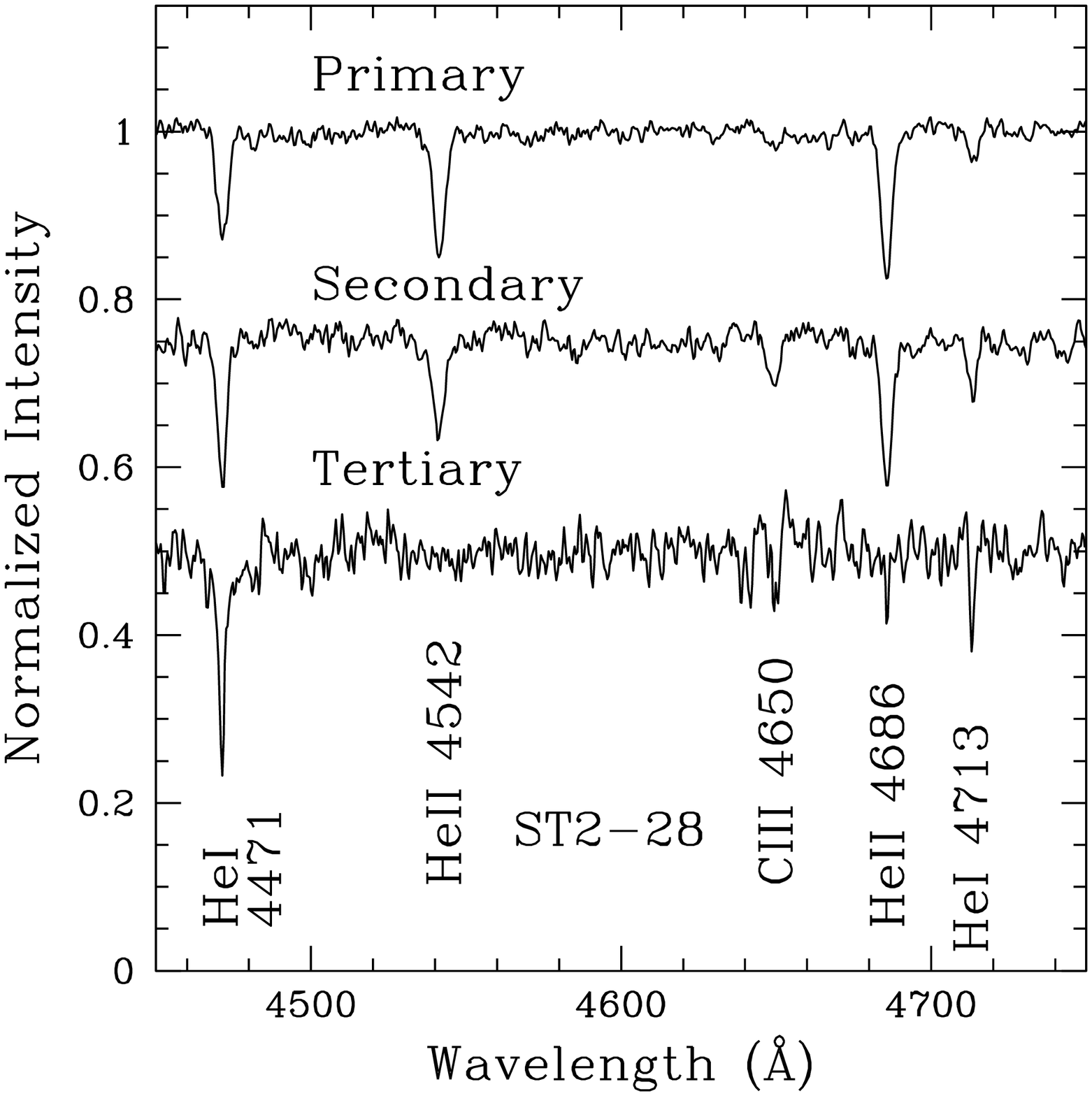}
\caption{\label{fig:ST2-28spec} Spectrum of the ST2-28 system.  A section of the spectral
region of primary importance for classification is shown.  {\it Left:} The data were taken 
with MagE on HJD 2455140.732 at a phase of 0.28, i.e., just a little over 
a quarter of a cycle after primary eclipse.  The brighter and slightly earlier-type star is blue shifted, while
the fainter and slightly later type star is red shifted. The double lines are well separated.
The third component is barely visible as a small blip
in the He I $\lambda 4471$ line.  
Only the He II lines were used for the orbit solution.  The original spectrum has
a S/N of 330 per 4-pixel spectral resolution element, and has been smoothed here by
a 3-point boxcar average for display purposes.  {\it Right:} The spectra of the individual components separated by tomography.  The normalized spectrum of the secondary has been
shifted downwards by 0.25 normalized units, and that of the tertiary has been shifted
downwards by 0.5 normalized units.}
\end{figure}

We show one of the double-lined spectra in Figure~\ref{fig:ST2-28spec} {\it left}. Schild \& Testor (1992) classify the star as O8~V.  We initially identified double lines by
observing the system at quadrature, and classified the components as O7~V and O8~V.
Careful subsequent inspection, however,  revealed a third component in the He I lines.
This component did not change in radial velocity, and it is not obvious in the He II lines.
The lack of He II indicates  
it is a B-type star, a result we confirm below in our
tomographic separation of the three components.
 Presumably it is either a line-of-sight companion,
or, more probably, a distant third member of the system. 

Our radial velocities were based upon fitting two Gaussians (and indeed the third component
is only barely visible at He I), and although the spectral
lines of each component appear to be well separated from each other and the stationary
third component, we were concerned that the third component could affect our
velocities.  Therefore we restrict our analysis to the He II lines, as no third component
appeared to be present in these lines, except for a faint hint at He II $\lambda 4686$;  
we confirm this below by our tomographic
analysis.    The radial velocities are given in Table~\ref{tab:ST228vel}, along with the
standard deviation of the means and the number of lines measured.

\subsubsection{Physical Parameters}

\begin{figure}
\epsscale{0.9}
\plotone{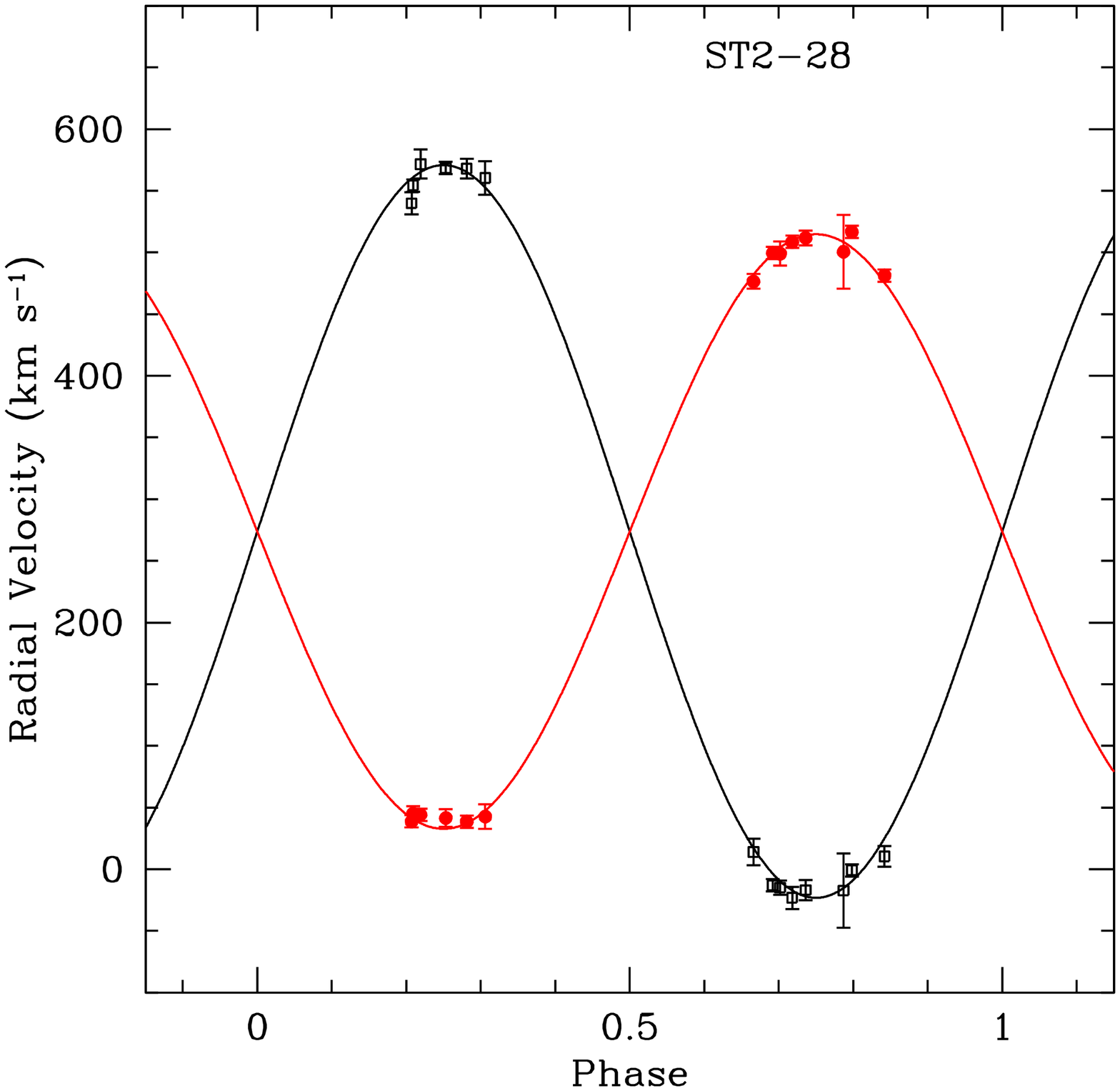}
\caption{\label{fig:ST2-28orbit} Velocity curve ST2-28.  The radial velocities of the primary (O7 V component) are
shown by filled red circles, and the radial velocities of the secondary (O8 V component) are shown by
black open squares.  The red and black  curves come for the best fit orbit solutions for the primary and
secondary, respectively.}
\end{figure}

Allowing the eccentricity to vary resulted in a best value $e=0.005\pm0.020$, and we
therefore adopted a circular orbit.   The orbit solution was particularly
well determined, and we show the parameters in Table~\ref{tab:st228physical} and
in Figure~\ref{fig:ST2-28orbit}.   The inferred minimum masses $m\sin^3i$ are $24.6\pm0.4$
for the O7~V primary, and $20.0\pm0.3$ for the O8~V secondary.   

\begin{figure}
\epsscale{0.9}
\plotone{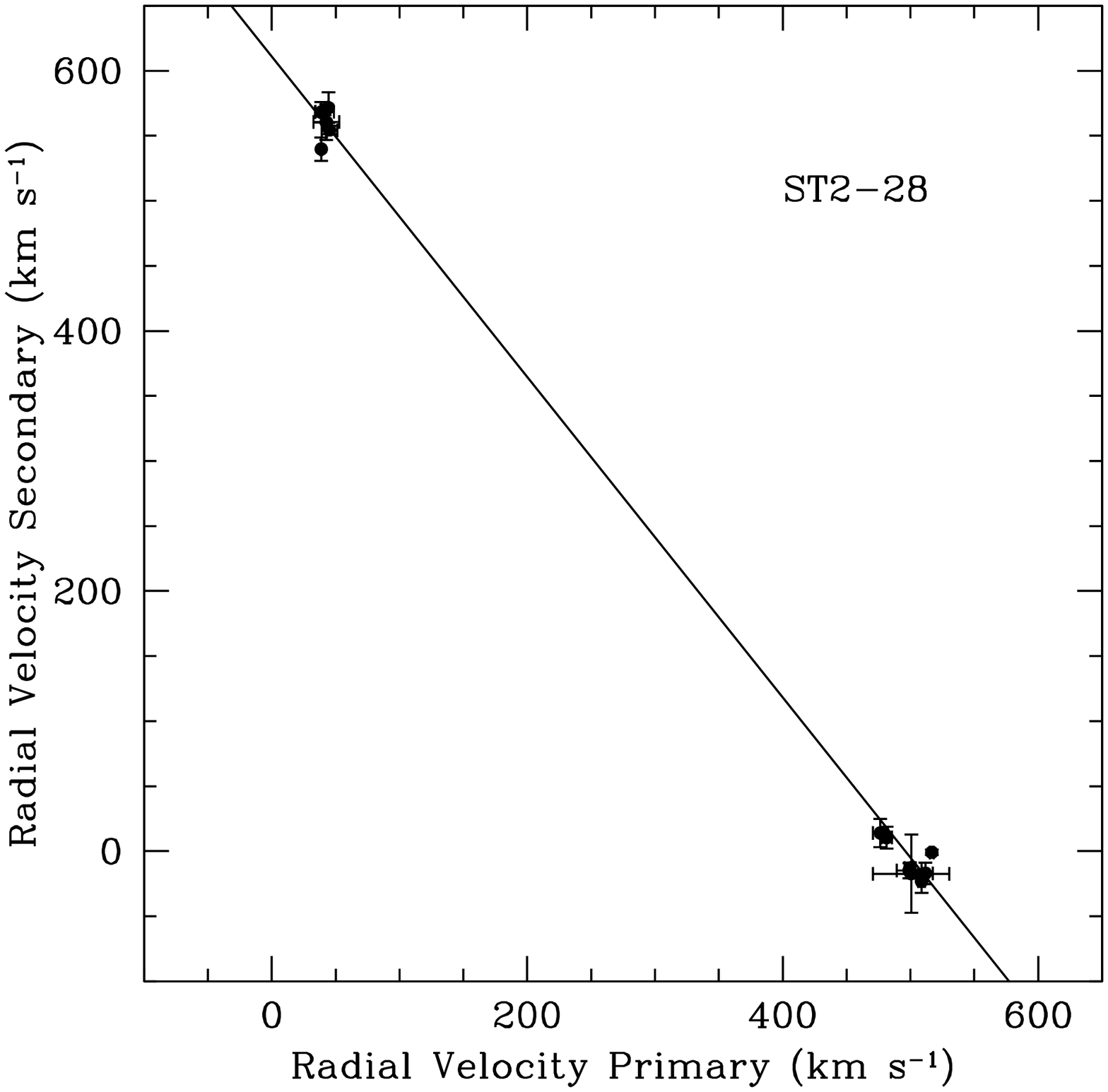}
\caption{\label{fig:st228wil} Wilson Diagram ST2-28.  The velocities of the two components are shown plotted against each other, along with the best line fit.
The slope and intercept are consistent with those derived from the orbit solution shown in Figure~\ref{fig:ST2-28orbit}.}
\end{figure}

Our Wilson plot for this star is shown in Figure~\ref{fig:st228wil}, and is compared
to the orbital parameters in Table~\ref{tab:wilson}.  The agreement both for the $\gamma$ velocity and the inverse mass ratio $K_1/K_2$ is excellent.  

Photometry of the star in the literature varies, doubtless due to some observations
having been made during eclipse. Massey et al.\ (2000) found $V=14.22$
and $B-V$=0.09, while Massey et al.\ (2002) found $V=14.18$, $B-V=0.06$.  
These values are considerably brighter than the $V=14.63$ found by Schild \& Testor (1992).
Our light-curve suggests $V=14.15$ outside of eclipse, in accord with the maximum
brightness $V=14.13$ found by Alcock et al.\ (1997) in the MACHO data.  We
adopt $V=14.15$ and $B-V=0.09$.  The intrinsic color from Martins \& Plez (2006) for 
an O8-O9~V star is $(B-V)_0=-0.27$, so $M_V=-5.5\pm0.3$.

The preliminary light curve solution using effective temperatures of 37,000 and 35,800 K
yielded an estimate of the flux ratio $F_{V_2}/F_{V_1}\sim 0.63$, and a contribution of the third component  ($F_{V_3}/(F_{V_1}+F_{V_2}$)) of
about 15\% in order to match the
observed absolute magnitude, although this is considerably uncertain.  Similar to our
analysis of LMC 172231 above, we used this with the radial velocity solution to 
perform a tomographic analysis using
a combination of 8 IMACs and 6 MagE spectra covering both double-lined quadratures. 
Note the model (computed with $e=0.00$) matches both the eclipse separations and eclipse durations, demonstrating that the orbit is indeed circular.

The tomographically separated spectra are shown in Figure~\ref{fig:ST2-28spec} {\it right}.
Comparison to the Sota et al.\ (2011) atlas suggests that the primary is
intermediate between that of O6.5~V and O7~V.  Measurement of the equivalent widths (EWs)
of He I $\lambda 4471$ and He II $\lambda 4542$ also indicate that the primary is barely on
the O7~V dividing line between the two subtypes (Conti \& Frost 1977),
with $\log W'=\log ({\rm EW} (\lambda 4471)/{\rm EW} (\lambda 4542))=-0.09$.  We
adopt O7~V as the type.  The tomographically separated spectrum of
the  secondary appears to be of O8~V, consistent with our estimation from the blended
spectra.  The third component contributes so little light to the
system that extracting a good spectrum
is difficult, and the spectrum is too poor to see either Si IV or Si III; Mg I $\lambda 4481$ is
only marginally visible, so we know this is an early B star (B0-2) and not later.  There is
a hint of He II $\lambda 4686$, which, if real, would suggest a B0 V type.  

\begin{figure}
\epsscale{0.32}

\plotone{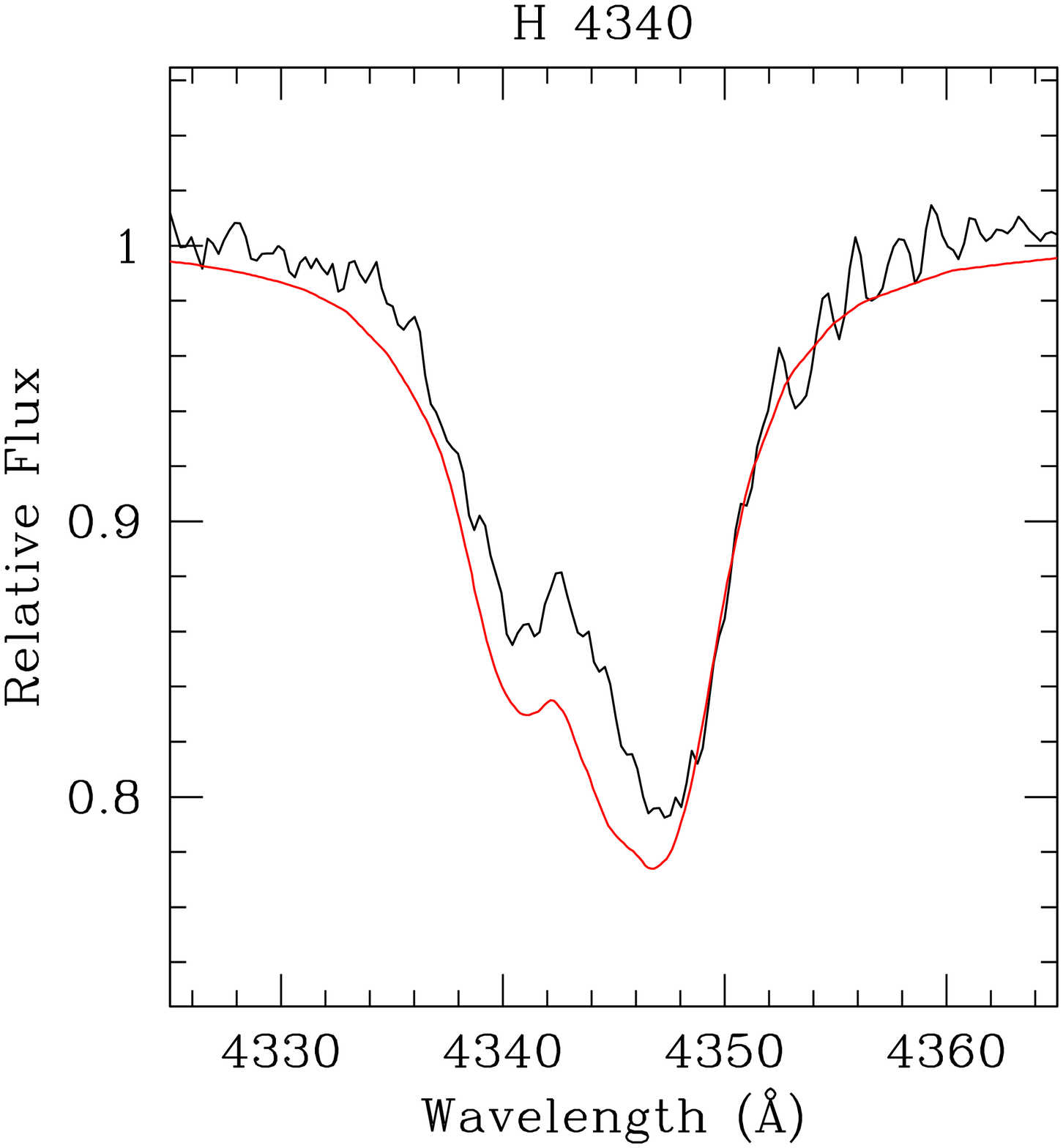}
\plotone{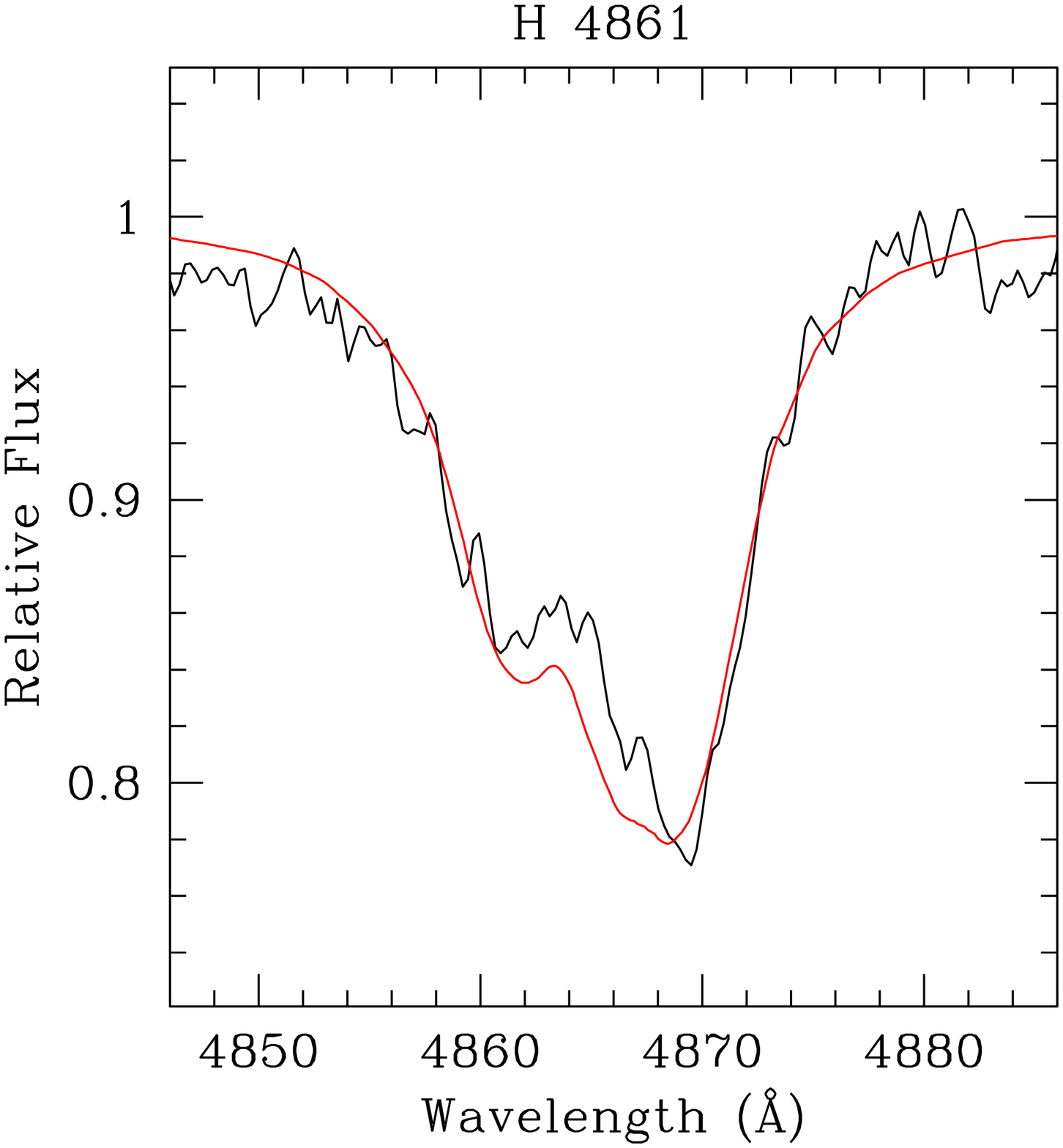}
\plotone{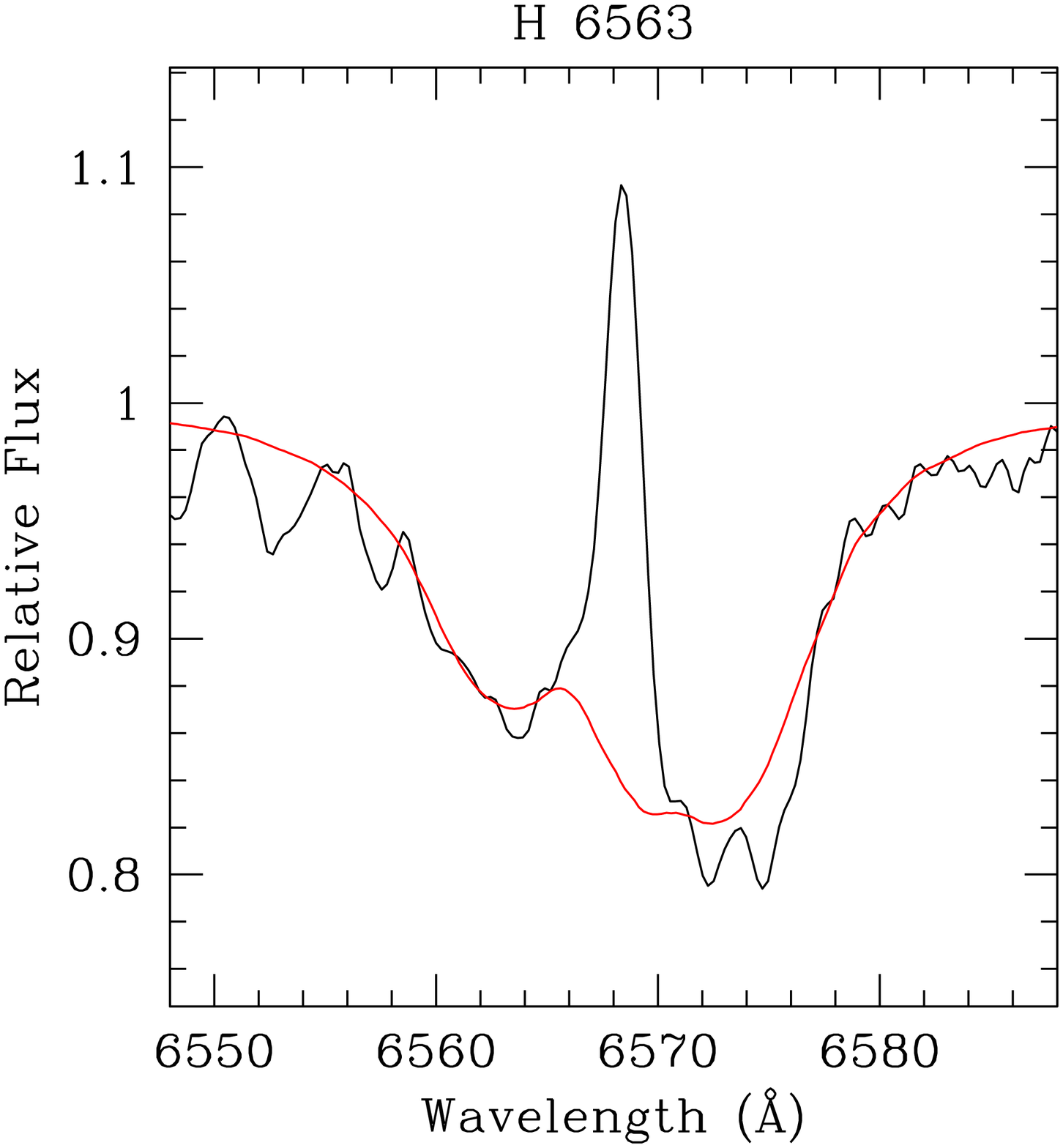}
\plotone{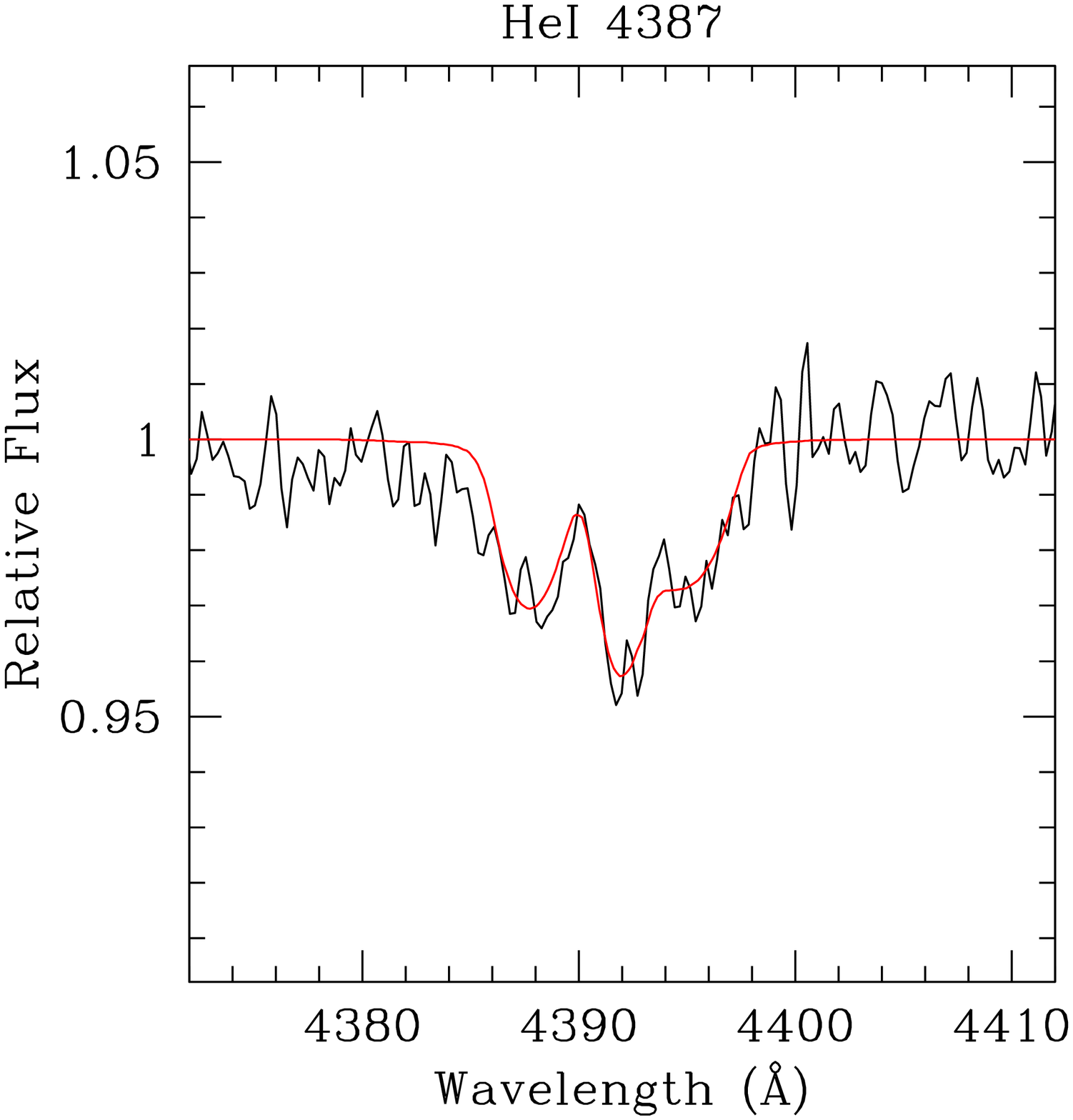}
\plotone{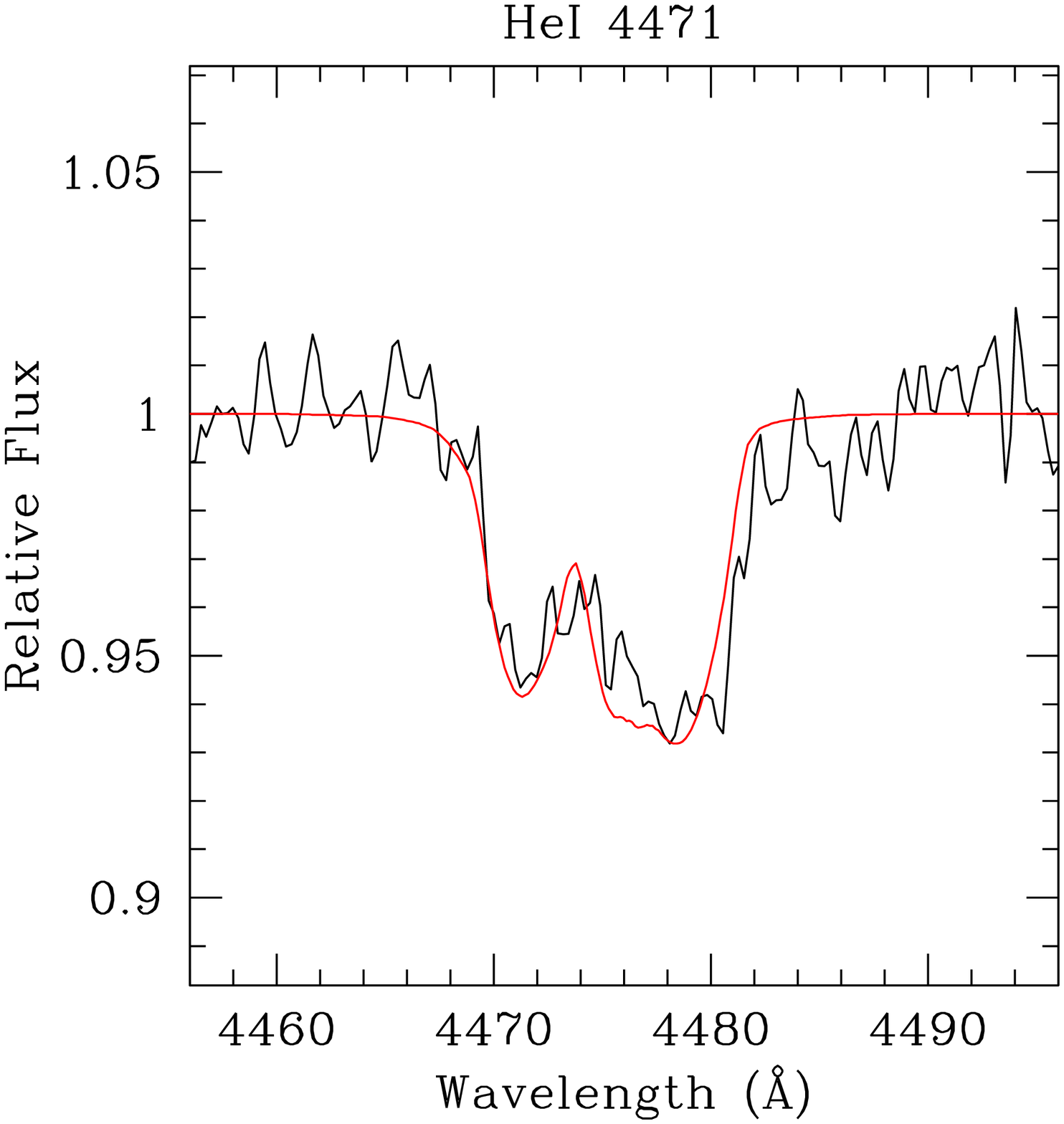}
\plotone{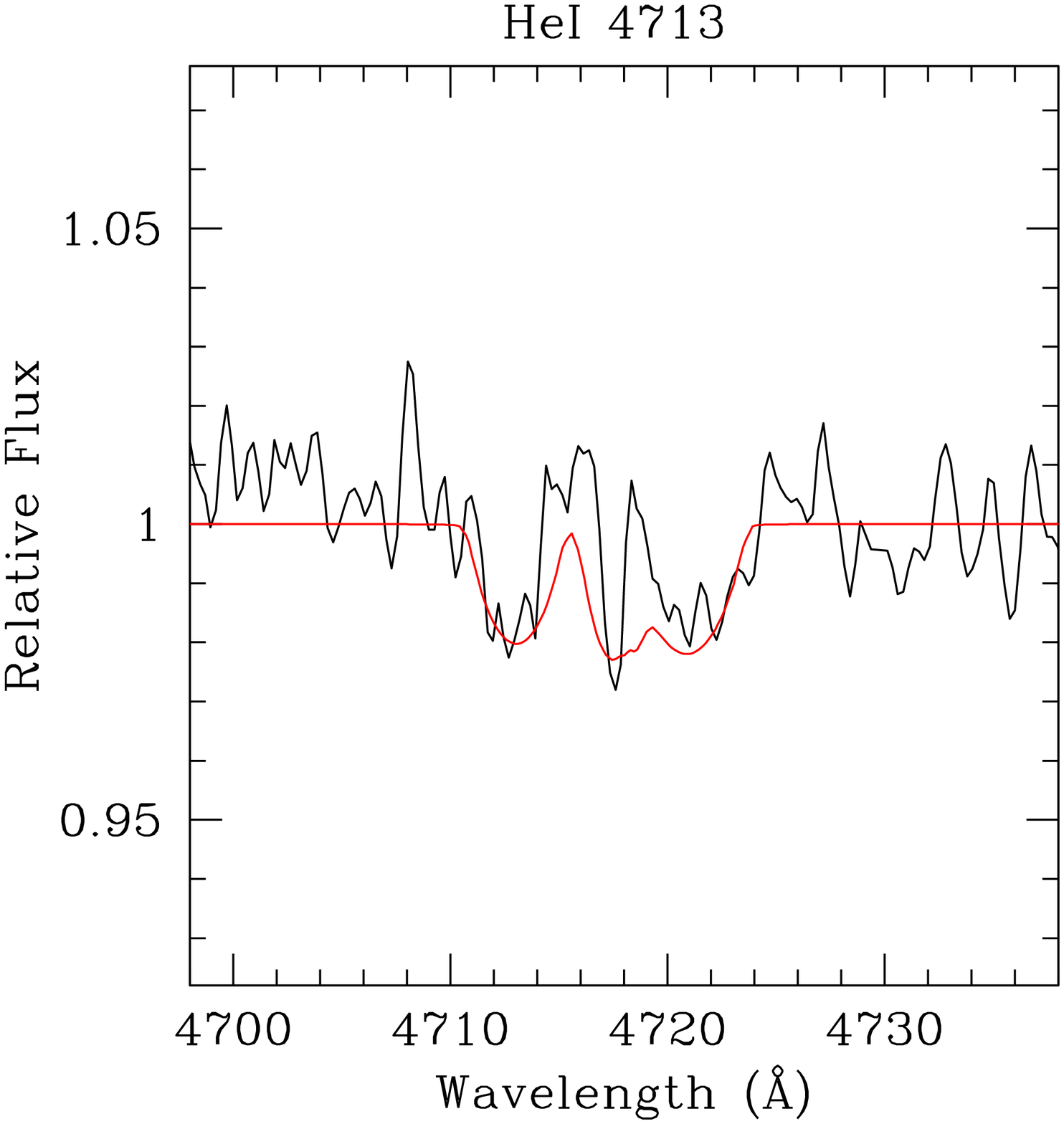}
\plotone{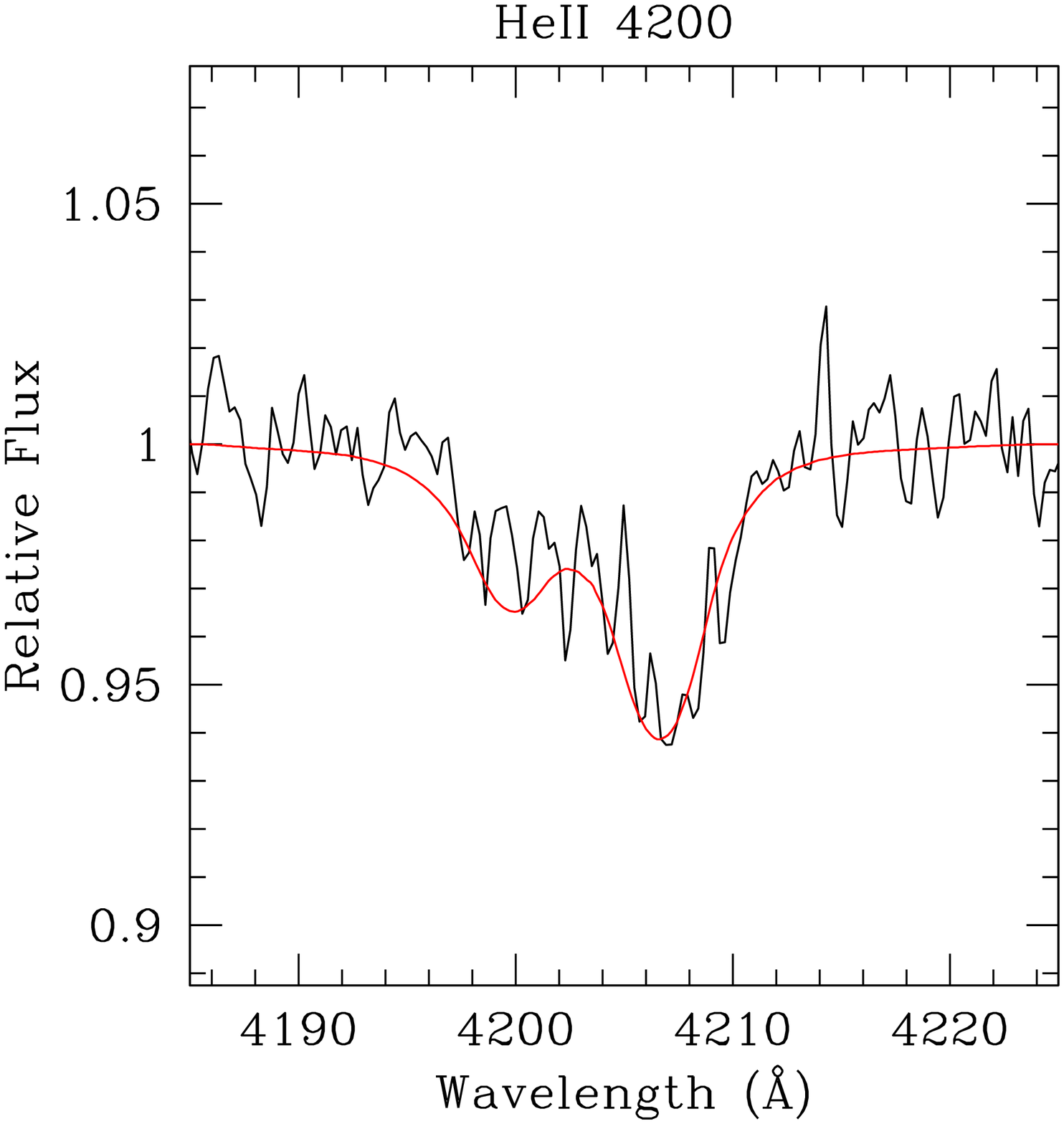}
\plotone{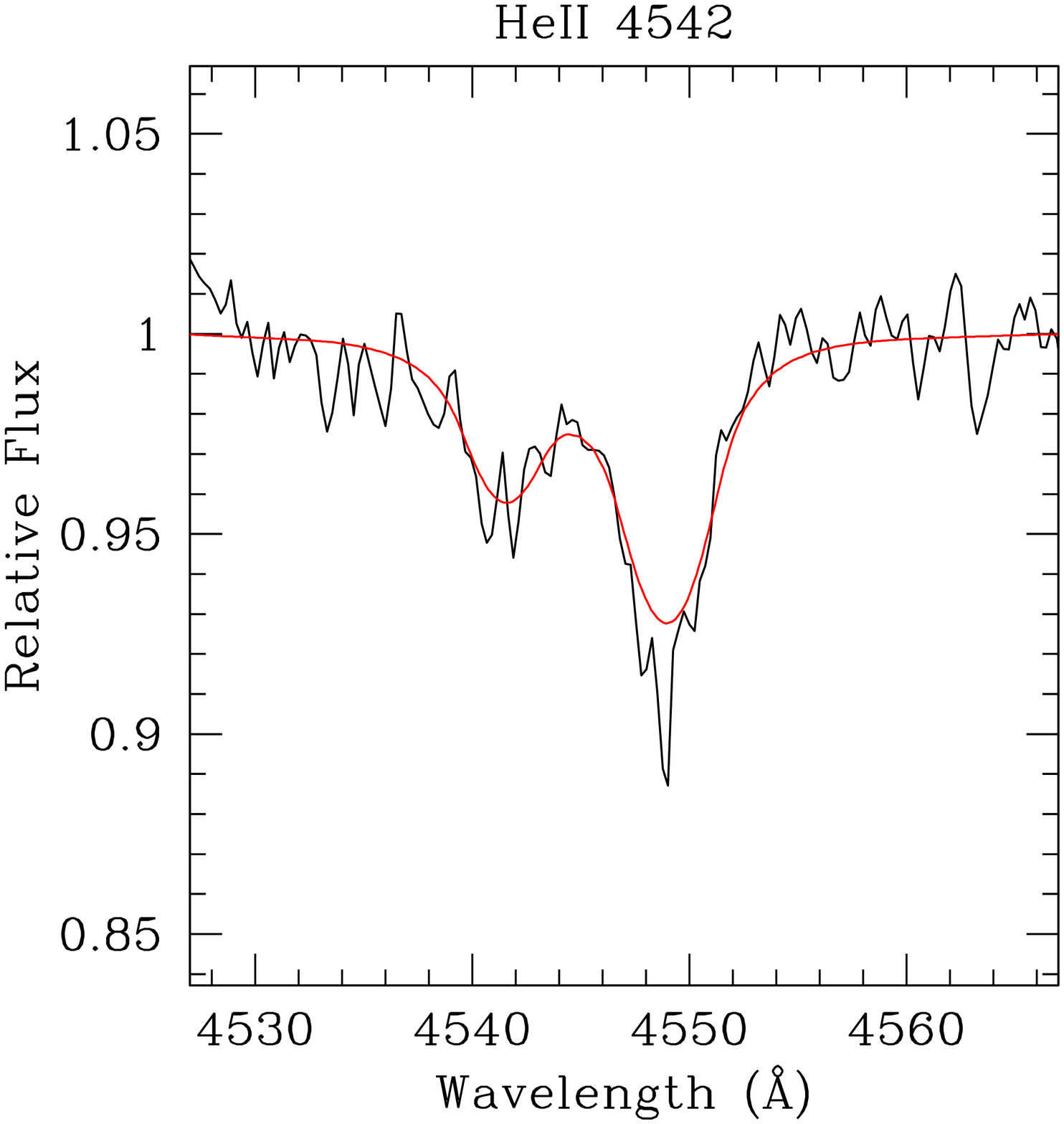}
\plotone{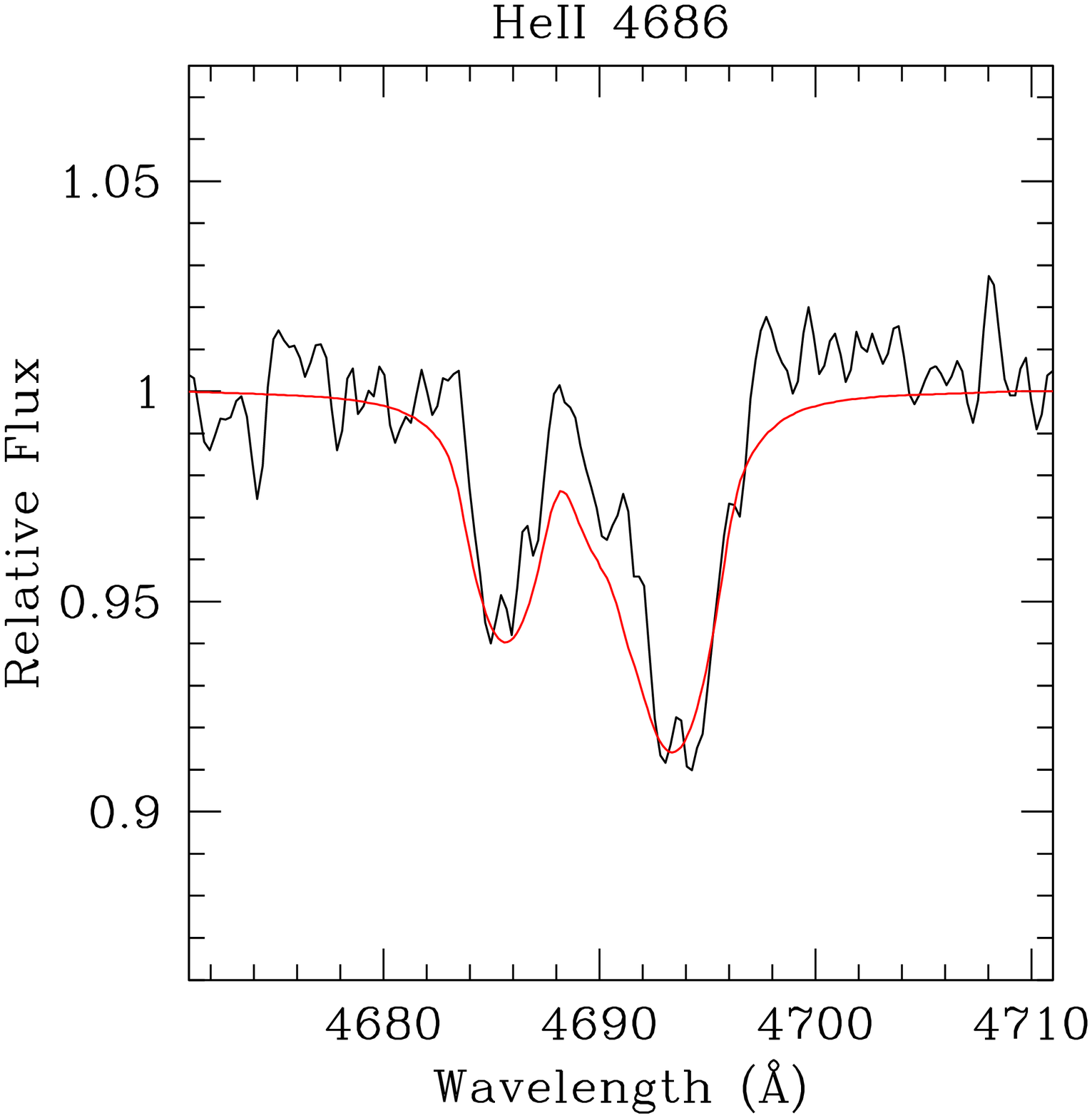}
\caption{\label{fig:ST228mods} LMC ST2-28 comparison with stellar atmosphere models.
We compare the combined  FASTWIND models (smooth, red curves) with the observed spectrum.
The spectrum used was taken on  2454876.697 and corresponds to a phase of 0.702.
The top row shows the H$\gamma$, H$\beta$, and H$\alpha$ lines. Note the presence of
remaining nebular emission at line center, albeit not as strong as in LMC 172231 (Figure~\ref{fig:LMC172231mods}).
The second row shows the  He I $\lambda4387$, He I $\lambda$4922,
and He I $\lambda4713$ lines.  The bottom row shows the He II $\lambda$ 4200, He II $\lambda 4542$, and He II $\lambda 4686$ lines. The latter is primarily sensitive to the 
mass-loss rate and wind law.
}
\end{figure}

Our FASTWIND fitting of the tomographically separated primary and secondary components
found slightly higher effective temperatures (38,500 and 36,500 K) than what we had
initially assumed.   For the purposes
of finding good values for the flux ratios, we adopted a third component model with an effective temperature of 
30,000 K, corresponding to that of a B0~V (Massey et al.\  2005).  From the depths of the
spectral lines we confirmed that the flux ratio $F_{V_2}/F_{V_1}$ must be close to the
0.63 value determined by the light curve, but that the contribution from the tertiary ($F_{V_3}/(F_{V_1}+F_{V_2})$) 
might be a little higher, probably 0.25\footnote{If instead we adopted an effective
temperature of 28,000 K for the third component, typical of a B1~V star (Trundle et al.\ 2007),
  we obtain similar results.}.
We then performed a second light
curve analysis, fixing the effective temperatures and third light-light component.
The final flux ratio $F_{V_2}/F_{V_1}$ from this fitting is $0.62\pm0.02$.
The light curve is shown in Figure~\ref{fig:ST2-28lc}, and the FASTWIND fits to the
spectra are shown
are Figure~\ref{fig:ST228mods}.   The third component is very obvious in the He I lines.
The agreement of the model fitting to the observed spectra is not perfect, but we judge it
adequate.  The greatest uncertainty is in the flux contribution of the third component.
We have adopted projected rotational velocities of 170 km s$^{-1}$ and 140 km s$^{-1}$ for the
primary and secondary, consistent with the synchronous values of $173\pm5$ km s$^{-1}$ and $141\pm5$ km s$^{-1}$.  The lines of the tertiary are sharper, and we adopted 100 km s$^{-1}$ 
for the fits.

The final masses
we derived were $24.9\pm0.4M_\odot$ and $20.2\pm0.4M_\odot$, with the slightly
larger errors due to the larger uncertainty in the orbital inclination $i$ (compared to that
of LMC 172231) due to the 
third light component.  The formal error on the inclination for the light curve analysis is
0\fdg5, but we quote 5\fdg0, as this is consistent with uncertainty in the third light component.
It is worth noting that the radii changed by only 1$\sigma$ from the original to final fits,
 despite the changes in the assumed effective temperatures, third light component,
 and resulting inclination changes.  The O7~V and O8~V components
  have $\log L/L_\odot \sim 5.3$
 and 5.0, respectively.

\section{The Mass-Luminosity Relationship and Comparison with Stellar Evolutionary Models}
\label{Sec-MLR}

In Section~\ref{Sec-intro} we argued that the goal of this project was to provide 
masses and luminosities that were so well determined that they could serve as {\it linchpins} 
in the upper end of the MLR.  In addition, we wished to see how well our parameters
agreed with the current generation of stellar evolutionary tracks. 

\begin{figure}
\epsscale{0.9}
\plotone{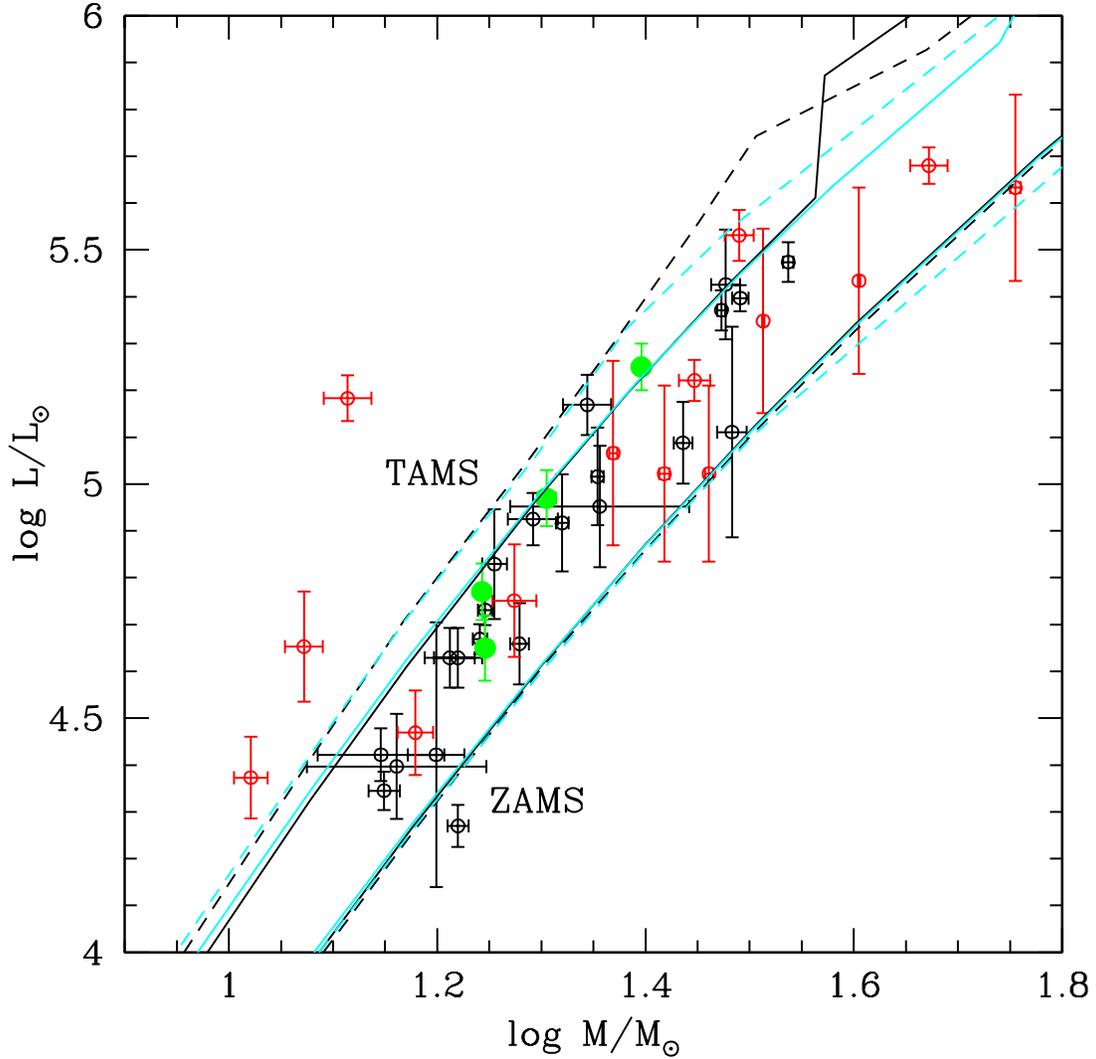}
\caption{\label{fig:mlr} The Mass-Luminosity Relationship for massive stars.   The colored points are all from
LMC binaries, with green presenting the present work, and red representing other data.  Black denotes
data for Milky Way massive stars.  Other than the present work, all the data come from Table 1 in 
Gies (2012).   The three LMC stars whose masses are too low for their luminosities are
the secondary component of SC1-105 (Bonanos 2009) and the two components of 78.6097.13
(Gonz\'{a}lez et al.\  2005).  Lines shows
the mass-luminosity relationship obtained from the latest Geneva models for solar
metallicity (black, Ekstr\"{o}m et al.\ 2011) and for LMC metallicity (cyan, Chomienne et al.\ 2011, in prep) for the zero-age main-sequence (ZAMS) and terminal-age main-sequence (TAMS). Solid lines denote the
mass-luminosity relationship determined from evolutionary models without rotation, while the dashed lines show
the theoretical mass-luminosity relationship for models with initial rotation velocities of 40\% of the critical velocity. }

\end{figure}

In Figure~\ref{fig:mlr} we compare the masses and luminosities of massive
stars in young detached eclipsing binary systems, where the data come from
Table 1 of Gies (2012) and references therein\footnote{We have updated the effective temperature
of SC1-105 to those given by Bonanos et al.\ (2011). }. Black points shows the data for the Milky
Way stars, while red points shows the data for previously studied LMC stars.  Our four
LMC stars are shown by  green points. 
The lines denote the expectations from the latest Geneva evolutionary models 
with solar metallicity  ($z=0.014$ using the Asplund et al.\ 2009 abundances) shown in black
and LMC metallicity ($z=0.006$) shown in cyan.  The solar metallicity tracks are 
from Ekstr\"{o}m et al.\ (2011) while the LMC metallicity tracks are from Chomienne et al.\ (2011, in prep)
\footnote{We are indebted to Georges Meynet for forwarding these
models and allowing us to use them in advance of publication.}.  The solid lines correspond to no initial
rotation, while the dashed lines correspond to an initial rotation of 40\% of the critical
velocity.    

The  models emphasize the point we made in Section~\ref{Sec-intro}, namely that
the mass-luminosity relationship really depends upon age for massive stars.  Although the error bars are large for
the entire data set, we find it reassuring nearly all of the points are found within the expectations of main-sequence
objects.    The exceptions are three red (LMC) outliers on the left, which are, in order of
decreasing luminosity, the secondary component of SC1-105 (Bonanos 2009; Bonanos et al.\ 2011) and the primary and secondary
components of 78.6097.13 (Gonz\'{a}lez et al.\ 2005). 

\begin{figure}
\epsscale{0.45}
\plotone{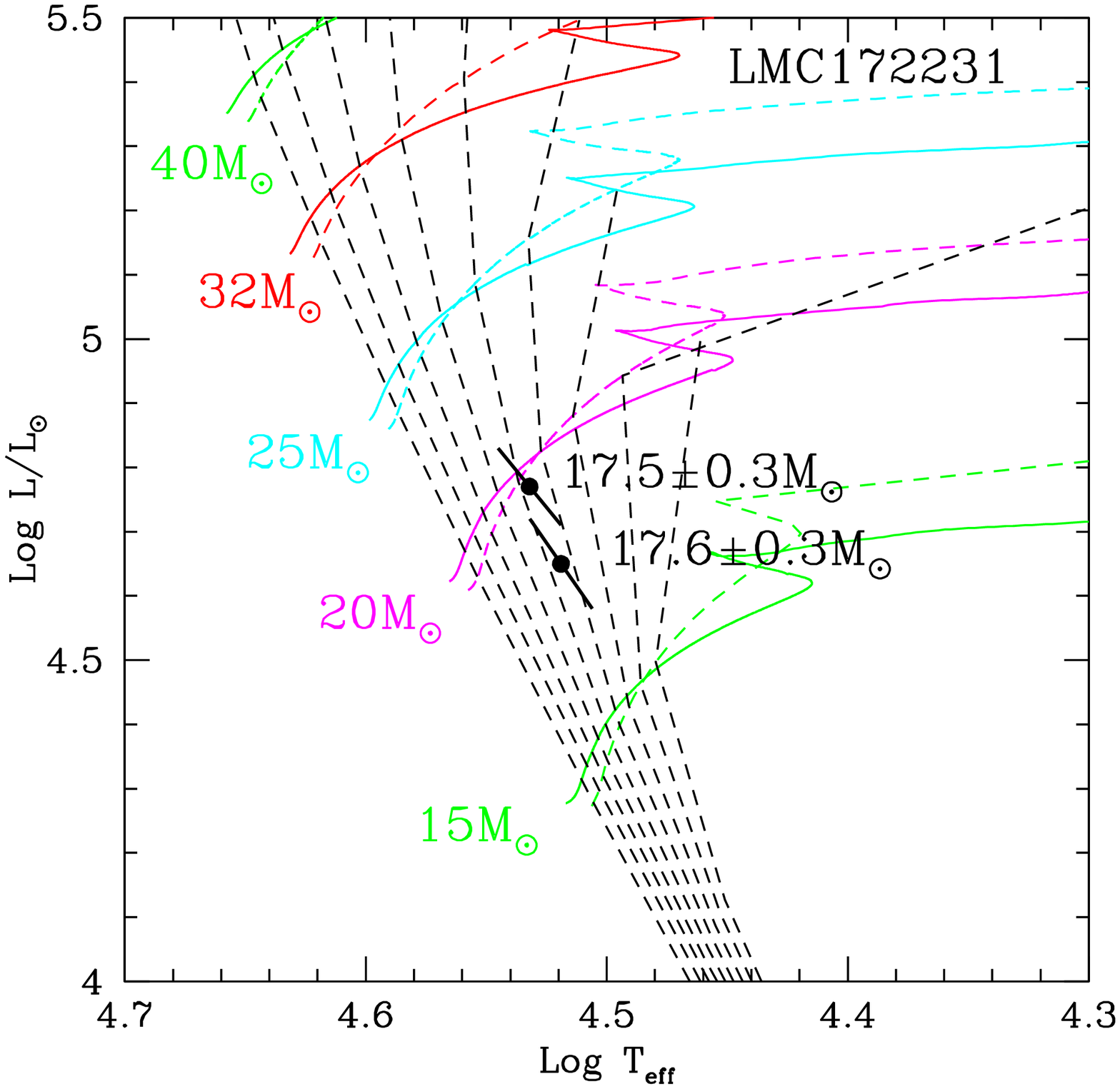}
\plotone{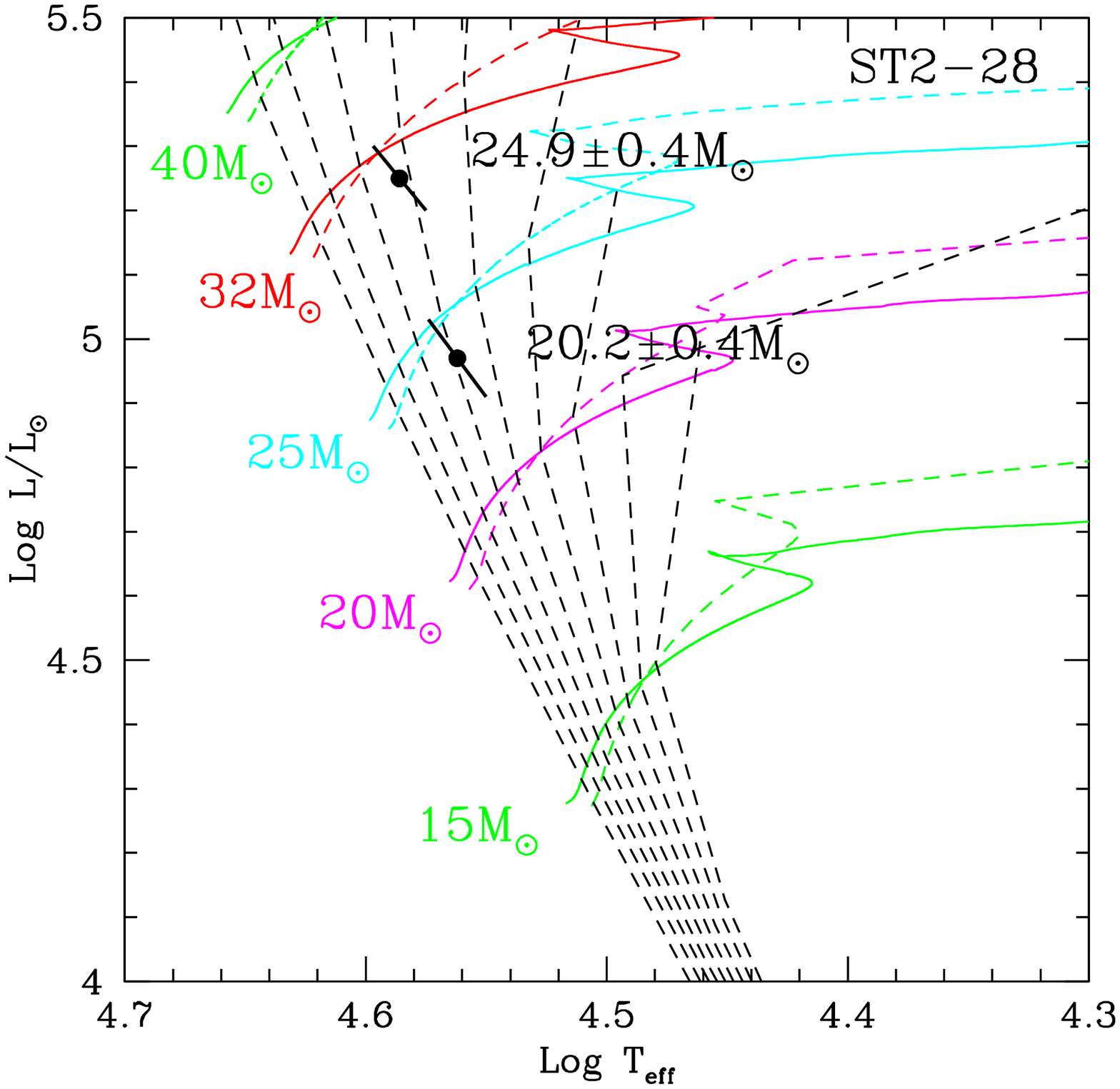}
\plotone{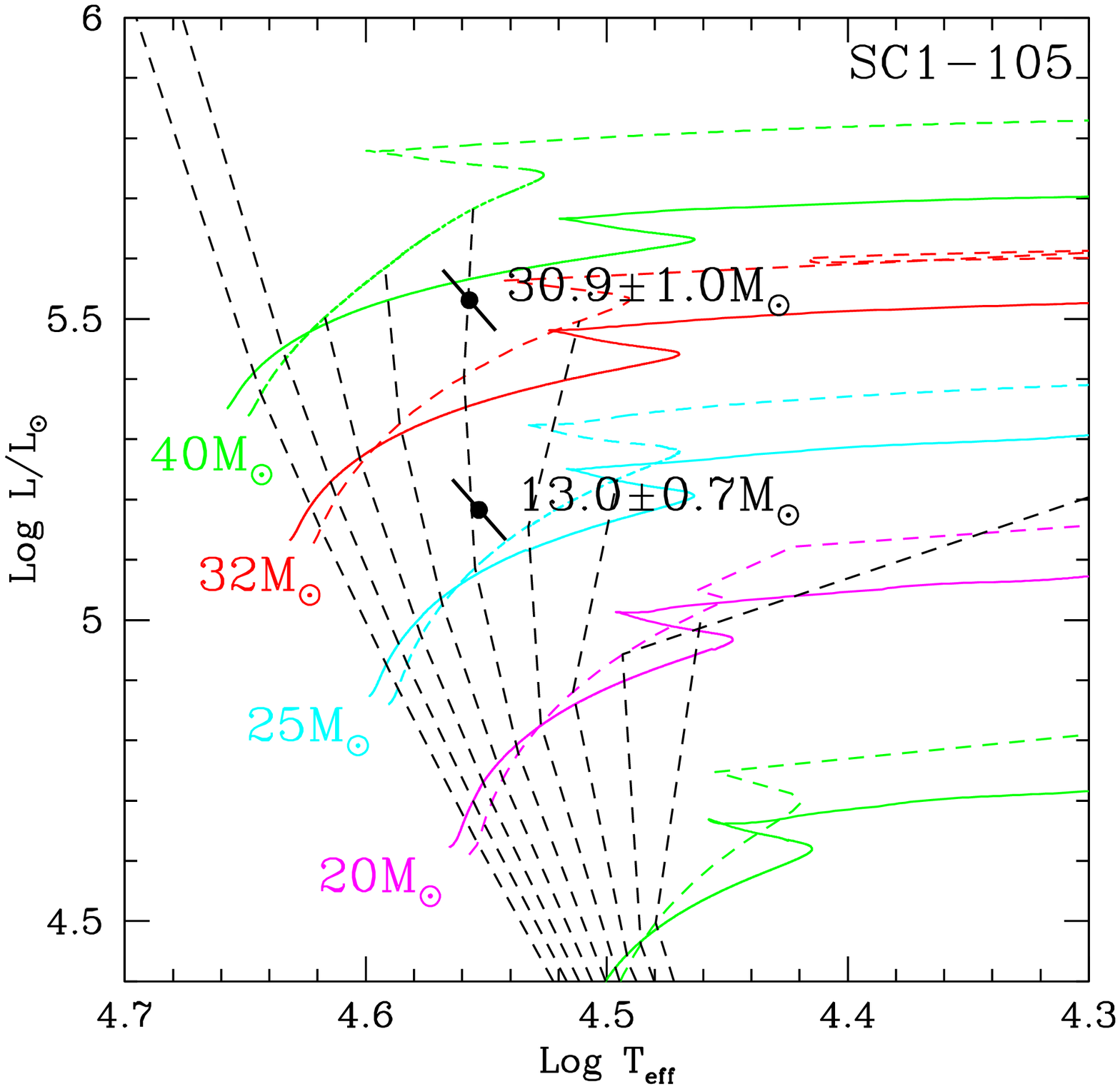}
\plotone{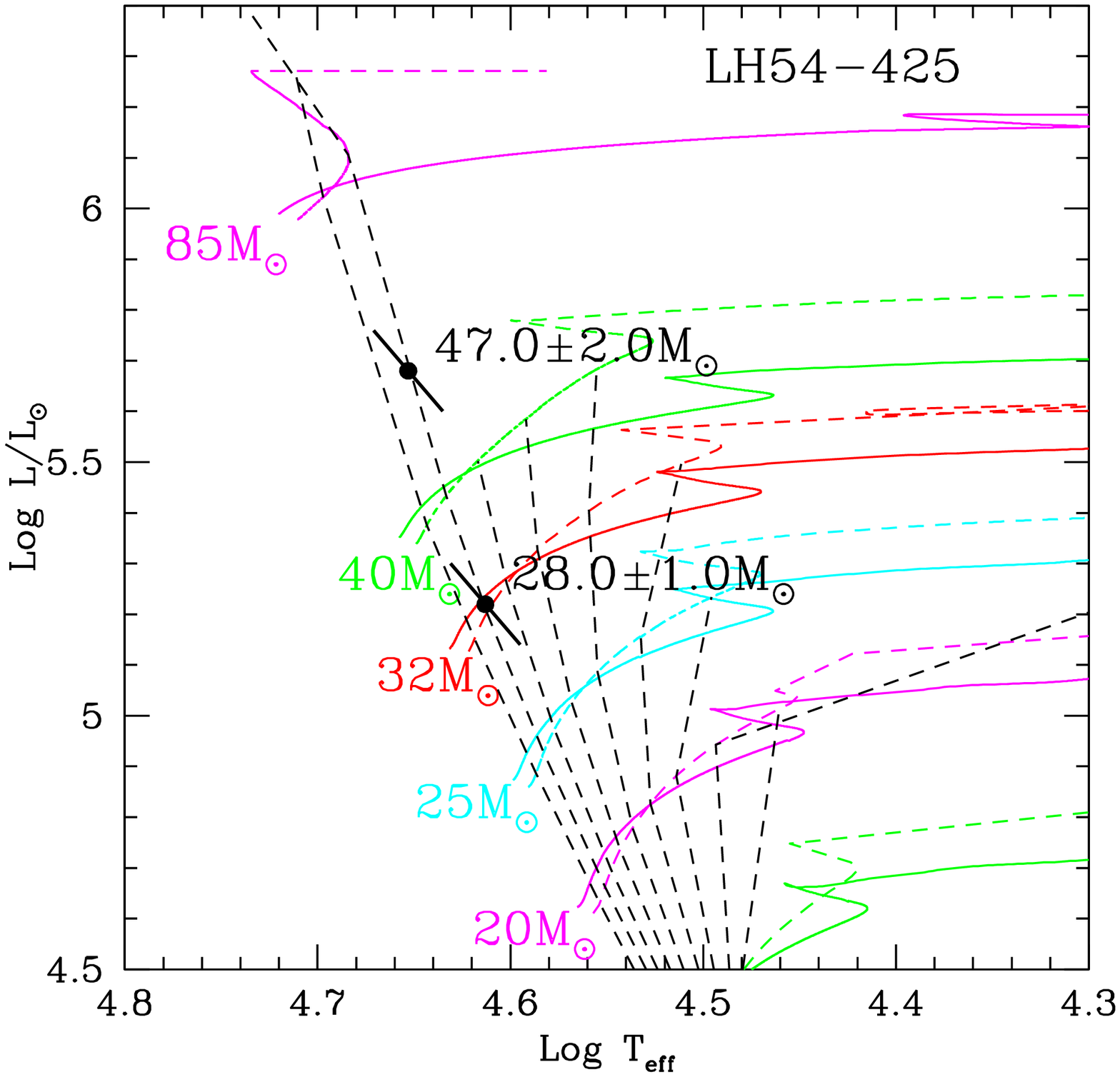}
\caption{\label{fig:hrd} Comparison to evolutionary tracks.  The location of the stars are indicated in the
H-R diagram as points; the associated errors in effective temperatures and luminosities are shown as
a diagonal line, as hotter temperatures would lead to higher bolometric luminosity.
The new Geneva $z=0.006$ tracks
are shown in colors, with the solid colored lines being the tracks for no rotation, and the dashed colored lines being the
tracks computed for an initial rotation of 40\% of the critical velocity.  
The dashed black lines correspond to isochrones of 1-9~Myr at 1~Myr intervals.  Note that only the main-sequence,
non-WR portion of the tracks (and isochrones) are shown for simplicity.
 {\it Upper:} The locations of our stars are shown, labeled with their dynamical masses. {\it Lower:} The locations of SC1-105 (Bonanos 2009; Bonanos et al.\ 2011) and LH54-425
(Williams et al.\ 2008) are shown, also labeled with their dynamical masses.
}
\end{figure}

While this provides good statistical agreement with the evolutionary models, it is not the critical
test.  In the upper two panels of Figure~\ref{fig:hrd}  we show the location of our four components
 on the H-R diagram with the dynamical masses indicated.   
(Note that because hotter effective temperatures imply high luminosities, error bars are diagonal lines in 
this diagram.) Overall the agreement is
encouraging.   For instance, the components in each
system appear to have the same age. We show isochrones at 1~Myr intervals, and
the components lie parallel to to these dashed lines, indicating that the physical parameters are
consistent with the same age for each component, which we would expect.  
Although the masses are approximately what we expect from their location in the HRD, 
it is intriguing that the masses for all four components are a bit too low compared
to the masses of the tracks.  We note that these differences are not due to
mass loss. For instance, in the
case of LMC 172231, a 5 Myr old star with an initial mass of 20$M_\odot$ would 
be expected to have a mass of 19.9$M_\odot$; i.e., the amount of mass loss expected
at these luminosities are small over these time scales.

We can quantify the size of the discrepancy by examining the differences in masses between
what we observe and what the models predict by interpolating along an isochrone. We list these
differences in Table~\ref{tab:comparisons}.  The ``observed" masses come 
from our orbit solution, and the ``Model" mass comes from using our ``observed" bolometric
luminosities to interpolate along the isochrone using a smooth, high order polynomial fit.
Both components of LMC 172231 and ST2-28 have differences
in the same sense, that the masses we observe are lower than the masses predicted by the
model.   Alternately we can ask what bolometric luminosities should correspond to our observed
masses.  We find (equivalently) that  all four components are slightly over-luminous for their
mass, by about 0.2~dex. 
While the discrepancy for the secondary of LMC 172231 is not significant, the others are all significant
at the 2-3$\sigma$ level.  Although these discrepancies are small, it is unsettling  that they are all in the same sense.
Are there underlying systematic issues that may be in play?

First, we consider our calculated bolometric luminosities.  
We have adopted the effective temperature determinations from FASTWIND,
and if these were systematically too large we would derive too high a bolometric
luminosity.  But, they would have be off by more than seems reasonable: to lower
the bolometric luminosities by 0.1-0.2~dex would require
temperatures that were cooler by about 6-12\%, or 2,000-4,000~K.    The precision in our fitting 
the spectra is
of order 500 K, and we have formally adopted errors of $\pm$1,000 K
in the propagation of errors in Tables~\ref{tab:lmc172231physical} and \ref{tab:st228physical}.
The agreement between FASTWIND
and the ``gold standard" hot star model atmosphere CMFGEN (Hillier \& Miller 1998) is
near perfect in terms of the effective temperatures (Neugent et al.\ 2010; Massey et al.\ 2011, in prep).
Alternatively, the radii would have to be over estimated by about 12-25\%.   This, too, seems beyond
any reasonable possibility: the radii are fixed by the widths of the eclipse depths and we
believe are
determined to 3-6 times better than this.  

Nevertheless, we have assumed spherical geometry in determining the bolometric luminosities, and used the
``volume radius" from the light curve analysis with our FASTWIND effective temperatures.  If
these stars were so close that tidal distortions were significant, this could be a potential problem, as
our effective temperatures really only correspond to those measured at quadrature.   Note though that a change of 
several thousand degrees would correspond to a change 1-2 spectral types during an orbit.  Such changes are not
seen even in contact systems.  In truth, the tidal distortions in our detached systems are very small.
Examining the light curves, the lack of significant ellipsoidal variations supports the spherical nature of the objects.  In addition GENSYN defines a fill-out factor, the ratio of the photospheric surface to the Roche surface.  Any value below unity indicates a detached component, and values below 0.90 correspond to essentially spherical stars.  The primary of ST2-28 has the highest fill-out, 0.87, and we can see slight evidence of this in the small ellipsoidal variations in its light curve. However we stress that the distortion of even this star is minimal. 
  As noted above, we have an alternative ``tidal model" solution for the bolometric luminosities as well given
in Table~\ref{tab:lmc172231physical} and \ref{tab:st228physical}.  The bolometric luminosities
implied by these are about 0.06-0.08~dex {\it larger} than the more conservative values we have been
using, which would exacerbate the differences with the evolutionary models. 

In the introduction we touted the advantage of studying systems in the Magellanic Clouds
rather than in the Milky Way as the distances are known: many methods yield essentially
the same answer for the distance of the LMC, and the distance is established to at least
10\% (van den Bergh 2000, and references therein).   This provides a reality check
on whether the luminosities we derive are reasonable or not.  In the
case of LMC 172231 the observed $M_V$ of the system is $-5.0\pm0.3$~mag, while our
physical parameters lead to a modeled $M_V$ of $-4.6\pm0.1$~mag.  The agreement is
within the errors,  but it suggests that if we had instead adopted effective temperatures that
matched the observed $M_V$ our problem would be worse by 0.1-0.2~dex, as matching the observed
$M_V$ would lead to a higher luminosity.  For
ST2-28 the situation is more complex, as the amount of light contributed by the third component adds a complication.  The modeled total luminosity of the system agrees with what we observe when we adopt
a third component contamination based upon our spectral fitting, but the uncertainties are large\footnote{We do note that as a cautionary tale to our colleagues attempting to derive distances to nearby
galaxies from such binaries that the errors associated are larger than are sometimes estimated, as we see in the comparisons here. For a contrasting view, see Bonanos et al.\ (2011).}.

\begin{figure}
\epsscale{0.45}
\plotone{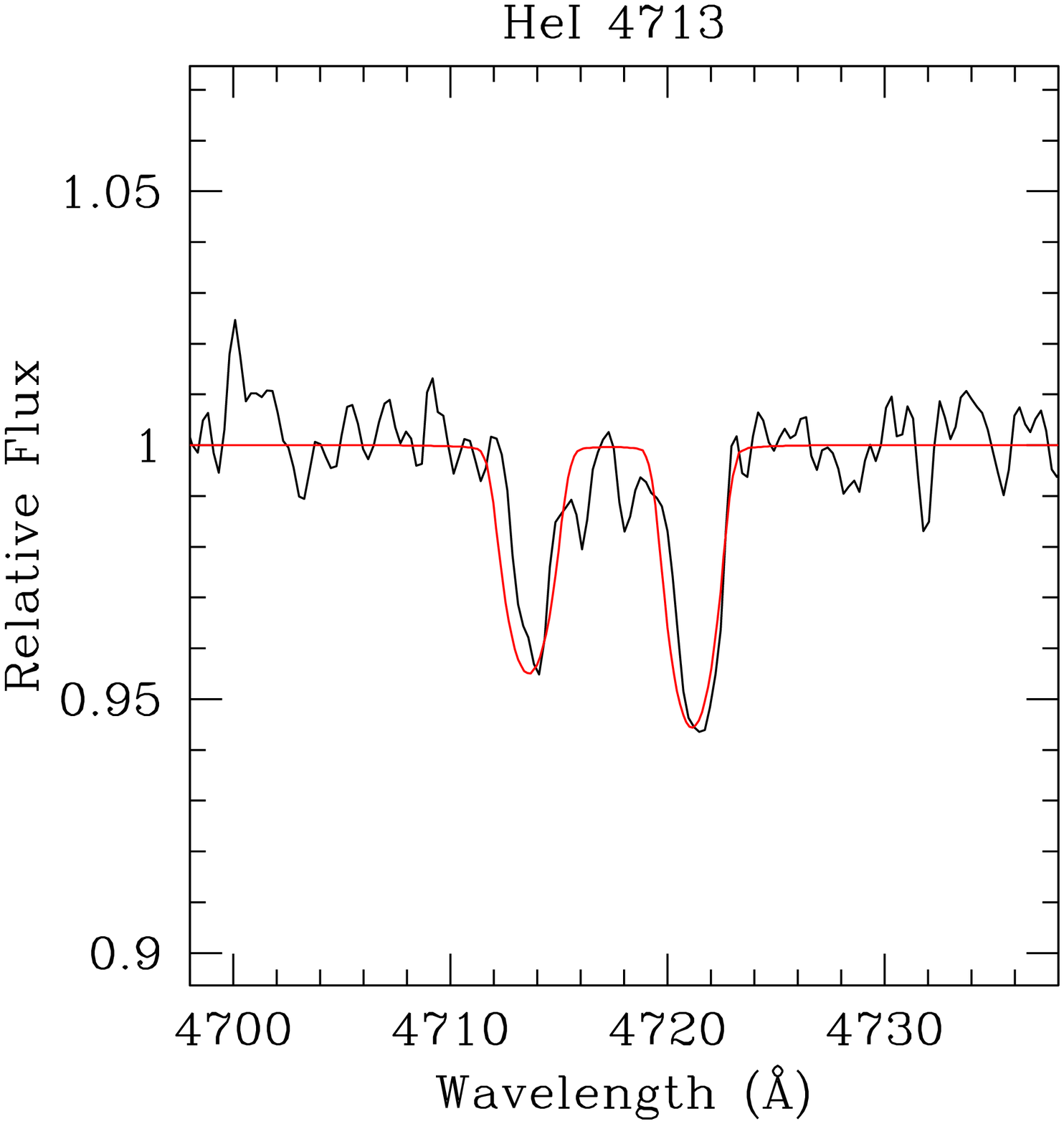}
\plotone{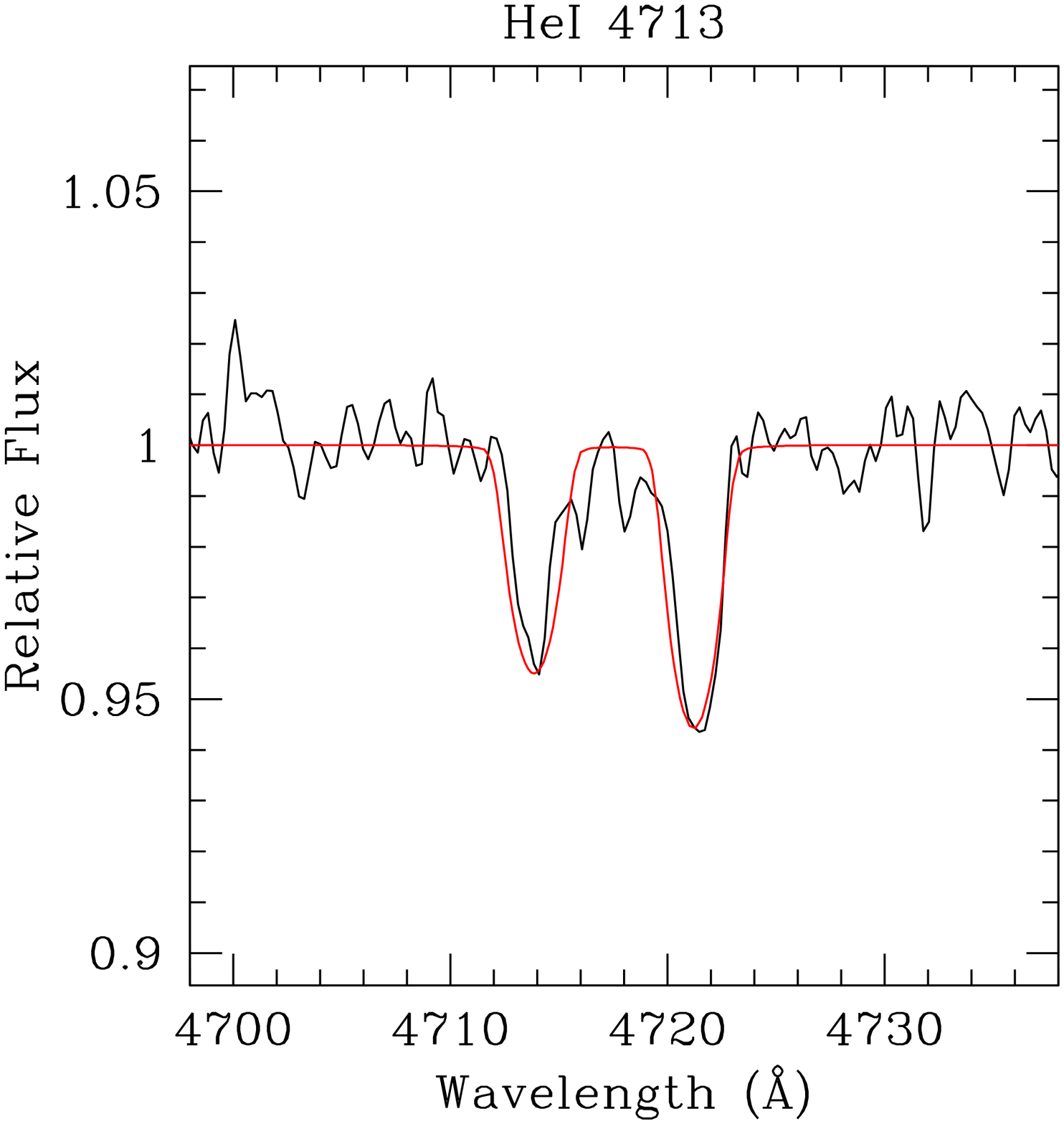}
\plotone{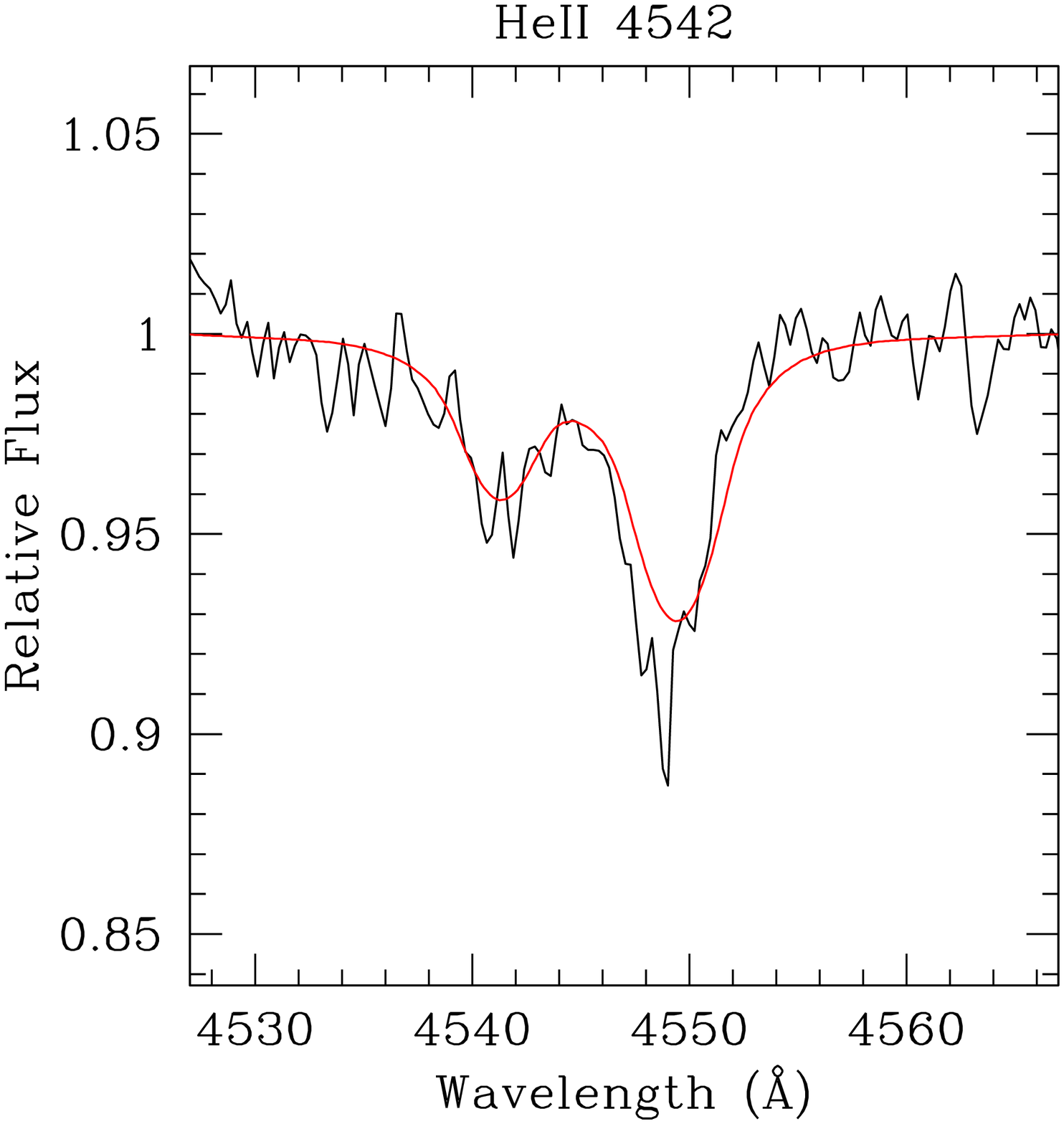}
\plotone{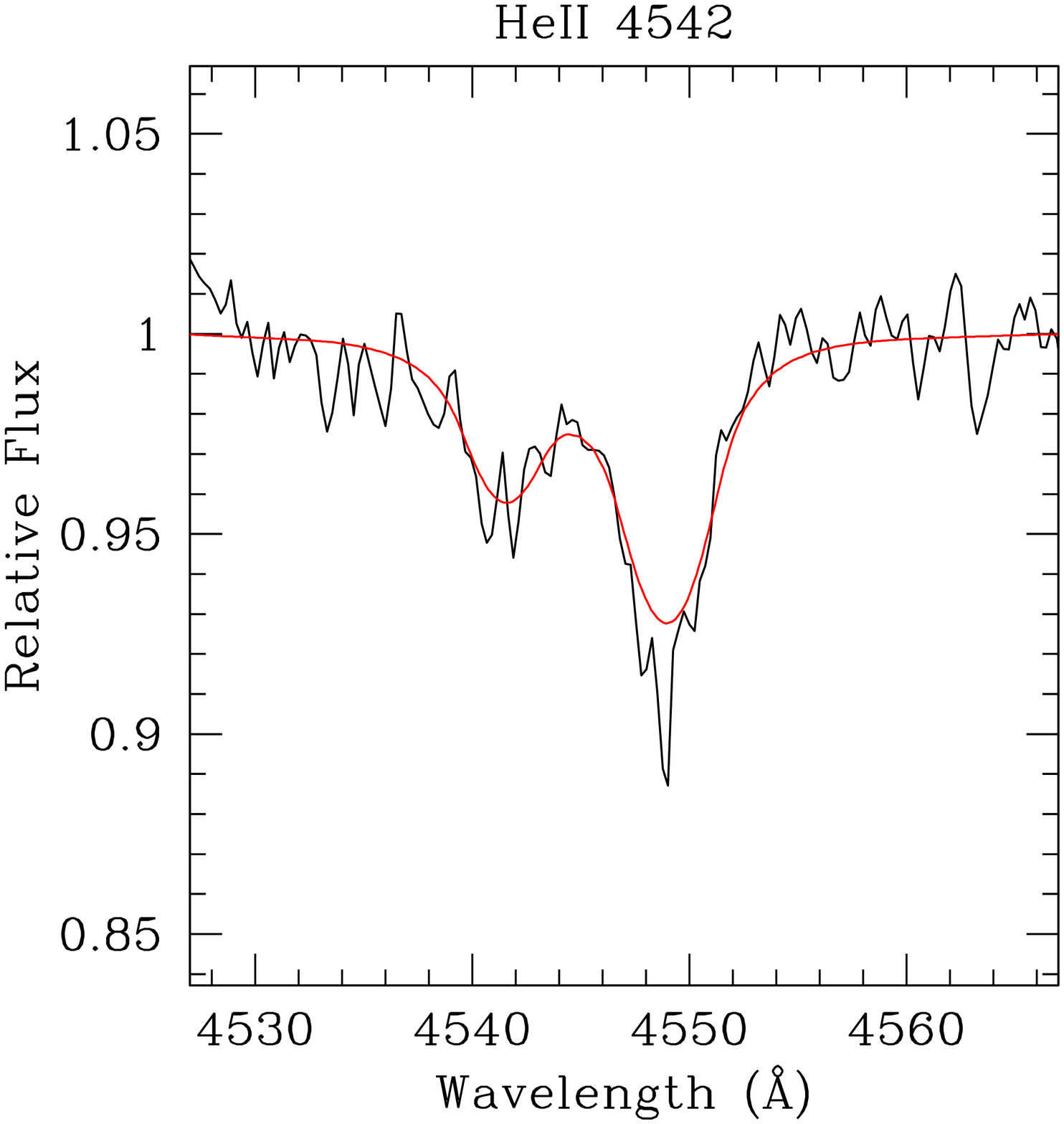}
\caption{\label{fig:compare} Comparison of velocity shifts.  {\it Upper:} The He I $\lambda 4471$ LMC 172331
FASTWIND model
lines have been shifted by the amount needed to match the required orbital semi-amplitudes if the evolutionary
masses were correct on the left.  On the right we show the shifts based upon our adopted orbital semi-amplitudes.
The underlying spectrum is that shown in Figure~\ref{fig:LMC172231mods}.  
 {\it Lower:} The same is shown for the He II $\lambda 4542$ FASTWIND model lines for ST2-28.  The underlying
 spectrum is that shown in Figure~\ref{fig:ST228mods}.
}
\end{figure}

Could the masses themselves be in error?   The masses would only have to be low by $\sim$10-12\%.  
Our reality checks have all agreed: Wilson's method (which is independent of all phase information)
has nearly identical $\gamma$-velocities and values for the {\it ratio} of the orbital semi-amplitudes as that found from our orbit
solutions.   But what if there were some systematic blending problem that was not being
properly accounted for by our measurement technique?  We can test this directly by
inverting the problem: what would the orbital amplitudes have to be in order to match the
masses from the evolutionary models, assuming that the orbital inclination is approximately
right.  For LMC 172231, matching the evolutionary masses would require that 
$K_{\rm prim}$=230.1 km s$^{-1}$ rather than 234.9 km s$^{-1}$
 (i.e., 4.8 km s$^{-1}$ smaller), and 
 $K_{\rm sec}$ would have to be 248.8 km s$^{-1}$, rather than our 233.2 km s$^{-1}$ (i.e., 15.6 km s$^{-1}$ larger).  
For ST2-28 the differences are even greater: to match the model
masses, $K_{\rm prim}$ would have to be 254.0 km s$^{-1}$ rather
than 241.0 km$^{-1}$ (i.e., 13.0 km s$^{-1}$ larger) and $K_{\rm sec}$ would have
to be 316.4 km s$^{-1}$
rather than 297.0 km $^{-1}$ (i.e., 19.4 km $^{-1}$ larger).  In Figure~\ref{fig:compare} we
show how how the required values would compare with an observed spectrum {\it vs} how the
measured values compared.  The agreement for He I $\lambda 4922$ in LMC 172231 is shown in the upper two
panels, with the shifts on the left being computed by what would be required to match the evolutionary masses,
while the shifts on the right are based upon our measured orbital amplitudes.  The difference is negligible for
the primary, but it is clear from the secondary that our adopted velocity is a better fit.  The
He II $\lambda 4542$ line in ST2-28 is shown in the bottom two panels.   Again on the left we show
the shifts required to bring the masses into agreement with the evolutionary values, while on the right
we  show the shifts based upon our adopted velocities.  The latter is clearly better. 

We were of course curious to see if other systems show similar discrepancies.
Comparisons at this level of precision between the binary and evolutionary masses
are usually not possible; it requires both very good mass determinations {\it and} careful effective
temperatures determinations for the bolometric luminosities.    Two studies that meet these criteria are that
of SC1-105\footnote{The star is more commonly known as W28-22 from Westerlund (1961)
or LH81-22 from Lucke (1972); see Table 2 in Massey et al.\ (2000).}  
(by Bonanos 2009) and of the LH54-425 (by Williams et al.\ 2008), both LMC binaries.
We compare their masses to those expected from the evolutionary tracks in the lower part of
Figure~\ref{fig:hrd}\footnote{In making the figure we have updated the effective temperatures to 
those found by Bonanos et al.\ 2011.}.

Bonanos (2009) describes the components of SC1-105 as highly
non-coeval, with ages
of 5~Myr (primary) and $>$10~Myr using the older (non-rotating) Geneva
evolutionary tracks, and cites this as evidence that there has been
significant
mass transfer.  We see in Figure~\ref{fig:hrd} that in fact
both components neatly fall along a single isochrone (5~Myr) with the
newer Geneva tracks
that include the effects of rotation, and are in fact quite coeval.
The components of SC1-105 are late O-type dwarfs, similar to those studied
here.
The secondary is highly over-luminous for its mass.  We agree that the
system is semi-detached, with the secondary filling its Roche lobe.
However, her fit of the light-curve significantly fails to match the
eclipse depths for both the primary and secondary eclipses by 0.1~mag or more,
a problem
she attributes to ``spots".   Rather, we believe this is indicative of a well-known
problem in analyzing
contact and semi-detached systems.  In such systems the orbital
semi-amplitudes are
significantly over-estimated using low-excitation optical lines, with
smaller $K$ values obtained
from the UV (see, for example, Penny et al.\ 2008) The physical
explanation is that
the side of the star in overflow that is facing its companion is producing
weaker optical lines because the temperature gradient is less steep there.
The center of light of the optical lines is thus skewed toward the outer
parts of the star, giving a spuriously large $K$.  This overestimates
the separation of the two stars, producing more shallow model eclipses
than what is actually
seen.
Unless one corrects for this by using only high-excitation lines one will
derive masses
that are too large, as well as spurious values for the other physical
parameters\footnote{We note that despite these problems Bonanos et al.\
(2011) successfully used this system to derive
a realistic value for the distance of the LMC.}.  Note that since the 5
Myr isochrone is almost vertical in the HRD,
a change in luminosity (due to an incorrect radius) would have little
effect on the derived ages as long
as the same effective temperatures were adopted.

LH54-425 consists of earlier O-type stars than what
we are studying here, but the sense and the size of the mass discrepancy is very similar to what
we find here (Figure~\ref{fig:hrd}, {\it lower right}): both components are slightly over-luminous for their masses (or, conversely, under-massive
for their luminosities) when compared to the evolutionary tracks.  Note we find
that the components
are highly coeval, according to the isochrones, with an age of 2.0~Myr, slightly larger than
the 1.5~Myr derived by Williams et al.\ (2008).   We have added the differences between the observations and the models to the end of Table~\ref{tab:comparisons}.   We can see that the
masses for LH54-425 are discrepant with theory in the same sense, and by the same amount, as what
we find here.  The result is also only marginally significant, however, when compared to the errors.

The mass-loss rates assumed in the evolutionary models are unlikely to be the culprit. The Geneva models
rely upon the Vink et al.\ (2001) mass-loss laws 
 for O stars, and for 20$M_\odot$ stars (similar to the mass of those discussed here), 
 the mass loss rate averaged over the first 5~Myr are indeed very low, about $0.02 \times 10^{-6}M_\odot$
yr$^{-1}$.
Puls et al.\ (1996) did not measure mass-loss in the LMC for stars as low luminosity as these,
but did find values $<0.1\times 10^{-6} M_\odot$ yr$^{-1}$ for LMC O-type stars that were considerably more luminous.  Similarly Massey et al.\ (2004, 2005, 2009) find mass-loss rates for
LMC O stars with $\log L/L_\odot$ of 5.0 to be $<0.1\times 10^{-6}$ yr$^{-1}$.  In order to account for the
discrepancy we find, the mass-loss rates would have to be about 0.2-0.5$\times 10^{-6} M_\odot$ yr$^{-1}$ for LMC 172231,
and $0.9-1.0 \times 10^{-6} M_\odot$ yr$^{-1}$ for ST2-28.  These high rates can be ruled out observationally for radiatively-driven winds. 

What if instead binary evolution (i.e., Roche-lobe induced mass-loss) has raised the mass-loss rates?
We see no evidence of this at present.  Certainly in the case of  LMC 172231 and
ST2-28 we see little current evidence: as discussed above,  the stars are not significantly distorted.
The residual nebular contamination at H$\alpha$ in the tomographically separated spectra precludes our 
making an estimate of the mass-loss rate directly.

Is this a problem restricted to that of the LMC?  Unfortunately none of the massive young detached Galactic
systems listed by Gies (2012) have undergone a similar sort of analysis; at best, the effective temperatures
have come from assigning spectral types, usually without tomographic separation.  The {\it masses} of
these systems are likely well determined, but their bolometric {\it luminosities}  are  not.  (This can be
inferred from an inspection of the error bars in Figure~\ref{fig:mlr}.)  

One intriguing possibility has to do with the metallicity we have assumed for the LMC, using a 
$z=0.006$ for the LMC based upon the Asplund et al.\ (2009) abundances.  What if the appropriate
abundance was a little bit higher or a little bit lower?  Would that be enough to explain our discrepancy?
To investigate the effect of abundances on the derived masses from the models, we used the solar metallicity
($z=0.014$) Geneva models for comparison, re-deriving the ages and constructing new isochrones.  In all cases the
evolutionary masses are about 7\% higher for a given luminosity.  (At a given age the mass for a particular luminosity
is actually {\it lower} for Galactic metallicity than for the LMC, but the isochrones are shifted to cooler temperatures
and so one derives a younger age for these stars, and that results in a higher expected mass.) The median change
needed in the models (Table~\ref{tab:comparisons}) is about 11\%.   There is no reason
to believe this effect is linear, but if it were, explaining the discrepancy by metallicity would require the appropriate
models to have much lower metallicities ($z=0.002$), lower than what is allowed by observations.
  Of course, given the complicated nature (isochrones shifting and
different luminosity for a given mass), it might be that even slightly lower  (or even slightly higher) metallicity
models might help alleviate the problem.
As more  models become available, it will be interesting to test this.

We note that the discrepancy between the dynamical masses and the evolutionary
masses we find here is in the same sense as that of the mass discrepancy discussed in the introduction, namely that
the masses inferred from the evolutionary tracks are larger masses than those inferred from the 
surface gravities derived from fitting model atmospheres.  But, there the size of the discrepancy is much larger, sometimes a factor of 2 or more,
when present.   We cannot, however, use our FASTWIND modeling to obtain a useful additional estimate of the mass.
 The uncertainty in fitting $\log g$ to the tomographically separated spectra is about 0.2~dex.
 An error of 0.2~dex translates to an uncertainty of 60\% in the masses, 
making this not useful for weighing in on the sort of 15\% differences we are seeing here.  In addition,
we note that FASTWIND produces values for $\log g$ that are systematically about 0.05-0.10~dex smaller
than those found using CMFGEN  (Neugent et al.\ 2010; Massey et al.\ 2011, in prep.). We note that
our FASTWIND modeling found a $\log g$ of 4.0~dex [cgs], which is consistent
with the $\log g$ values we list in Tables~\ref{tab:lmc172231physical} and \ref{tab:st228physical}, derived
from the masses and radii from our analysis.

We believe this emphasizes the need for such fundamental data such as ours.  With queue or robotic
observing on small imaging telescopes, and modern spectrographs on large apertures, obtaining
excellent photometry and radial velocities and hence very accurate masses is now standard.  However,
simply relying upon spectral types to assign effective temperatures do not result in bolometric luminosity
determinations that are accurate enough to contribute to the discussion.

If the trend is real, what then could be the problem with the models?  Either a higher initial rotation than assumed, or convective overshooting,
would extend the size of the convective core in the model, resulting in higher luminosity at
a given mass.   Ribas et al.\ (2000) have argued that convective overshooting is more
significant at higher masses, and this may be consistent with what we observe here.
(We are indebted both to the anonymous referee and to Georges Meynet for correspondence
on this point.)

Although our empirical data do show a slight ``mass discrepancy" between the Keplerian
masses, perhaps the real emphasis should be on how well the observations do in fact confirm
the evolutionary masses.   This is perhaps the most stringent comparison made for high mass stars,
and we should be encouraged that the agreement is as good as it is in the masses.  It is also encouraging
that observation and analysis tools have advanced to the point where one can worry about a discrepancy
of order 10\% in the masses of these stars. 
 Analysis in future papers will either confirm or refute this as a real trend.

\acknowledgments

We are grateful 
to the various observers who were involved in taking imaging as part of our photometric
monitoring, and to the SMARTS  queue managers Jenica Nelan and Michelle Buxton.  
It was also a pleasure to work with Suzanne Tourtellotte, the SMARTS data manager,
who kept the data pipeline flowing, helped us retrieve missing images whenever we made 
mistakes, and provided some of the most challenging password clues we've ever encountered.  
A number of undergraduate students were involved in the early stages of examining the photometry data,
including Erin Darnell, Corey Ritter, Shaye Storm, and Yelena Tsitkin.  
Darnell also performed a preliminary reduction of
the early Swope data.  Brian Skiff looked after the data transfer of the images from Yale 
daily for several observing seasons, and offered good advice.
Vincent Chomienne and Georges Meynet generously
made the Geneva models available for our use.  Larry Wasserman was kind enough to
run our velocity data through his orbit program in order to ascertain whether or not
the velocity data
were consistent with a circular orbit.
 Alceste Bonanos, Georges Meynet,
Noel Richardson, and Stephen Williams
all offered comments on an early draft of the manuscript.
We thank an anonymous referee for helping us improve the paper.
This work was supported by the National Science Foundation
under awards AST 0506541, 0506573, 0506577, 0506749, 0946314, and 1008020.  We are grateful
for the excellent support we received at Las Campanas during the course of this project; it is always a pleasure and privilege
to observe on that mountaintop.  We also gratefully acknowledge 
the patience and encouragement of the Carnegie and Arizona TACs who were generous with time
allocation on the two Magellan telescopes.   

Facilities: \facility{Magellan:Baade}, \facility{Magellan:Clay}, \facility:{Swope}, \facility{CTIO:1.0m}, \facility{CTIO:1.3m}

\clearpage

\clearpage

\clearpage

\clearpage

\clearpage

\clearpage

\begin{deluxetable}{l l c c c c c c}
\tabletypesize{\scriptsize}
\tablecaption{\label{tab:telescopes} Telescopes Used for Photometry}
\tablehead{
\colhead{Index}
&\colhead{Telescope}
&\colhead{Camera}
&\colhead{Observatory}
&\colhead{Scale}
&\colhead{FOV} 
&\colhead{Median}
&\colhead{\#}
\\
& 
& & & \colhead{(\arcsec/pixel)}
&
&\colhead{(Seeing (\arcsec)}
&\colhead{Images}
}
\startdata
1& Swope  1.0-m             &SITe\#3      &LCO   &0.435   &15\arcmin x23\arcmin\tablenotemark{a}  & 1.60 & 2970 \\
2 & SMARTS Yale 1.0-m  &Y4KCam   &CTIO  &0.289   &20\arcmin x20\arcmin  & 1.67 & 1836\tablenotemark{b} \\
3 & SMARTS 1.3-m          &ANDICAM &CTIO  &0.369    &6\arcmin x6\arcmin     & 1.61 & 1493\tablenotemark{b}  \\
\enddata
\tablenotetext{a}{Usually formatted to smaller region.}
\tablenotetext{b}{These include back-to-back exposures, so independent measurements are half of these.}
\end{deluxetable}

\begin{deluxetable}{l c c c c c c}
\tablecaption{\label{tab:clusters} Photometric Monitoring of OB Associations\tablenotemark{a}}
\tablehead{
\colhead{Season}
&\colhead{NGC 346}
&\colhead{NGC 602c}
&\colhead{NGC 1910}
&\colhead{NGC 2044}
&\colhead{NGC 2074}
&\colhead{R136}
}
\startdata
2003/2004      &       1                  &       1               &       1            &       \nodata    &  \nodata  &  \nodata \\
2004/2005      &       1                 &        1                &      1          &         \nodata   &   \nodata   &   1        \\
2005/2006      &       2                 &        2               &       2           &        2              &      2         &   1,2     \\
2006/2007      &       2                 &        2               &       2           &        2              &      2         &    2     \\
2007/2008      &       3                 &   \nodata         &       3           &        3              &      3         &    3       \\
2008/2009      &   \nodata           &   \nodata         &  \nodata       &   \nodata        &  \nodata    & \nodata   \\
2009/2010      &      3                  &   \nodata            &     3            &         3             &       3        &     3      \\ 
2010/2011      &    \nodata            &    \nodata          &   1,3           &          1,3         &     1,3      &     1,3   \\
\enddata
\tablenotetext{a}{Telescope/instrument indices from Table~\ref{tab:telescopes}.}
\end{deluxetable}

\begin{deluxetable}{l c c c l l l l}
\rotate
\setlength{\tabcolsep}{0.05in} 
\tabletypesize{\tiny}
\tablecaption{\label{tab:winners} Identified Periodic Variables}
\tablehead{
\colhead{Star}
&\colhead{Ref.\tablenotemark{a}}
&\colhead{$\alpha_{\rm 2000}$}
&\colhead{$\delta_{\rm 2000}$}
&\colhead{Sp.\ Type}
&\colhead{Period (d)}
&\colhead{Status}
&\colhead{Comments}
}
\startdata
\cutinhead{SMC NGC 346}                   \\
NGC 346 MPG 088  & 1 & 00 58 28.41 & -72 12 34.3 & B2 III &  3.61520 &  Dropped---insufficient spectroscopy & OGLEJ005828.46-721234.1 P=3.615130\tablenotemark{b} \\
NGC 346 MPG 342 & 1 & 00 59 00.01 & -72 10 37.8 & O5 + O7&  2.35482  & Dropped---triple lines \\
NGC 346 MPG 372 & 1 & 00 59 01.88 & -72 10 21.3 & B 0.2 V\tablenotemark{c} & 1.19620   &Dropped---not double lined\\                     
NGC 346 MPG 644 & 1 & 00 59 14.95 & -72 11 35.0 & B1 V &      3.10434&   Dropped---insufficient spectroscopy  & OGLE SMC-SC8 160725   P=3.104360\tablenotemark{b}  \\
SMC 042367  &  2 & 00 58 07.49 & -72 15 48.0 & B1.5 V  &   1.41727  & Dropped---not double lined & OGLE SMC-SC8 52827 P=1.41713\tablenotemark{b,d}  \\
SMC 044626 &  2 & 00 58 51.29 & -72 05 10.3 & B1-2 V   &  1.52297   &Dropped---insufficient spectroscopy                & NGC 346 ELS 38  \\
SMC 046456 &  2 & 00 59 30.35 & -72 09 09.4 & B1-2     &   4.20415  & Dropped---not double lined                          & NGC 346 MPG 782 \\
SMC 047161 &  2 & 00 59 46.62 & -72 05 32.3 & B1 V\tablenotemark{e}      &    3.75305  &  Dropped---not double lined   &  NGC 346 ELS 35  \\
\cutinhead{NGC 604c}
\nodata  & \nodata & \nodata & \nodata & \nodata & \nodata \\
\cutinhead{LMC NGC 1910}   \\
LH41-51             & 3  & 05 18 13.81 & -69 12 01.2 & O9.7 I(f) + O9 V & 6.838932   &  Paper IV  \\
LH41-52             & 3  & 05 18 09.81 &-69 11 36.0 &B+B  & 6.55180 & Dropped--blended double lines\\
LH41-55             & 3  & 05 18 25.75 & -69 12 12.8  &B       & 1.06203 & Dropped--not double lined \\
LH41-58             & 3  & 05 18 23.75  & -69 11 01.5 & O8 Iabf + O7 V   & 4.738649 & Paper IV 
& LMC 110852 \\
\cutinhead{LMC NGC 2044}
BAT99-77           & 4 & 05 35 59.02 &-69 11 47.8 &WN7+O3 &    3.003084   &  Paper III                                        &HD 269828\\
Sk -69$^\circ$ 212 & 5 & 05 36 06.52 & -69 11 47.4 & O5 III &   2.39929     &  Dropped---not double lined      \\
LMC 161594      & 2  & 05 34 59.88 & -69 16 52.9 &Early B      &  1.64093 & Dropped---blended double lines  & MACHO 82.8888.31 P=1.64085\tablenotemark{f}\\
LMC 163970      & 2  & 05 35 55.21 & -69 08 54.9 & O9 V       &    3.91221     &  Dropped---blended double lines& MACHO 82.9011.7 P=3.91248\tablenotemark{f}  \\
LMC 164325      & 2  & 05 36 02.64 & -69 18 21.4& O8 V + O9 V & 3.12120  & Dropped---blended double lines  & MACHO 82.9008.11 P=3.12011\tablenotemark{f}\\
LMC 164717      & 2  & 05 36 10.61 & -69 23 09.5& B star              & 4.77805  &  Dropped---not double lined & MACHO 82.9128.35 2P=4.77738\tablenotemark{g}\\
LMC 165507      & 2  & 05 36 26.14 & -69 19 28.7 &M star            & 1.98748  &  Dropped---not early-type star \\
LMC 165792      & 2 & 05 36 32.37 & -69 20 51.3 &O9.5 I + dwarf & 8.10475 & Dropped---blended double  lines  & MACHO 82.9129.7 P=8.09837\tablenotemark{f} \\
LMC 165885     & 2  & 05 36 34.54 &-69 15 21.4& B0 V & 1.36110 &  Dropped---not double lined & MACHO 82.9130.20 P=1.36097\tablenotemark{f}\\
LMC 166580     & 2  & 05 36 48.82 & -69 16 59.2 & B      &  3.85386 &  Dropped---blended double lines & MACHO 82.9130.25 P=3.75353\tablenotemark{h}\\
ST2-28              & 6  & 05 35 50.94& -69 12 00.4& O7 V + O8 V   & 2.762456 &  This paper & MACHO 82.9010.36 P=2.76245\tablenotemark{f,i} \\
ST2-42              & 6 & 05 36 00.01&  -69 12 08.6& O8 V + B0 III    & 4.363011 & Paper IV & LMC 164202  \\
ST2-63              & 6  &05 36 11.14& -69 11 01.4 & O9                   & 2.43522   & Dropped---not double lined \\
LMC 169415     & 2  & 05 37 51.11 & -69 10 59.9 & O6 V + O9.5 V & 1.777586 &  Paper IV \\
\cutinhead{LMC NGC 2074}
LMC 171676    &  2 & 05 38 44.88 & -69 24 40.1 & ?        & 1.25973 & Dropped---image multiple & LHA 120-N 158B\\
LMC 172231    &  2  & 05 38 58.23 & -69 30 11.4 &  O9 V + O9.5 V & 3.225414 & This Paper  & [ST92] 5-67 \\
LMC 173064    &  2 & 05 39 18.68 & -69 28 46.0 &B0 III & 2.50972 & Dropped---not double lined & MACHO 81.9611.8 P=2.50988\tablenotemark{f}\\
LMC 173712     & 2 & 05 39 39.49 & -69 29 05.6 & O9 III + O9 V & 4.912573 & Paper IV & MACHO 81.9611.4 P=4.91250\tablenotemark{h,j}\\
W4-6                 & 7 &05 39 54.95  &-69 24 10.2  &O7 III & 1.71603 & Dropped---not double lined & LMC 174056 \\
Br 95                 & 8 & 05 40 07.64 & -69 24 31.9& WN3 + O7dbl & 1.552931 & Paper III & HD 269956\\
LMC 174250    &  2 & 05 40 04.15 & -69 27 07.5 & B0.7 IV + B1 V & 1.483261 & Paper IV & MACHO 81.9732.32 P=1.48327\tablenotemark{f} \\
LMC 174491    &  2 &05 40 14.94 & -69 27 58.8 & B1 III + B1 IV-V  & 2.04390 & Paper IV \\
LMC 174510    & 2 &05 40 15.68 &-69 30 28.4& B2 V + B2 V   & 3.02517     & Dropped---blended double lines &MACHO 76.9731.1562 P=3.02596\tablenotemark{f}\\

LMC 174734    & 2 &05 40 25.14 & -69 24 32.7 & B0.5 IV + B1 V & 2.319493 & Paper IV \\
\cutinhead{LMC R136}
LMC 168477  &    2      & 05 37 30.85 & -69 05 17.5   & O8.5 V + O9 V & 2.333363 &Paper IV \\
LMC 169782   & 2   &05 37 59.57 & -69 09 01.4     & O4 V + O5 V  & 1.855280 & Paper II \\
LMC 171520   & 2   & 05 38 41.26 & -69 02 58.4    & O6 V + O6.5 V & 2.875275 & Paper II \\
R136-015        &  9 & 05 38 43.18 & -69 05 46.9& O3 If*      & 4.69880 & Dropped---too crowded \\
R136-038        & 9, 10 & 05 38 42.10 & -69 06 07.9 & O3 V+ O6 V & 3.38845 & Dropped---too crowded plus previous orbit & P=3.39\tablenotemark{k} \\
Mel 50             & 11  &05 38 38.56 & -69 06 21.9    & O9 I & 6.89228 & Dropped---blended double lines \\
$[$P93$]$ 467      & 12   &05 38 35.63 &-69 06 06.7    & O8.5 V              & 4.27588   & Dropped---not double lined, nebular contam.\\
$[$P93$]$ 661      & 12   &05 38 38.81 & -69 06 13.2&    O4 V                 & 1.58952   & Dropped---not double lined \\
$[$P93$]$ 729      & 12   &   05 38 39.86 & -69 06 08.7 & O6-7 dbl           & 1.58053   & Dropped---too crowded\\
$[$P93$]$ 921      & 12      &  05 38 42.18 & -69 05 45.5 &  O5 III(f) + O5 V & 2.389321 & Paper II \\
$[$P93$]$ 1024    & 12      &05 38 43.21 &-69 04 13.1& O9 V                 & 4.15991   & Dropped---not double lined & Mel 22\tablenotemark{l}\\
RR Dor           & \nodata   & 05 39 53.32 & -69 15 34.7 & O9.5 III + B0 III & 2.149363  & Paper IV \\
\enddata
\tablenotetext{a}{References for object identification: 
1--Massey et al.\ 1989; 
2--Massey 2002; 
3--Lucke 1972; Massey et al.\ 2000;
4--Breysacher et al.\ 1999;
5--Sanduleak 1970;
6--Shield \& Testor 1992; Massey et al.\ 2000;
7--Westerlund 1961;
8--Breysacher 1981;
9--Massey \& Hunter 1998;
10--Massey et al.\ 2002;
11--Melnick 1985; 
12--Parker 1993}
\tablenotetext{b}{Wyrzykowski et al.\ 2004}
\tablenotetext{c}{Spectral type from Evans et al.\ 2006}
\tablenotetext{d}{Also MACHO 207.16490.12 P= 1.41713, from Faccioli et al.\ 2007 .}
\tablenotetext{e}{Spectral type from Hunter et al.\ 2008}
\tablenotetext{f}{Derekas et al.\ 2007}
\tablenotetext{g}{Alcock et al.\ 1996 quoted in Vizier II/247}
\tablenotetext{h}{Faccioli et al.\ 2007}
\tablenotetext{i}{Also LMC 163763}
\tablenotetext{j}{Position of MACHO source differs by 1\farcs7 with ours.}
\tablenotetext{k}{Massey et al.\ 1998}
\tablenotetext{l}{Melnick 1985}
\end{deluxetable}

\begin{deluxetable}{l c c c c c c c c}
\setlength{\tabcolsep}{0.05in} 
\tablecaption{\label{tab:lcsummary} Summary of Light Curve Data}
\tablehead{
\colhead{Star}
&\multicolumn{2}{c}{\#Data Frames}
&&\multicolumn{2}{c}{Time Coverage}
&\colhead{Period\tablenotemark{a}}
&\colhead{$\theta$\tablenotemark{b}}
&\colhead{T\tablenotemark{c}} \\ \cline{2-3} \cline{5-6} 
&\colhead{Raw}
&\colhead{Indep.\tablenotemark{d}}
&&\colhead{First}
&\colhead{Last}
&\colhead{(days)}
}
\startdata                           
LMC 172231& 544 & 285 && 2453591.9 & 2455618.7 &   3.225414 (30)  & 0.07  &  2453591.469 \\
ST2-28         & 552 & 279 && 2453591.9  & 2455605.7 &  2.762456 (10)  & 0.07  &  2453590.217 \\
\enddata
\tablenotetext{a}{The value in parentheses denotes the uncertainty in the last digits of the period.}
\tablenotetext{b}{The parameter $\theta$ is a measure of the reliability of the period derived from
the Lafler \& Kinman (1965) method, ranging from 0 to 1, with 0 being the most reliable, and 1 having
no better significance than random.}
\tablenotetext{c}{HJD of primary eclipse.}
\tablenotetext{d}{The number of independent measures, combining the back-to-back photometry into
a single value.}
\end{deluxetable}

\begin{deluxetable}{c c c c c}
\tablecaption{\label{tab:LMC172231phot} LMC 172231 Photometry}
\tablehead{
\colhead{HJD}
&\colhead{$V$}
&\colhead{$\sigma_V$}
&\colhead{Telescope\tablenotemark{a}} 
&\colhead{Phase\tablenotemark{b}}
}
\startdata
2453606.819&14.040&0.015&2&0.759\\
2453599.876&14.076&0.015&2&0.607\\
2453646.710&14.111&0.019&2&0.127\\
2453679.782&14.090&0.004&2&0.380\\
2453987.845&14.038&0.008&2&0.892\\
2453691.714&14.124&0.015&2&0.080\\
2454061.635&14.029&0.010&2&0.769\\
2453968.828&14.585&0.013&2&0.995\\
2453682.720&14.010&0.004&2&0.291\\
2453643.898&14.079&0.011&2&0.255\\
\enddata
\tablecomments{Table \ref{tab:LMC172231phot} is published in its entirety in the electronic edition.
A portion is shown here for guidance regarding its form and content.}
\tablenotetext{a}{Telescope and instrument combination index as defined in Table~\ref{tab:telescopes}.}
\tablenotetext{b}{Based upon $P=3.225414$ and $T=2453591.469$.}
\end{deluxetable}

\begin{deluxetable}{l r r r r r r}
\tablecaption{\label{tab:lmc172231vel}Radial Velocities LMC 172231}
\tablehead{
\colhead{HJD}
&\colhead{Phase\tablenotemark{a}}
&\colhead{$v_1$}
&\colhead{$\sigma_{v_{\rm pri}}$}
&\colhead{$v_2$}
&\colhead{$\sigma_{v_{\rm sec}}$}
&\colhead{\#\tablenotemark{b}}
}
\startdata
2454809.842 & 0.742 &   455.9 & \nodata &    30.1 & \nodata &   1 \\
2454811.595 & 0.285 &    47.0 &   2.4 &   502.3 &   7.4 &   6 \\
2454814.710 & 0.251 &    44.0 &   2.1 &   512.0 &   8.8 &   4 \\
2454877.619 & 0.755 &   520.1 &   3.1 &    48.4 &   4.0 &  10 \\
2455138.693 & 0.698 &   452.8 &   3.2 &     2.7 &   3.9 &   3 \\
2455143.685 & 0.245 &    29.1 &   6.0 &   479.0 &   4.8 &  10 \\
2455143.825 & 0.289 &    15.6 &   1.9 &   487.1 &   5.3 &  10 \\
2455248.576 & 0.766 &   519.4 &   5.5 &    66.6 &   2.3 &   8 \\
2455248.599 & 0.773 &   523.7 &   4.0 &    62.4 &   3.8 &  10 \\
2455249.734 & 0.125 &   137.5 &   6.1 &   475.1 &   4.6 &  10 \\
2455527.606 & 0.276 &    30.7 &   3.1 &   485.1 &   5.3 &  17 \\
\enddata
\tablenotetext{a}{Based upon $P=3.225414$ and $T=2453591.469$.}
\tablenotetext{b}{Number of spectral lines used.}
\end{deluxetable}

\begin{deluxetable}{l c c}
\tablecaption{\label{tab:lmc172231physical}Orbital and Physical Parameters LMC 172231}
\tablehead{
\colhead{Parameter}
&\colhead{Primary}
&\colhead{Secondary}
}
\startdata
Period (days) & \multicolumn{2}{c}{$3.225414\pm0.000030$} \\
Time Primary is eclipsed $T$ & \multicolumn{2}{c}{2453591.469}\\
Spectral Types & O9 V & O9.5 V \\
Orbital Semi-amplitude $K$ (km s$^{-1})$& $234.9\pm 1.7$ & $233.2\pm 1.7$ \\
Center of Mass Velocity $\gamma$ (km s$^{-1})$ & \multicolumn{2}{c}{$271.5\pm 1.2$}\\
Residuals from fit $\sigma$ (km s$^{-1}$)& 22.6 & 27.9 \\
$m \sin^3i$ ($M_\odot$) & $17.1\pm 0.3$ & $17.2\pm0.3$\\
Orbital inclination $i$ & \multicolumn{2}{c}{$83\fdg0 \pm 0\fdg5$} \\
Effective temperature $T_{\rm eff}$ & 34,000$\pm$1,000 & 33,000$\pm$1,000\\
Stellar radius $R$ ($R_\odot$) &$7.0\pm0.3$ & $6.5\pm 0.3$ \\
Surface gravity from analysis $\log g$ [cgs] & $3.99\pm0.04$ & $4.06\pm0.04$ \\
Absolute visual magnitude $M_V$ (observed) & \multicolumn{2}{c}{$-5.0\pm0.3$} \\
Absolute visual magnitude $M_V$ (total, spherical model) & \multicolumn{2}{c}{$-4.55\pm0.05$} \\
Absolute visual magnitude $M_V$ (total, tidal model) & \multicolumn{2}{c}{$-4.72\pm0.08$} \\
Visible light flux ratio $F_{V_2}/F_{V_1}$ & \multicolumn{2}{c}{$0.80\pm0.03$} \\
Absolute visual magnitude $M_V$ (individual, spherical model)&  $-3.89\pm0.07$ & $-3.69\pm0.07$\\
Absolute visual magnitude $M_V$ (individual, tidal model) & $-4.08\pm0.08$ & $-3.84\pm0.08$\\
Bolometric luminosity $\log L/L_\odot$ (spherical model)& $4.77\pm0.06$ & $4.65\pm0.07$ \\
Bolometric luminosity $\log L/L_\odot$ (tidal model) &        $4.84\pm0.05$ & $4.71\pm0.05$ \\
age\tablenotemark{a} $t$ (Myr)                 & $5.2\pm0.8$ & $5.5\pm0.5$ \\
mass $m$ ($M_\odot$) & $17.5\pm0.3$& $17.6\pm0.3$ \\
\enddata
\tablenotetext{a}{From evolutionary models; see text.}
\end{deluxetable}

\begin{deluxetable}{l c c c c c}
\tablecaption{\label{tab:wilson}Comparison of Orbit Solution with Wilson's Method}
\tablehead{
\colhead{Star}
&\multicolumn{2}{c}{Orbit Solution}
&
&\multicolumn{2}{c}{Wilson's Method} \\ \cline{2-3} \cline{5-6}
&\colhead{$\gamma$}
&\colhead{$K_{\rm pri}/K_{\rm sec}$}
&
&\colhead{$\gamma$}
&\colhead{$K_{\rm pri}/K_{\rm sec}$}
}
\startdata
LMC 172231    & $271.5\pm1.2$ &  $1.01\pm0.01$  && $269.5\pm 1.9$ &   $1.03\pm0.01$\\
ST2-28             & $273.8\pm1.2$ &  $0.81\pm0.01$  &&   $273.8\pm2.1$ &   $0.81\pm0.01$ \\
\enddata
\end{deluxetable}

\begin{deluxetable}{c c c c c}
\tablecaption{\label{tab:ST2-28phot} ST2-28 Photometry}
\tablehead{
\colhead{HJD}
&\colhead{$V$}
&\colhead{$\sigma_V$}
&\colhead{Telescope\tablenotemark{a}} 
&\colhead{Phase\tablenotemark{b}}
}
\startdata
2453679.716&14.136&0.009&2&0.398\\
2453679.734&14.188&0.003&2&0.405\\
2453691.708&14.245&0.011&2&0.739\\
2453599.872&14.592&0.013&2&0.495\\
2453991.844&14.198&0.025&2&0.388\\
2453600.907&14.210&0.025&2&0.870\\
2453682.711&14.530&0.007&2&0.482\\
2453968.825&14.355&0.010&2&0.055\\
2454061.630&14.153&0.008&2&0.650\\
2454080.670&14.398&0.011&2&0.542\\
\enddata
\tablecomments{Table \ref{tab:ST2-28phot} is published in its entirety in the electronic edition.
A portion is shown here for guidance regarding its form and content.}
\tablenotetext{a}{Telescope and instrument combination index as defined in Table~\ref{tab:telescopes}.}
\tablenotetext{b}{Based upon $P=2.762456$ and $T=2453590.217$.}
\end{deluxetable}

\begin{deluxetable}{l r r r r r r r}
\tablecaption{\label{tab:ST228vel}Radial Velocities ST2-28 (He II Lines Only)}
\tablehead{
\colhead{HJD}
&\colhead{Phase\tablenotemark{a}}
&\colhead{$v_1$}
&\colhead{$\sigma_{v_{\rm pri}}$}
&\colhead{$v_2$}
&\colhead{$\sigma_{v_{\rm sec}}$}
&\colhead{\#\tablenotemark{b}}
}
\startdata
2454810.665 & 0.798 &   516.7 &   1.5 &    -0.9 &   1.9 &   3 \\
2454811.801 & 0.209 &    45.1 &   6.1 &   554.4 &   3.1 &   2 \\
2454813.547 & 0.842 &   481.4 &   4.0 &    10.4 &   8.4 &   2 \\
2454814.557 & 0.207 &    38.8 &   0.8 &   539.8 &   9.0 &   2 \\
2454814.683 & 0.253 &    41.5 &   7.2 &   568.5 &   2.6 &   3 \\
2454814.831 & 0.306 &    42.7 &  10.0 &   560.5 &  13.6 &   2 \\
2454876.697 & 0.702 &   499.0 &   9.8 &   -14.9 &   5.8 &   3 \\
2455136.603 & 0.787 &   500.5 &  \nodata &   -17.3 &  \nodata &   1 \\
2455137.798 & 0.219 &    44.3 &   0.2 &   571.7 &  11.7 &   2 \\
2455140.732 & 0.281 &    38.6 &   4.7 &   568.1 &   8.0 &   4 \\
2455141.794 & 0.666 &   476.5 &   5.9 &    14.0 &  10.8 &   4 \\
2455249.601 & 0.692 &   499.6 &   4.3 &   -13.0 &   4.2 &   3 \\
2455249.723 & 0.736 &   511.7 &   5.8 &   -17.0 &   8.1 &   4 \\
2455528.681 & 0.718 &   508.6 &   2.7 &   -23.2 &   8.9 &   4 \\
\enddata
\tablenotetext{a}{Based upon $P=2.762456$ and $T=2453590.217$.}
\tablenotetext{b}{Number of spectral lines used.}
\end{deluxetable}

\begin{deluxetable}{l c c}
\tablecaption{\label{tab:st228physical}Orbital and Physical Parameters ST2-28}
\tablehead{
\colhead{Parameter}
&\colhead{Primary}
&\colhead{Secondary}
}
\startdata
Period (days) & \multicolumn{2}{c}{$2.762456\pm0.000010$} \\
Time Primary is eclipsed $T$ & \multicolumn{2}{c}{2453590.217}\\
Spectral Types & O7~V & O8~V \\
Orbital Semi-amplitude $K$ (km s$^{-1})$& $241.0\pm 1.6$ & $297.0\pm 1.9$ \\
Center of Mass Velocity $\gamma$ (km s$^{-1})$ & \multicolumn{2}{c}{$273.8\pm 1.2$}\\
Residuals from fit $\sigma$ (km s$^{-1}$)& 6.0 & 8.6 \\
$m \sin^3i$ ($M_\odot$) & $24.6\pm 0.3$ & $20.0\pm0.3$\\
Orbital inclination $i$ & \multicolumn{2}{c}{$85\fdg0 \pm 2\fdg0$} \\
Effective temperature $T_{\rm eff}$ & 38,500$\pm$1,000 & 36,500$\pm$1,000\\
Stellar radius $R$ ($R_\odot$) &$9.5\pm0.3$ & $7.7\pm 0.3$ \\
Surface gravity from analysis $\log g$ [cgs]& $3.88\pm0.03$ & $3.97\pm0.03$ \\
Absolute visual magnitude $M_V$ (observed) & \multicolumn{2}{c}{$-5.5\pm0.3$} \\
Absolute visual magnitude $M_V$ (total, spherical model) & \multicolumn{2}{c}{$-5.49\pm0.10$} \\
Absolute visual magnitude $M_V$ (total, tidal model) & \multicolumn{2}{c}{$-5.68\pm0.08$} \\
Absolute visual magnitude $M_V$ (binary only, spherical model)& \multicolumn{2}{c}{$-5.24\pm0.04$}\\
Absolute visual magnitude $M_V$ (binary only, tidal model) & \multicolumn{2}{c}{$-5.44\pm0.08$}\\
Visible light flux ratio $F_{V_2}/F_{V_1}$ & \multicolumn{2}{c}{$0.62\pm0.02$} \\
Third light component $F_{V_3}/(F_{V_1}+F_{V_2})$ & \multicolumn{2}{c}{$0.25\pm0.10$} \\
Absolute visual magnitude $M_V$ (individual, spherical model)&  $-4.72\pm0.05$ & $-4.20\pm0.06$\\
Absolute visual magnitude $M_V$ (individual, tidal model)& $-4.92\pm0.08$ & $-4.40\pm0.08$\\
Bolometric luminosity (spherical) $\log L/L_\odot$ & $5.25\pm0.05$ & $4.97\pm0.06$ \\
Bolometric luminosity (tidal model)  $\log L/L_\odot$ & $5.33\pm0.04$ & $5.05\pm0.05$\\
age\tablenotemark{a} $t$ (Myr)                 & $3.6\pm0.5$ & $4.0\pm0.5$ \\
mass $m$ ($M_\odot$) & $24.9\pm 0.4$& $20.2\pm0.4$ \\
\enddata
\tablenotetext{a}{From evolutionary models; see text.}
\end{deluxetable}

\begin{deluxetable}{l c c c}
\tablecaption{\label{tab:comparisons} Comparison with Evolutionary Models}
\tablehead{
\colhead{Parameter}
&\colhead{Observed}
&\colhead{Models}
&\colhead{Difference}
}
\startdata
\cutinhead{LMC 172231, $t=5.3$ Myr}
Mass$_{\rm pri}$  & $17.5\pm0.3M_\odot$ &   $19.6\pm0.8M_\odot$&  $-2.1\pm1.1M_\odot$\\
Mass$_{\rm sec}$  & $17.6\pm0.3M_\odot$ &   $18.0\pm0.8M_\odot$&  $-0.4\pm1.1M_\odot$\\
$(\log L/L_\odot)_{\rm pri}$ &$4.77\pm0.06$& $4.61\pm0.02$ &$0.16\pm0.08$\\
$(\log L/L_\odot)_{\rm sec}$ &$4.65\pm0.07$& $4.62\pm0.02$ &$0.03\pm0.09$\\
\cutinhead{ST2-28, $t=3.8$ Myr}
Mass$_{\rm pri}$  &$24.9\pm0.4M_\odot$    &$29.8\pm1.2M_\odot$    & $-4.9\pm1.6M_\odot$\\
Mass$_{\rm sec}$  &$20.2\pm0.4M_\odot$    &$24.0\pm1.1M_\odot$    & $-3.8\pm1.5M_\odot$\\
$(\log L/L_\odot)_1$  &$5.25\pm0.05$&$5.02\pm0.02$& $0.23\pm0.07$\\
$(\log L/L_\odot)_2$  &$4.97\pm0.06$&$4.75\pm0.03$& $0.22\pm0.09$\\
\cutinhead{LH54-425, $t=2.0$ Myr}
Mass$_{\rm pri}$  & $47\pm2M_\odot$ & $50.2\pm2.1M_\odot$  & $-3.2\pm4.1$ \\
Mass$_{\rm sec}$  & $28\pm1M_\odot$ & $32.2\pm1.1M_\odot$  & $-4.2\pm2.1$ \\
$(\log L/L_\odot)_{\rm pri}$ &$5.68\pm0.04$& $5.61\pm0.05$ & $0.07\pm0.09$ \\
$(\log L/L_\odot)_{\rm sec}$ &$5.22\pm0.04$& $5.06\pm0.03$ & $0.16\pm0.07$ \\
\enddata
\end{deluxetable}


\clearpage

\begin{references}

\reference {} Alcock et al.\ 1996, ApJ, 470, 583

\reference {} Alcock et al.\ 1997, AJ, 114, 326

\reference {} Asplund, M., Grevesse, N., Sauval, A. J., \& Scott, P. 2009, ARA\&A, 47, 481

\reference {} Bagnuolo, W. G., \& Gies, D. R. 1991, ApJ, 376, 266

\reference {} Bagnuolo, W. G., \& Gies, D. R. 1992, in Complementary Approaches to Double and Multiple Star Research, ASP Conf.\ Ser.\ 32, ed.\ H. A. McAlister and W. I. Hartkopf
(San Francisco: ASP), 140

\reference {} Batten, A. H. 1973, {\it Binary and Multiple Systems of Stars}, (Oxford: Pergamon Press), 17

\reference {} Bevinginton, P.  1969, {\it Data Reduction and Error Analysis for the Physical Sciences} (NY: McGraw-Hill),
110

\reference {} Binnendijk, L. 1960, {\it Properties of Double Stars} (Oxford: Oxford University Press), 69

\reference{} Bohannan, B., \& Conti, P. S. 1976, ApJ, 204, 797

\reference {} Bonanos, A. Z. 2009, ApJ, 691, 407

\reference {} Bonanos, A. Z., et al.\ 2006, ApJ, 652, 313

\reference {} Bonanos, A. Z., Castro, N., Macri, L. M., \& Kudritzki, R.-P. 2011, ApJ, 739, 9

\reference {} Breysacher, J. 1981, A\&AS, 43, 203

\reference {} Breysacher, J. Azzopardi, M., \& Testor, G. 1999, A\&AS, 137, 117

\reference {} Burkholder, V., Massey, P., Morrell, N. 1997, ApJ, 490, 328

\reference {} Claret, A. 2000, A\&A, 3636, 1081

\reference {} Coluzzi, R. 1993, BICDS, 43, 7

\reference {} Conti, P. S., Ebbets, D., Massey, P., \& Niemela, V. S. 1980, ApJ, 238, 184

\reference {} Conti, P. S., \& Frost, S. A. 1977, ApJ, 212, 728

\reference {} Conti, P. S., Leep, E. M., \& Lorre, J. J. 1977, ApJ, 214, 759

\reference{} Conti, P. S. \& Walborn, N. R. 1976, ApJ, 207, 502

\reference {} Derekas, A., Kiss, L. L., \& Bedding, T. R. 2007, ApJ, 663, 249

\reference {} Eddington, A. S. 1924, MNRAS, 84, 208

\reference {} Ekstr\"{o}m, S. et al.\ 2011, A\&A, submitted

\reference {} Evans, C. J., Lennon, D. J., Smartt, S. J., \& Trundle, C. 2006, A\&A, 456, 623

\reference {} Faccioli, L., Alcock, C., Cook, K., Prochter, G. E., Protopapas, P., \& Pyphers, D. 2007, AJ, 134, 1963

\reference {} Fitzpatrick, E. L., et al.\ 2003, ApJ, 587, 685


\reference {} Gies, D. R. 2003, in {\it A Massive Star Odyssey, from Main Sequence to Supernova,
IAU 212}, ed.\ K. A. van der Hucht, A. Herrero, \& C. Esteban (San Francisico: ASP), 91

\reference {} Gies, D. R. 2004, in Spectroscopically and Spatially Resolving the Components of the Close Binary Stars, ASP Conf.\ Ser.\  318 (San Francisco: ASP), 61

\reference {} Gies, D. R. 2012, in Four Decades of Research on Massive Stars: A Scientific Meeting in Honour of Anthony
F. J. Moffat, (ASP Conf.\ Ser.\ vol.), ed.\ C. Robert, N. St-Louis, \& L. Drissen (San Francisco: ASP), in press

\reference {} Gonz\'{a}lez, J. F., Ostrov, P., Morrell, N., \& Minniti, D. 2005, ApJ, 624, 946

\reference {} Guinan, E. F., et al.\ 1998, ApJ, 509, L21

\reference {} Hamuy, M. et al.\ 2006, PASP, 118, 2

\reference {} Harries, T. J., Hilditch, R. W., \& Howarth, I. D. 2003, MNRAS, 339, 157

\reference {} Henry, T. J., Franz, O. G., Wasserman, L. H., Benedict, G. F., Shelus, P. J., Ianna, P. A.,
Kirkpatrick, J. D., \& McCarthy, Jr., D. W. 1999, ApJ 512, 864

\reference {} Herrero, A., Kudritzki, R. P., Vilchez, J. M., Butler, K., \& Haser, S. 1992, A\&A, 261, 209

\reference {} Hilditch, R. W., Howarth, I. d., Harries, T. J. 2005, MNRAS, 357, 304

\reference {} Hillier, D. J., \& Miller, D. 1998, ApJ, 496, 407

\reference {} Hubeny, I., \& Lanz, T. 1995, ApJ, 439, 875

\reference {} Hutchings, J. B. 1968a, MNRAS, 141, 219

\reference {} Hutchings, J. B. 1968b, MNRAS, 141, 329

\reference {} Hutchings, J. B. 1969, MNRAS, 144, 235

\reference {} Hutchings, J. B. 1970a, MNRAS, 147, 161

\reference {} Hutchings, J. B. 1970b, MNRAS, 147, 367

\reference {} Huchings, J. B. 1979, in {Mass Loss and Evolution of O-type Stars, IAU Symp.\ 83}, ed.\ P. S. Conti \& C. W. H. de Loore (Dordrecht: Reidel), 3


\reference {} Hunter, I., Lennon, D. J., Dufton, P. L., Trundle, C., Simon-Diaz, S., Smartt, S. J., Ryans, R. S. I., \& Evans, C. J. 2008, A\&A, 479, 541

\reference {} Lafler, J., \& Kinman, T. D. 1965, ApJS, 11, 216

\reference {} L\'{o}pez-Morales,  \& Clemens, J. C. 2004, PASP, 116, 22

\reference {} Lucke, P. B. 1972, PhD Thesis, Univ.\ Washington

\reference {} Lucke, P. B., \& Hodge, P. W. 1970, AJ, 75, 171

\reference {} Mace, G. N., Prato, L., Wasserman, L. H., Schaefer, G. H., Franz, O. G.,\
\& Simon, M. 2009, AJ, 137, 3487

\reference {} Maeder, A., \& Meynet, G. 2000, ARA\&A, 38, 143

\reference {} Martins, F., \& Plez, B. 2006, A\&A, 457, 637


\reference {} Massey, P. 2002, ApJS, 141, 81

\reference {} Massey, P., Bresolin, F., Kudritzki, R.-P., Puls, J., \& Pauldrach, A. W. A. 2004, ApJ, 608, 1001

\reference {} Massey, P., \& Conti, P. S. 1977, ApJ, 218, 431

\reference {} Massey, P., \& Hanson, M. M. 2012, in {\it Planets, Stars and Stellar Systems, Vol 2: Astronomical Techniques and Standards,} ed. H. Bond (New York: Springer), in press; arXiv:1010.5270

\reference {} Massey, P., \& Hunter, D. A. 1998, ApJ, 493, 180

\reference {} Massey, P.,  Olsen, K. A. G., Hodge, P. W.,  Jacoby, G. H.; McNeill, R. T.,  Smith, R. C.,  Strong, Shay B. 2007, AJ, 133, 2393

\reference {} Massey, P.,  Olsen, K. A. G.,  Hodge, P. W., Strong, S. B.; Jacoby, G. H.; Schlingman, W.; Smith, R. C. 2006, AJ, 131, 2478

\reference {} Massey, P., Parker, J. W., \& Garmany, C. D. 1989, AJ, 98, 1305

\reference {} Massey, P., Penny, L. R., \& Vukovich, J. 2002, ApJ, 565, 982

\reference {} Massey, P., Puls, J., Pauldrach, A. W. A., Bresolin, F., Kudritzki, R. P., \& Simon, T. 2005,
ApJ, 627, 477

\reference {} Massey, P., Waterhouse, E., DeGioia-Eastwood, K. 2000, AJ, 119, 2214

\reference {} Massey, P., Zangari, A. M., Morrell, N. I., Puls, J., DeGioia-Eastwood, K., Bresolin, F.,
\& Kudritzki, R.-P. 2009, ApJ, 692, 618

\reference {} Melnick, J. 1985, A\&A, 153, 235

\reference {} Meynet, G., \& Maeder, A. 2000, A\&A, 361, 101

\reference {} Mochnacki, S. W., \& Doughty, N. A. 1972, MNRAS, 156, 51

\reference {} Moore, C. E. 1972, {\it A Multiplet Table of Astrophysical Interest, Revised Edition} (Boulder: National Bureau of Standards)

\reference {} Morrell, N. I., Gonzalez, J. F., Ostrov, P. G., \& Minniti, D. 2007, in Massive Stars in
Interacting Binaries, ASP Conf.\ 367, ed. N. St-Louis \& A. F. J. Moffat (San Francisco: ASP), 77

\reference {} Morrell, N. I., Ostrov, P., Massey, P., \& Gamen, R. 2003, MNRAS, 341, 583

\reference {} Morrison, N. D., \& Conti, P. S. 1978, ApJ, 224, 558

\reference {} Morrison, N. D., \& Conti, P. S. 1980, ApJ, 239, 212

\reference {} Neugent, K., Massey, P., \& Hillier, D. J. 2010, BAAS, 42, 344

\reference {} Niemela, V. S., \& Morrell, N. I. 1986, ApJ, 310, 715

\reference {} Niemela, V. S., Morrell, N. I., Fern\'{a}ndez Laj\'{u}s, E., Barb\'{a}, R.,
Albacete Colombo, J. F., \& Orellana, M. 2006, MNRAS, 367, 1450

\reference {} North, P., Gauderon, R., Barblan, F., \& Royer, F. 2010, A\&A, 520, A74

ed.\ F. Bresolin, P. A. Crowther, \& J. Puls (Cambridge: Cambridge University Press), 
71

\reference {} Paczynski, B 1997, in The Extragalactic Distance Scale, ed. M.\ Livio, M., 
Donahue, \& N. Panagia (Cambridge: Cambridge University Press), 273

\reference {} Parker, J. W. 1993, AJ, 106, 560

\reference {} Penny, L. R., Gies, D. R., Wise, J. H., Stickland, D. J., \& Lloyd, C. 2002, ApJ, 565, 1050

\reference {} Penny, L. R., Ouzts, C., \& Gies, D. R. 2008, ApJ, 681, 554

\reference {} Petrie, R. M. 1962, Publ. Dom. Astrophys. Obs., 12, 111

\reference {} Prato, L. 2007, ApJ, 657, 338

\reference{} Press, W. H., Teukolsky, S. A., Vetterling, W. T., \& Flannery, B. P. 1997, {\it Numerical Recipes in Fortran 77}, (Cambridge: Cambridge Univ Press), 660

\reference {} Puls, J. et al.\ 1996, A\&A, 305, 171

\reference {} Puls, J., Urbaneja, M. A., Venero, R., Repolust, T, Springmann, U., Jokuthy, A., \& Mokiem, M. R. 2005, A\&A 435, 669

\reference {} Rauw, G., Crowther, P. A., Eenens, P. R. J., Manfroid, J., \& Vreux, J.-M. 2002, A\&A, 392, 563

\reference {} Rauw, G., Sana, H., Antokhin, I. I., Morrell, N. I., Niemela, V. S., Albacete Colombo, J. F., Gosset, E., \& Vreux, J.-M. 2001, MNRAS, 326, 1149

\reference {} Ribas, I., Jordi, C., \& Gim\'{e}nez, \'{A}. 2000, MNRAS, 318, L55

\reference {} Ribas, I., et al. 2005, ApJ, 635, L37

\reference {} Rosero, V., Prato, L., Wasserman, L. H., \& Rodgers, B. 2011, AJ, 141, 13

\reference {} Sanduleak, N. 1970, Contrib.\ Cerro Tololo  Inter-American Obs., No.\ 89

\reference {} Schaefer, G. H., Simon, M., Prato, L., \& Barman, T. 2008, AJ, 135, 1659

\reference {} Schild, H., \& Testor, G.  1992, A\&AS, 92, 729

\reference {} Schwarzenberg-Czerny, A. 1989, MNRAS, 241, 153

\reference {} Sota, A., et al.\  2011, ApJS, 193: 24

\reference {} Stickland, D. J., Lloyd, C., \& Penny, L. R. 1997, Obs, 117, 213

\reference {} Testor, G., \& Niemela, V. 1998, A\&AS, 130, 527

\reference {} Trundle, C., Dufton, P. L., Hunter, I., Evans, C. J., Lennon, D. J., Smartt, S. J., \& Ryans, R. S. I. 2007, A\&A, 471, 625

\reference {} Udalski, A., Soszynski, M., Kubiak, M., Petrzynski, G., Wozniak, P., \& Zebrun, K. 1998, AcA, 48, 563

\reference {} van den Bergh, S. 2000, {\it The Galaxies of the Local Group}, (Cambridge:
Cambridge Univ.\ Press)

\reference {} Vink, J. S., de Koter, A., \& Lamers, H. J. G. L. M. 2001, A\&A, 369, 574

\reference {} Vilardell, F., Ribas, I., Jordi, C., Fitzpatrick, E. L., \& Guinan, E. F.  2010, A\&A, 509, A70

\reference {} Williams, S. J. et al.\ 2008, ApJ, 682, 492

\reference {} Wilson, O. C. 1941, ApJ, 93, 29

\reference {} Westerlund, B. 1961, Uppsala Astron.\ Obs.\ Ann., 5, 1 

\reference {} Wyrzykowski, L., et al.\ 2004 Acta Astron. 54, 1

\end{references}
\end{document}